\documentclass[prb,reprint,superscriptaddress,showkeys,twocolumn,amsmath,amssymb,nofootinbib]{revtex4-2}

\pdfoutput=1

\usepackage[T1]{fontenc}
\usepackage[utf8]{inputenc} 
\usepackage{dsfont}
\usepackage{amsmath}
\usepackage{amssymb}
\usepackage{graphicx}
\usepackage{comment}
\usepackage[normalem]{ulem}
\usepackage{cancel}
\usepackage{booktabs} 
\usepackage{cellspace} 
\usepackage{soul}

\makeatletter

\usepackage{pifont}
\usepackage[dvipsnames]{xcolor}
\usepackage[section]{placeins}
\usepackage[pdfborder={0 0 0}]{hyperref}
\makeatother
\usepackage{babel}

\definecolor{dk}{rgb}{0.54, 0.17, 0.89} 

\begin{document}

\title{Interplay between evanescent scattering modes and finite dispersion in superconducting junctions}
\author{D. Kruti and R.-P. Riwar}
\affiliation{Peter Gr\"unberg Institute, Theoretical Nanoelectronics, J\"ulich Research Centre, D-52425 J\"ulich, Germany}



\begin{abstract}

\noindent
Superconducting junctions are essential building blocks for quantum hardware, and their fundamental behaviour remains a highly active research field. The behaviour of generic junctions is conveniently described by Beenakker's determinant formula, linking the subgap energy spectrum to the scattering matrix characterising the junction. In particular, the gap closing between bound and continuum states in short junctions follows from unitarity of the scattering matrix, and thus, from probability conservation. In this work, we critically reassess two assumptions: that scattering in short junctions is approximately energy-independent and dominated by planar channels. We argue that strongly energy-dependant scattering follows from finite dispersion of the conductor electrons even when they spend little time within the scattering region, and show that evanescent modes play a central role when cross-channel scattering is important. By generalising Beenakker's equation and performing a mapping to an effective Hamiltonian, we show that the gap closing is linked to a chiral symmetry. While finite energy dependence in the scattering breaks the chiral symmetry, we show two distinct mechanisms preserving the gap closing, each connected to new types of constraints on energy-dependant scattering matrices beyond unitarity. If the dispersive mode is planar, the gap closing is still preserved through a time-dependant probability conservation analysis of the scattering process. If the dispersive channel is evanescent, we derive a constraint which, notably, cannot follow from probability conservation. We thus demonstrate that Andreev physics reveal a much wider variety of properties of normal-metal scattering than commonly expected. We expect that our findings will have an impact on the dissipative behaviour of driven junctions, and offer a new perspective on fundamental properties of scattering matrices.

\end{abstract}

\maketitle

\section{Introduction}

\noindent Superconducting junctions are both of fundamental interest and a key circuit element for quantum hardware. In particular, weak links where one or many channels have a high transparency have been a focal point of research activity in the past decades both theoretically~\mbox{\cite{Beenakker_1991,Gorelik_1995,Wendin_1996,Shumeiko_1997,Johansson_1999, Cuevas_1999,Bezuglyi_2004,Zazunov_2005, Michelsen_2008,Michelsen_2010,Heck_2011, van_Heck_2012, van_Heck_2014, Bretheau_2014a, Olivares_2014, Zazunov_2014, Riwar_2015, Souto_2016, Pientka_2017, Sticlet_2017, Riwar_2016,  Riwar_2016b, Xie_2017, Meyer_2017, Eriksson_2017, Xie_2018, Houzet_2019, Istas_2018,Setiawan_2022, Davydova_2022,Pillet_2023, Maiellaro_2024, Ness_2022}} and experimentally~\mbox{\cite{Samuelsson_2004,Zgirski_2011, Bretheau_2013, Bretheau_2014b, Levenson_2014, Janvier_2015, Bretheau_2017, Tosi_2019, Park_2020,  Matute_2022,Koelzer_2021, Koelzer_2023, behner_2023}}. The ground state of a superconductor consists of a coherent many-body condensate, out of which single quasiparticles can only be excited above the gap energy~$\Delta$, where states exist as an incoherent, delocalised continuum. Yet, if superconductors are linked by a non-superconducting bridge, coherent single-particle excitations can occupy energy levels below the gap, in the form of bound quasiparticle states, so-called Andreev bound states (ABS). As Beenakker showed~\cite{Beenakker_1991}, the formation of these bound states can be directly linked to the scattering properties of the junction. This formalism allows for a convenient and straightforward description of a large variety of systems, such as atomic break junctions~\mbox{\cite{Muller_1992, Scheer_1997, Scheer_1998, Rodrigo_2006, DellaRocca_2007, Chauvin_2007, Zgirski_2011, Bretheau_2013b, Janvier_2015, Senkpiel_2020}}, multiterminal devices~\mbox{\cite{van_Heck_2012, van_Heck_2014, Riwar_2015, Riwar_2016, Xie_2017, Meyer_2017,Eriksson_2017, Xie_2018,Riwar_2016b, Houzet_2019, Koelzer_2021, Koelzer_2023}}, and heterostructures involving topological materials and superconductors~\mbox{\cite{Heck_2011, Xie_2019, Houzet_2019, Koelzer_2021, Kenawy_2022, Koelzer_2023, behner_2023,Kenawy_2024}}, to name a few.
On a more formal level, the phenomenology of Andreev bound states can be connected to fundamental scattering properties, such as probability conservation (guaranteeing a gap closing between sub- and supergap states in short junctions) or random matrix theory~\mbox{\cite{Stone_1991,Brouwer_1997,Pichard2001,Nazarov_Blanter_2009}}, impacting, e.g., the transport properties of chaotic cavities~\mbox{\cite{Beenakker_1992b,Agam_2000,Oberholzer_2002}}, the AC response of short diffusive junctions~\cite{Riwar_2015}, or the statistical probability to find Weyl points in multiterminal junctions~\cite{Riwar_2016}.

For very short junctions, it is the currently predominant consensus that the scattering matrix can be approximated to be energy-independent, as a finite energy-dependence is commonly related to the dwelling time of electrons within the weak link (i.e., a finite Thouless energy
\footnote{\label{note1}Note that the Thouless energy is often explicitly defined in the context of diffusive transport, whereas we here use the term to represent a general inverse dwelling time that also includes, e.g., ballistic propagation, scaling with the Fermi velocity~${ v_\mathrm{F} }$.}). Moreover, while evanescent modes (waves with complex wave vectors) have to some extent been included in recent works~\cite{Guigou_2010,Recher_2011, Mandal_2024}, these efforts were limited to single channel transport, where evanescent contributions to the wave function can only emerge due to special (non-quadratic) dispersion relations. There exists as of now no treatment for generic multichannel conductors, where both strong cross-channel coupling and Andreev reflection of purely evanescent channels are important. In our work, we revisit and extend the Beenakker formalism for generic multi-terminal and multi-channel junctions, with a particular focus on the ABS properties near $\Delta$. In particular, we show that small dwelling time is only a necessary, but not a sufficient condition for scattering to be energy-independent. Whenever the chemical potential is tuned across a change of the channel number, the scattering matrix changes strongly as a function of energy, even for infinitesimally small junctions. By exploiting a mapping to an effective subgap Hamiltonian similar to Ref.~\cite{van_Heck_2014}, we relate the energy-dependence to the breaking of a type of chiral symmetry, resulting (at least in a first approximation) in a detaching of the ABS spectrum from the quasiparticle continuum.
Importantly, we show that there are two distinct processes, which allow to effectively restore the gap closing, and thus the chiral symmetry -- revealing an intricate interplay between Andreev processes, virtual (evanescent) channels, and constraints on the scattering matrix beyond unitarity. 

When the dominant dispersion comes from a regular planar channel, the gap closing is restored due to finite normal backscattering at the superconductor-normal metal (SN) interface, in conjunction with an energy-dependant constraint on the scattering processes which is derived through time-resolved probability conservation considerations. If instead the dispersive channel is evanescent, the gap is closed through regular Andreev reflection -- but between evanescent modes. Here, we show that the resulting scattering matrix constraint can no longer be derived from probability conservation, due to the generic impossibility to normalise wave function ansatzes containing diverging evanescent modes. Overall, we thus show that Andreev bound state physics are capable to unravel nontrivial and unexpected constraints on energy-dependant scattering matrices. We explicitly demonstrate the validity of our general concepts and constraints using the example of a ballistic L-shaped junction, a geometry of high experimental relevance~\mbox{\cite{Koelzer_2021, Koelzer_2023, behner_2023}}.

Almost all existing analytic calculations describe the subgap transport at an ideal SN-interface within the so-called Andreev approximation~\cite{andreev_1964,Beenakker_1992}, where finite dispersion within the conductor electrons is neglected, resulting in pure Andreev reflection (ideal conversion from electron to hole and vice versa). While there is precedence for going beyond the Andreev approximation~\cite{Setiawan_2022} (where a finite dispersion leads to a nonzero normal reflection process for planar modes), we find that it is  crucial to \textit{additionally} include resulting strong energy-dependencies in the normal scattering matrix~$S$, since the resulting ABS spectrum near the gap is highly sensitive to the interplay of both SN-reflection and scattering. Moreover, the behaviour of evanescent modes at the SN-interface has not been considered at all in the existing literature. In this work, we show that finite dispersion renders the evanescent analogue of Andreev reflection nonzero, which is at the origin of the above summarised gap closing phenomenon, and resulting nontrivial scattering matrix constraints.

Overall, our work generalises the state of the art $S$-matrix formalism (Beenakker framework) as a powerful and convenient tool to analyse ABS physics for generic multiterminal and multichannel junctions. We expect that the here uncovered spectral properties close to the superconducting gap are of importance for future projects to improve our understanding of dissipative properties of driven weak links (building on, e.g., Refs.~\mbox{\cite{Zazunov_2005,Michelsen_2010}}). Finally, the explicit inclusion of evanescent modes in the scattering problem (and our demonstration of their importance) might provide new impulses within the field of random matrix theory~\cite{Stone_1991,Brouwer_1997,Pichard2001}: namely, as of now, the literature only describes the distribution of scattering coefficients for planar waves (e.g., the Dorotkov distribution for short diffusive wires~\cite{Nazarov_Blanter_2009}), whereas the distribution of evanescent coefficients has to the best of our knowledge not been studied. Moreover, the understanding of constraints relating scattering matrices at different energies could potentially even provide an alternative, simpler access to the prediction of probability distributions of scattering coefficients as a function of energy.

This work is organised as follows. Section~\ref{sec_status_quo} recapitulates the standard Beenakker formalism. In Section~\ref{sec:dispersion_vs_probability} the impact of finite dispersion on energy dependant scattering and ABS is explored. Section~\ref{sec:recipe} introduces the revised Beenakker framework with evanescent modes and energy dependant scattering fully included. In Section~\ref{sec_constraint_evanescent} evanescent scattering and Andreev reflection is related to a constraint on the scattering matrix. In Section~\ref{sec_L_junction} the revised framework is applied to a junction with non-trivial geometry and the effects on the ABS spectra are verified by explicit numerical calculation.

\section{Summary of standard Beenakker formalism and main results}\label{sec_status_quo}

%
\begin{figure}
\includegraphics[width=1\columnwidth,height=1\paperheight,keepaspectratio]{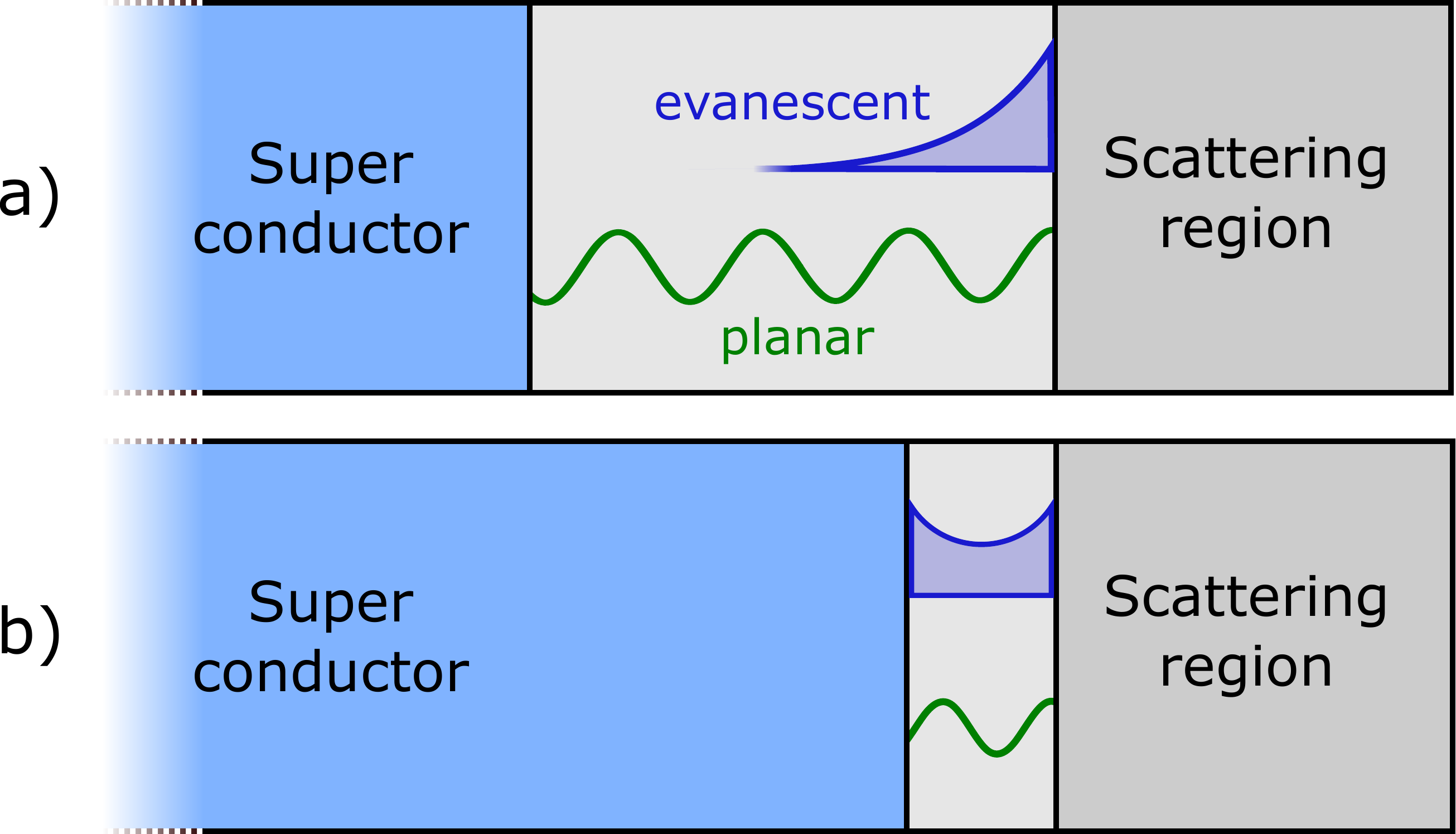}
\caption{\label{fig:junction_limits} 
Importance of evanescent modes for the bound states spectrum in short vs. long conductor arms. The figure depicts a conductor arm (light grey region) connecting the scattering centre (dark grey region) to one of the superconducting terminals (blue region). In the most general case, the solution of the central scattering problem involves (virtually occupied) evanescent modes due to energetically forbidden channels.  Panel~a): For long conductor arms, conventional Andreev reflection at the SN-interface and normal scattering at the centre result in planar standing wave configurations of electron and hole excitations (green). Potential evanescent modes exiting the scattering region (dark blue real exponential) decay before reaching the SN-interface. Panel~b):~For short conductor arms, evanescent modes penetrate the SN-interface such that evanescent Andreev reflection becomes important. In this situation, the standing waves in the conductor arms form superpositions consisting of both decaying and diverging real exponentials, in addition to planar modes.
}
\end{figure}
%

\noindent 
To provide an overview of the standard $S$-matrix formalism, let us consider a generic junction with $M$ superconducting terminals with a superconducting gap $\Delta$. Let them be joined by a central region consisting either of a normal metal or a semiconductor material (Fig.~\ref{fig:junction_limits} depicts a conductor arm connecting the scattering centre with one of the superconducting terminals). Electrons and holes are assumed to be subject to elastic scattering within the central region. Focusing on energies below~$\Delta$, electrons and holes undergo Andreev reflection~\cite{andreev_1964} at the interface to the superconductor (SN-interface). 

The elastic scattering process can be cast into the form of the boundary condition 
\begin{align}
\psi^{\text{out}}= & S\cdot\psi^{\text{in}},
\label{eq:Scattering}
\end{align}
where the vector $\psi^{\text{in}}$ ($\psi^{\mathrm{out}}$) denotes the amplitudes of modes incident
(outgoing) with respect to the scattering centre. The scattering matrix $S$ is block diagonal 
\begin{equation}
S=\left(\begin{array}{cc}
S_{\mathrm{e}} & 0\\
0 & S_{\mathrm{h}}
\end{array}\right)\text{ ,} \label{eq:scat_matrix}
\end{equation}
where $S_{\mathrm{e}}$ and $S_{\mathrm{h}}$ represent the scattering
matrices for electron and hole component, respectively. In the absence of many-body interactions, the electron and hole submatrices exhibit particle-hole symmetry
\begin{equation}
S_{\mathrm{h}}\left[E\right]=S_{\mathrm{e}}^{\ast}\left[-E\right]\text{ .}\label{eq:part_hole_symmetry}
\end{equation}
Note that here and throughout this article, $E$ is defined as the energy with respect to the chemical potential $\mu$.

The finite cross section area of the conductor arms joining at the centre of the normal region leads to the formation of channels. Assuming ballistic transport in these arms, their physics can be described by the dispersion relation
\begin{align}
    \xi_{k,n}=\epsilon_{k}+\epsilon_{n}-\mu , \label{eq:dispersion_NC}
\end{align}
where the transversal component gives rise to a discrete level contribution $\epsilon_{n}$ to the energy. For the specific example of a free particle in the (longitudinal) propagation direction 
\begin{align}
    \epsilon_{k}=\frac{k^{2}}{2m} ,
    \label{eq:quadratic_dispersion}
\end{align}
with the effective mass~$m$ and the longitudinal wave vector~$k$. Each channel, thus, has in general a different group velocity at energy~$E$, given as~$v_n=k_n(E)/m$ with~$k_n(E)$ being the solution of~${E=\xi_{k,n}}$. For~${E\ll \mu}$, these are the Fermi wave vectors~${k_{F,n}=k_n(0)}$, and Fermi velocities~$v_{F,n}$, respectively. When considering the single channel case in some of the sections below, we will simply drop the $n$-index for notational simplicity. The number of occupied channels depends on the chemical potential~$\mu$. We further note that we will occasionally refer to an effective chemical potential for each channel, defined as~$\mu_n=\epsilon_n-\mu$. Since energies are defined with respect to~$\mu$ as outlined above,~${ \mu_n=\xi_{k=0,n} }$ in fact corresponds to the minimum of the $n$th channel in the dispersion.

Overall, at a given energy $E$ (that is, for $E\ll \mu$, at a given chemical potential) there are a certain number of real-valued $k$ that solve $E=\xi_{k,n}$. Within this first (review) section, this total number of solutions shall refer to the number of available channels (the main part of this work is of course about including also imaginary solutions for $k$, as detailed further below). We can provide the scattering matrices with a substructure referring to these channels (and the corresponding terminals). The electron scattering matrix $S_{\mathrm{e}}$ (the hole matrix $S_{\mathrm{h}}$ has analogous structure) entails
all~${n_\mathrm{P}}$ available channels in the scattering process as
\begin{equation}\label{eq:Se}
S_{\mathrm{e}}=\left(\begin{array}{cccc}
S_{\mathrm{e}}^{1,1} & S_{\mathrm{e}}^{1,2} & \dots & \text{ }\\
S_{\mathrm{e}}^{2,1} & S_{\mathrm{e}}^{2,2} & \dots & \text{ }\\
\vdots & \vdots & \ddots & \text{ }\\
\text{ } & \text{ } & \text{ } & S_{\mathrm{e}}^{n_\mathrm{P},n_\mathrm{P}}
\end{array}\right)\text{ ,}  
\end{equation}
where $S_{\mathrm{e}}^{m,n}$ encodes scattering from the $n$th to
the $m$th channel. Each of these submatrices has the structure
\begin{equation}\label{eq:Se_sub}
S_{\mathrm{e}}^{m,n}=\left(\begin{array}{ccc}
R_{\mathrm{1,1}}^{m,n} & T_{\mathrm{1,2}}^{m,n} & \dots\\
T_{\mathrm{2,1}}^{m,n} & R_{\mathrm{2,2}}^{m,n} & \dots\\
\vdots & \vdots & \ddots
\end{array}\right)\text{ ,}
\end{equation}
where $T_{\mathrm{\mu,\nu}}^{m,n}$ denotes transmission from the $\nu$th to the $\mu$th terminal. Note that we assume here for simplicity that each terminal has the same number of channels (symmetric conductor arms). The formalism is however very easily generalised to the case where different terminals may have different channel numbers.

At this stage it is important to point out, that in the present work, we choose to stick to a slightly different definition of the multi-channel scattering matrix. In the ansatz of the wave function for the multi-channel regime, it is common practice~\cite{Blanter_2000} to add a normalisation prefactor related to the group velocity,~${ \sim\sqrt{v_n} }$ in front of the wave function component of channel $n$. For our purposes, we find it more convenient to omit this prefactor, as it simplifies some analytic proofs and the numerical computations below. Moreover, such a normalisation choice seems even more impractical with the inclusion of evanescent modes (virtual occupations of higher energy channels with imaginary $k$), as they have no meaningful definition of a group velocity. At any rate, for the purely planar mode description, our scattering matrix~$S$ is related to the more commonly used literature definition~$\widetilde{S}$ through~${ \widetilde{S}=\sqrt{\mathbf{v}}S\sqrt{\mathbf{v}}^{-1} }$, where~$\mathbf{v}$ is a diagonal matrix with the group velocities of the corresponding channels as its entries. Note that for the single-channel case, $\widetilde{S}=S$, so this distinction is only important for multi-channel scattering problems (with distinct group velocities). While~$\widetilde{S}$ is unitary due to probability conservation, the matrix~$S$ satisfies~$S^\dagger\mathbf{v}S=\mathbf{v}$. Similarly, the presence of time-reversal symmetry (TRS),~$\widetilde{S}^T=\widetilde{S}$ here manifests as~$S^T\mathbf{v}=\mathbf{v}S$. Some important identities however remain the same, such as the combined properties of probability conservation and TRS, yielding~$\widetilde{S}^*\widetilde{S}=S^* S=1$. 

The structure in Eqs.~\eqref{eq:Se} and~\eqref{eq:Se_sub} represents the most general case for an arbitrary scattering potential $V(x,y,z)$. Let us consider the special case of a system that has a separable potential with respect to the main propagation direction (which we here choose to be the $x$-axis, without loss of generality), that is, ${V(x,y,z)=V(x)+V(y,z)}$. In that case, the wave function can be cast into product form
\begin{align}
    \psi(x,y,z)=\psi_\mathrm{long}(x) \times \psi_\mathrm{trans}(y,z) , \label{eq:product_ansatz}
\end{align}
and scattering reduces to a set of independent one-dimensional~(${ 1\mathrm{D} }$)
problems. As a consequence, there is no longer any cross-channel scattering,~${ S_{e,h}^{m,n}=0 }$ for~${ n\neq m }$ (also meaning that here,~${ \widetilde{S}=S }$). Conversely, dropping this separability assumption leads to interactions between different channels. Cross-channel scattering will play an integral part in this work, as we detail further below. 

The superconducting part of the conductor arms has the dispersion relation
\begin{align}
    \mathcal{E}_{k,n}=\pm\sqrt{\Delta^{2}+\xi_{k,n}^{2}} . \label{eq:dispersion_SC}
\end{align}
Consequently, at the SN-interfaces, single electrons can only be absorbed into the superconductor if energy is above the superconducting gap, ${\left| E \right|>\Delta}$, where they are translated into quasi particle excitations~\cite{Nazarov_Blanter_2009}. Below the energy gap, ${\left| E \right|<\Delta}$, the superconductor only accepts integer multiples of two elementary charges as Cooper pairs. Thus, incident electrons with subgap energies are ejected as hole excitations back into the normal conductor. This process is referred to as Andreev reflection, which is described by the condition 
\begin{align}
\psi^{\text{in}}= & R\cdot\psi^{\text{out}}\text{ ,}
\label{eq:Andreev_ref}
\end{align}
where $R$ is the Andreev reflection matrix. It is usually assumed~\cite{Setiawan_2022}, that for a perfectly transparent interface and if~${ \left| \mu \right | \gg \Delta}$, no normal electron-electron backscattering is taking place, which is called the \textit{Andreev approximation}~\cite{andreev_1964,Beenakker_1992}. In this case, the reflection matrix takes block off-diagonal form 
\begin{equation}
R=\alpha \left(\begin{array}{cc}
0 & \text{e}^{\text{i}\widehat{\phi}}\\
\text{e}^{-\text{i}\widehat{\phi}} & 0
\end{array}\right) \text{ ,} \label{eq:planar_Andreev_matrix}
\end{equation}
with ${ \text{e}^{-\text{i}\widehat{\phi}} }$ (${ \text{e}^{\text{i}\widehat{\phi}} }$) governing electron-hole
(hole-electron) reflection of the $n$th channel
where
\begin{align}
\alpha= & \frac{E}{\Delta}-\mathrm{i}\sqrt{1-\frac{E^{2}}{\Delta^{2}}}\text{ .}\label{eq:alpha_factor_plane}
\end{align}
For a junction with $M$ terminals,
\begin{align}
\mathrm{e}^{\mathrm{i}\widehat{\phi}}= & \mathds{1}_{n_\mathrm{P}\times n_\mathrm{P}} \otimes \left(\begin{array}{ccc}
\mathrm{e}^{\text{i}\phi_{1}} &  & \\
 & \ddots &  \\
 &  &  \mathrm{e}^{\text{i}\phi_{M}}
\end{array}\right)\text{ ,}
\label{eq:phase_operator}
\end{align}
where  $\phi_{l}$ is the phase of the $l$th superconducting terminal. Under the above assumptions, the reflection coefficient~$\alpha$ is identical for each channel~${n\leq n_\mathrm{P}}$ involved in the process, such that we only need the unit matrix~$\mathds{1}_{n_\mathrm{P}\times n_\mathrm{P}}$ (here, actually spanning across the number of planar channels~$n_\mathrm{P}$, which will become important later).
Assuming that the spatial dependence of the superconducting pairing is mainly along the propagation direction ($x$), Andreev reflection can be treated for each channel independently, in the same spirit as for the discussion after Eq.~\eqref{eq:product_ansatz}. For simplicity, we continue to uphold this assumption throughout this work. Consequently, unlike for normal scattering, cross-channel interactions will not be incorporated in $R$. As a consequence, our previously discussed choice to not include group velocities explicitly in the ansatz of the wave functions would have no impact here, and our definition of $R$ remains consistent with the standard literature definition. Note that matters are different, once we go beyond the Andreev approximation. Here, electrons and holes have (slightly) different group velocities, such that here, different normalisation conventions could impact the final form of $R$. Again, we avoid such complications by not performing a normalisation with respect to group velocities altogether.

From the perspective of single particle excitations within the normal lead, the physics of tunnel junctions below the superconducting gap closely resembles that of a particle confined in a box potential. The boundary conditions, as defined in Eqs.~\eqref{eq:Scattering} and~\eqref{eq:Andreev_ref}, describe a scenario where electrons undergo Andreev reflection into holes at the SN-interfaces, which are then scattered at the centre, and vice versa. In order to maintain a standing wave configuration, this Andreev loop~\cite{Kroemer_1998, Zhang_2022} needs to create a coherent superposition of particle and hole excitations. By combining the boundary conditions, Eqs.~\eqref{eq:Scattering} and~\eqref{eq:Andreev_ref}, the corresponding interference condition can be formulated as the eigenvalue problem~\cite{Beenakker_1991}
\begin{align}
\det\left[\mathds{1}-R\cdot S\right]= & 0\text{ ,}\label{eq:interference_condition}
\end{align}
where both scattering and Andreev-reflection is energy dependant ($S\left[E\right], \ R\left[E\right]$)
in general. If $R$ is of the form given in Eq.~\eqref{eq:planar_Andreev_matrix}, the interference condition assumes the form
\begin{align}
\det\left[\mathds{1}-\alpha^{2} \mathrm{e}^{\mathrm{i}\hat{\phi}}S_{\mathrm{e}}  \mathrm{e}^{-\mathrm{i}\hat{\phi}} S_{\mathrm{h}}  \right]=0 & \text{ .}\label{eq:Beenakker_eq}
\end{align}
To simplify this equation even further, one usually invokes the notion of a short junction limit. In the existing literature, this limit is characterised by the Thouless energy, which is in turn defined by the inverse dwelling time ${ E_\mathrm{Th} \equiv 1/ \tau_\mathrm{dw} }$. The dwelling time ${\tau_\mathrm{dw}}$ represents the time scale the particle resides within the scattering centre. For a ballistic scattering region
\textsuperscript{\ref{note1}}, this time can be estimated as ${ \tau_\mathrm{dw}\sim v_F/L }$ where $L$ indicates the size of the region and $v_F$ is the Fermi velocity.

Its inverse, the Thouless energy, is commonly associated with the energy scale at which significant changes occur within the scattering matrix components. Consequently, for $E_\text{Th}\gg \Delta$, the scattering matrix is usually approximated to be energy independent $S{\left[ E \right] \approx S\left[ E=0 \right]}$, since Eq.~\eqref{eq:Beenakker_eq} deals with the formation of subgap bound states, $\vert E \vert <\Delta$. As a consequence, the particle hole symmetry relation, Eq.~\eqref{eq:part_hole_symmetry}, simplifies to
\begin{equation}
S_{\mathrm{h}}\overset{E=0}{\approx}S_{\mathrm{e}}^{\ast}\text{ .}
\label{eq:energy_symmetry}
\end{equation}
This leads to an important implication for the Andreev bound states that can be directly inferred from the Beenakker equation, Eq.~\eqref{eq:Beenakker_eq}. Namely, there always exist solutions for $E=\pm\Delta$ for special points in the space of phases $\phi_j$. In other words, the bound state spectrum must always connect to the band edge of the quasiparticle continuum. In the absence of magnetic fields (the default assumption for the remainder of this work), we have TRS present. As already indicated above, we thus have $S_e^{} S_e^*=1$, such that the Beenakker equation reduces to 
\begin{align}\label{eq_alpha}
1-\alpha^{2}=0 & \text{ ,}
\end{align}
for ${\hat{\phi}=0}$. Due to Eq.~\eqref{eq:alpha_factor_plane}, one sees that the ABS-spectrum needs to obey ${E=\pm\Delta}$. As argued in Ref.~\cite{Riwar_2016}, for a general multiterminal junction, the band touching the continuum ($E=\pm\Delta$) not only occurs at the origin, $\hat{\phi}=0$. Instead, for $M$ terminals, with a space spanned by the $M-1$ independent phase differences, the band touchings occur on an $M-2$ dimensional sheet, anchored to $\hat{\phi}=0$.

With energy dependence neglected, Eq.~\eqref{eq:energy_symmetry}, and for the specific case of a two-terminal junction with only one planar channel involved in subgap scattering, the Beenakker equation can be solved to obtain the well known~\cite{Nazarov_Blanter_2009} analytic expression
\begin{align}
E\left[\phi\right]= & \Delta\sqrt{1-\left|T\right|^{2}\mathrm{sin}^{2}\left[\phi/2\right]}\text{ ,}\label{eq:conv_ABS_spectrum}
\end{align}
for the ABS-spectrum, where ${ \phi=\phi_{\mathrm{I}}-\phi_{\mathrm{II}} }$ is the phase difference of the two superconducting terminals and ${ \left|T\right|^{2} }$ is the transmission amplitude. For this specific example, the ABS-spectrum touching the energy continuum at ${ \phi=0 }$ can be directly read off the explicit relation.

Crucially (and as foreshadowed), in the above described approach, there is an additional simplifying assumption, which has, to the best of our knowledge, not been explicitly stated in the existing literature (except in the aforementioned special cases in Refs.~\mbox{\cite{Guigou_2010,Recher_2011, Mandal_2024}}\footnote{In contrast to our work, in Ref.~\cite{Recher_2011}, real exponentials do \textit{not} originate from separate higher channels which become virtually occupied upon entering an energetically forbidden region. Rather, the origin lies in a quartic term in the dispersion relation giving rise to four solutions, of which two may correspond to imaginary momenta in certain regimes. Consequently, parts of the wave function ansatz may consist of both planar as well as evanescent contributions \textit{to the same channel}. The same applies similarly to Ref.~\cite{Mandal_2024}, where the dispersion relation is non-quadratic. However, full evanescent Andreev reflection~\mbox{--}~in the sense that channels initially consisting of real exponentials, which hit the SN interface, are once again reflected into evanescent components -- has not been considered so far.})~\mbox{--}~and instead was so far always tacitly taken for granted. Namely, the scattering matrix~$S$ as well as the reflection matrix~$R$ appearing in the Beenakker equation, Eq.~\eqref{eq:Beenakker_eq}, only take into account planar waves, explicitly (that is, real-valued solutions of the wave vector $k$). As indicated above, $E=\xi_{k,n}$ also has imaginary solutions for $k$, giving rise to evanescent modes, which can be thought of as virtual occupations of nominally energetically forbidden channels. In fact, solving the scattering problem (i.e., explicitly calculating $S$) in general necessitates the inclusion of these evanescent modes in the wave function ansatz (see for instance Refs.~\cite{Setiawan_2017, Sticlet_2017}), as otherwise the conditions from continuity and differentiability of the scattering problem are ill-defined. For instance, an incoming planar wave not only scatters to other planar modes (the amplitudes of which are captured in $S$), but also to evanescent modes that decay sufficiently far away from the scattering region (which corresponds to an entire subblock of additional scattering amplitudes that are not represented in $S$). Consequently, neglecting evanescent scattering amplitudes is in principle only correct in the asymptotic limit where the superconductors are situated far away from the central scattering region. If the superconductors are placed near the scattering region, one can in general not exclude evanescent modes, as they come into contact with the SN-interfaces. 

In this work, we explicitly extend the framework to account for evanescent modes and identify specific conditions under which they become of importance. We find in particular that when one or several of these channels have finite dispersion, they cause two related effects which both impact the ABS spectrum. First, finite dispersion in conjunction with cross-channel coupling yields a strong energy dependence of the scattering matrix, \textit{notably} even if the Thouless energy is large. Secondly, Andreev reflection of evanescent channels becomes particularly relevant, such that not only $S$, but also $R$, needs to be extended to accommodate evanescent modes. While strong energy-dependence of $S$ alone would provide a detaching of the bound states from the continuum (as it invalidates the conditions leading to Eq.~\eqref{eq_alpha} above), the evanescent Andreev reflection allows to exactly compensate this detachment. Curiously, this special combination of mechanisms culminates in a formulation of energy-dependant constraints on the full scattering matrix (including planar and evanescent modes) which, crucially, cannot be derived from probability conservation arguments. 
To summarise, our main achievements are that
\begin{enumerate}
    \item we extend Eq.~\eqref{eq:interference_condition} to include evanescent modes.
    \item we demonstrate that $S$ can strongly depend on energy even for short junctions.
    \item we show that evanescent modes can strongly impact the ABS spectrum, especially for energies close to~$\Delta$.
    \item we derive constraints on the $S$-matrix, which extend beyond conservation of probability arguments.
\end{enumerate}
The above points are developed in detail in the following sections. After developing some general findings regarding the ABS spectrum and the energy dependence of the normal metal scattering process (cf. Sec.~\ref{sec:dispersion_vs_probability}), we setup the new formalism (cf. Sec.~\ref{sec:recipe}), and subsequently apply it to a specific, experimentally relevant junction geometry (Sec.~\ref{sec_L_junction}).

\section{Impact of energy-dependant scattering on Andreev bound-state spectrum} \label{sec:dispersion_vs_probability}

\subsection{Detaching from continuum\label{sec:detaching_from_continuum}}

\noindent To set the stage, let us first consider a generic case of an energy-dependant scattering matrix, and show how this energy-dependence is noticeable within the ABS spectrum, at least within the above described standard literature formalism.

For this purpose, we focus in particular on the previously discussed fact that the entire ABS energy spectrum approaches the gap $\Delta$ for zero phase differences,~${\widehat{\phi}=0}$, and thus touches the continuum. As already explained, for a symmetric scattering matrix, the origin of the~${E=\Delta}$ solution can be conveniently seen on the level of Eq.~\eqref{eq:Beenakker_eq}, where for~${E=\Delta}$ and~${\widehat{\phi}=0}$, ${\alpha^2 S_e e^{i\widehat{\phi}} S_h e^{-i\widehat{\phi}}=S_e^{}S_e^*=1}$. If we now include a finite energy dependence in the scattering, we instead get the term $S_e(E)S_e^*(-E)$ (due to particle-hole symmetry), which in general no longer reduces to one. As a consequence, the ABS spectrum detaches from the continuum at $\widehat{\phi}=0$.

It is interesting to note that this simple observation can be cast into a different language. Namely, the touching of the energy spectrum can be understood as being protected by a type of chiral symmetry, and the detaching due to a finite energy dependence of $S$ as a breaking of that symmetry. To see this, we rely on a trick similar to the one proposed in Ref.~\cite{van_Heck_2014}, whereby one can express the  eigenvalue problem for $E$ in terms of an effective (Hermitean) Hamiltonian. For this purpose, we start from the eigenvalue problem $R\cdot S \psi=\psi$ (which follows from the combination of Eqs.~\eqref{eq:Scattering} and~\eqref{eq:Andreev_ref} with $\psi=\psi^\text{in}$) and transform it into two equivalent variants
\begin{align}\label{eq_variant_1}
    \alpha^*\psi&=\alpha^*R\cdot S \psi\\\label{eq_variant_2}
    \alpha\psi&=\alpha S^\dagger\cdot R^\dagger \psi\ .
\end{align}
If we subtract Eq.~\eqref{eq_variant_2} from Eq.~\eqref{eq_variant_1}, we arrive at an eigenvalue problem with an effective, Hermitean Hamiltonian, $h_\text{eff}(E)\psi=\sqrt{\Delta^2-E^2}\psi$, with
\begin{equation}\label{eq_Heff_Etilde}
h_{\text{eff}}(\widehat{\phi})=\frac{\Delta}{2i}\left(\begin{array}{cc}
0 & h_S(\widehat{\phi})\\
h_S^\dagger(\widehat{\phi}) & 0
\end{array}\right)\ ,
\end{equation}
where
\begin{equation}
h_S(\widehat{\phi})=S_{e}^*(E)e^{i\widehat{\phi}}-e^{i\widehat{\phi}}S_{e}^*(-E)\ .
\end{equation}
This Hamiltonian is special in two regards. First, if the scattering matrix depends on $E$, the above is an explicitly nonlinear problem (as the Hamiltonian depends on its own eigenvalues). This will be the pivotal property which we examine in more detail in the subsequent sections of this work.

The second peculiarity is that it does not return the eigenenergies directly, but instead, it provides what we refer to as the pseudoenergy $\widetilde{E}=\sqrt{\Delta^2-E^2}$. This is because the subtraction of Eq.~\eqref{eq_variant_2} from Eq.~\eqref{eq_variant_1} yields $\text{Im}[\alpha]=-\sqrt{1-E^2/\Delta^2}$ on the right hand side. Obviously, the addition of Eqs.~\eqref{eq_variant_1} and~\eqref{eq_variant_2} (which was the original procedure of Ref.~\cite{van_Heck_2014}) would instead yield $\text{Re}[\alpha]=E/\Delta$ and would thus be much closer to the generally more conventional form of an effective Schr\"odinger equation. But, the form chosen above has a central advantage: it allows for the interpretation of the touching of the ABS spectrum with the continuum at $E=\Delta$ as a topological feature. Namely, for energy-independent scattering, we find the symmetry 
\begin{equation}
h_\text{eff}(-\widehat{\phi})=-\Gamma h_\text{eff}(\widehat{\phi}) \Gamma\ ,
\end{equation}
with the self-adjoint chiral transformation
\begin{equation}
    \Gamma =\left(\begin{array}{cc}
0 & S_{e}^{*}\\
S_{e} & 0
\end{array}\right)\ .
\end{equation}
Evidently, this symmetry guarantees that for $\widehat{\phi}=0$ all eigenvalues of $h_\text{eff}$ go to zero, and thus $E=\pm \Delta$. The above statement is true for the here considered conventional s-wave superconducting contacts. Note that if we were to replace the contacts by topological p-wave superconductors, the Andreev reflection coefficient $\alpha$ would acquire an additional $\pi/2$ phase shift (see, e.g., Refs.~\cite{Xie_2017,Xie_2018,Houzet_2019}). This shift gives rise to the well-known correspondence of the ABS spectra between conventional and topological superconducting junctions, where energies $E$ for conventional superconductors map onto energies $\sqrt{\Delta^2-E^2}$ for topological superconductors (and vice versa). Consequently, for \textit{topological} superconductors, the eigenvalues of $h_\text{eff}$ would return the actual ABS energies $E$ instead of $\sqrt{\Delta^2-E^2}$. By the same token, for topological superconductors, the very same chiral symmetry guarantees $E=0$ (due to the presence of Majorana zero energy states) instead of $E=\Delta$.

\begin{figure}
\begin{centering}
\includegraphics[width=1\columnwidth,height=1\columnwidth,keepaspectratio]{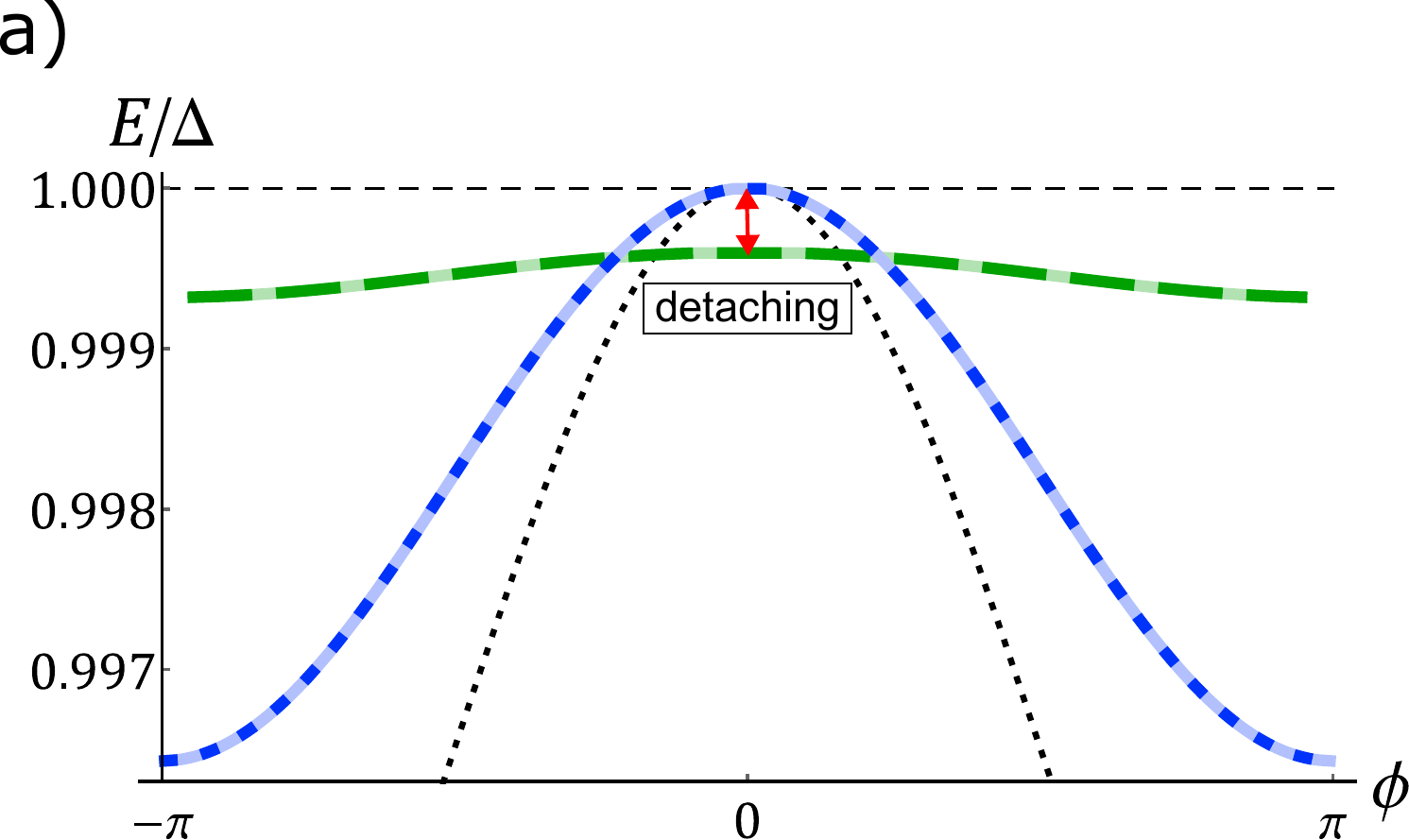} \\
\vspace{0.4cm}
\includegraphics[width=1\columnwidth,height=1\columnwidth,keepaspectratio]{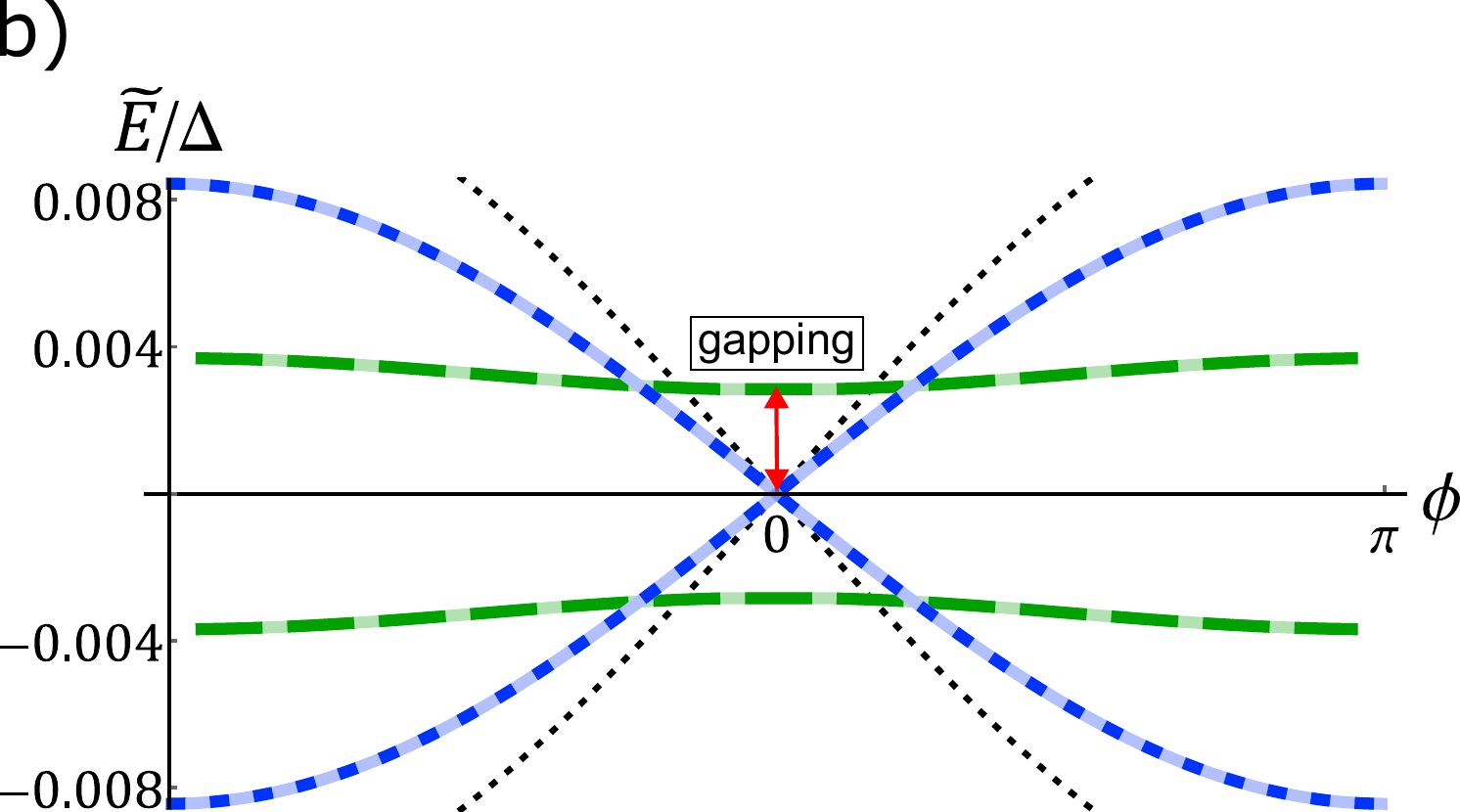}
\par\end{centering}\caption{\label{fig:ABS_spectrum_nonlinear}
Impact of finite energy dependence in the scattering $S$ and nonzero backscattering in $R$, due to finite dispersion, on the ABS spectrum in a single-channel device. The spectrum shown above is obtained using the specific example of an $L$-junction geometry as discussed in detail in Sec.~\ref{sec_L_junction} (see also Appendix~\ref{app_fig2}).  Panels a) and b) show the energy $E$ and pseudo-energy $\widetilde{E}$, respectively, for the same system parameters. With the standard framework employed -- where energy dependence in~$S$ is omitted and~$R$ is applied as in Eq.~\eqref{eq:planar_Andreev_matrix} -- the spectrum (black dotted curve) touches the gap ${E=\Delta}$ or ${\widetilde{E}=0}$, respectively, at ${\phi=0}$. When including energy dependant scattering~$S\left[ E \right]$ (while keeping the standard $R$) we observe a detaching from~${E=\Delta}$ and a gapping at~${\tilde{E}=0}$ in the pseudoenergy, respectively [green dashed curves in a) and b)]. If the energy-dependence in $S$ were due to a finite dwelling time, the green curve would be exact. Yet, for the parameters chosen here, the finite energy dependence is attributed to finite dispersive effects (whereas the dwelling time remains negligible). In this case, nonlinear dispersion effects in both the $R$-~and~$S$-matrices lead to a restoration of the gap closing (blue dotted curve), as argued in Sec.~\ref{sec_gap_closing}.
}
\end{figure}

Crucially, this correspondence ceases to be valid for the nonlinear case, that is, when $S_e$ is a function of $E$. This can be seen already in first order, $S_e(E)\approx S_e+E\delta S_e$, where the effective Hamiltonian (at $\widehat{\phi}=0$) simplifies to
\begin{equation}\label{eq_h_eff_nonchiral}
    h_\text{eff}\approx - i\Delta\left(\begin{array}{cc}
0 & -E \delta S_{e}^{*} \\
E \delta S_{e} & 0 
\end{array}\right) \ . 
\end{equation}
For the here considered conventional superconductors, a possible solution at ${\widetilde{E}=0}$ gets mapped to ${E=\pm\Delta}$, which has to be self-consistently reinserted into $h_\text{eff}$. But this nonlinear correction breaks the chiral symmetry, leading to a gapping of $\widetilde{E}$, or equivalently, to a detaching of $E$ from the continuous bands at $\pm\Delta$. Note that for topological superconductors, an energy-dependant scattering matrix does not lead to a similar breaking of the chiral symmetry, because $E$-dependant corrections to $S_e$ obviously vanish at ${E=0}$. For illustration purposes, we explicitly show this detaching mechanism in Figs.~\ref{fig:ABS_spectrum_nonlinear}a (for $E$) and~\ref{fig:ABS_spectrum_nonlinear}b (for $\widetilde{E}$). For concreteness, the energy spectra are computed with the explicit $L$-junction geometry, which is presented in Sec.~\ref{sec_Ljunction_scattering}. Details for the calculation of Fig.~\ref{fig:ABS_spectrum_nonlinear} can be found in Appendix~\ref{app_fig2}. 

Due to the above findings, one might think now that energy-dependant scattering inevitably causes a detaching to occur. However, it will turn out that in general this is not necessarily the case. By investigating various \textit{distinct} origins of energy dependant scattering, we discover in what follows a mechanism by which the detaching is compensated for in certain limits. In particular, while a strong nonlinearity in the dispersion relation leads to a strongly energy-dependant scattering matrix, it, importantly, cannot open a gap at~${ E=\Delta }$ (cf. blue dotted curve in Figure~\ref{fig:ABS_spectrum_nonlinear}). This is essentially due to the fact that the strong energy-dependence here does not come from a long dwelling time of the electrons inside the scattering region (i.e., the Thouless energy still remains large, $E_\text{Th}\rightarrow \infty$). While this finding may seem intuitively plausible, showing it rigorously on a mathematical level turns out to be a nontrivial task. In particular, in order to see the gap closing, and an effective restoring of chiral symmetry, one needs to go beyond the commonly made approximations (detailed in Sec.~\ref{sec_status_quo}). This is especially important for multi-channel conductors with strong cross-channel scattering.

In what follows, we first consider a precursor to that mechanism for the single channel case. But this reasoning will ultimately lead to the insight that evanescent modes (i.e., virtual occupations of higher energy channels, commonly neglected in the scattering problem) play a pivotal role for the proper description of the ABS spectrum, especially for energies close to~$\Delta$.

%
%

%
\subsection{Finite dwelling time versus finite dispersion}\label{sec_dwelling_time_vs_dispersion}

%
\begin{figure}
\centering{}\includegraphics[width=1\columnwidth,height=1\paperheight,keepaspectratio]{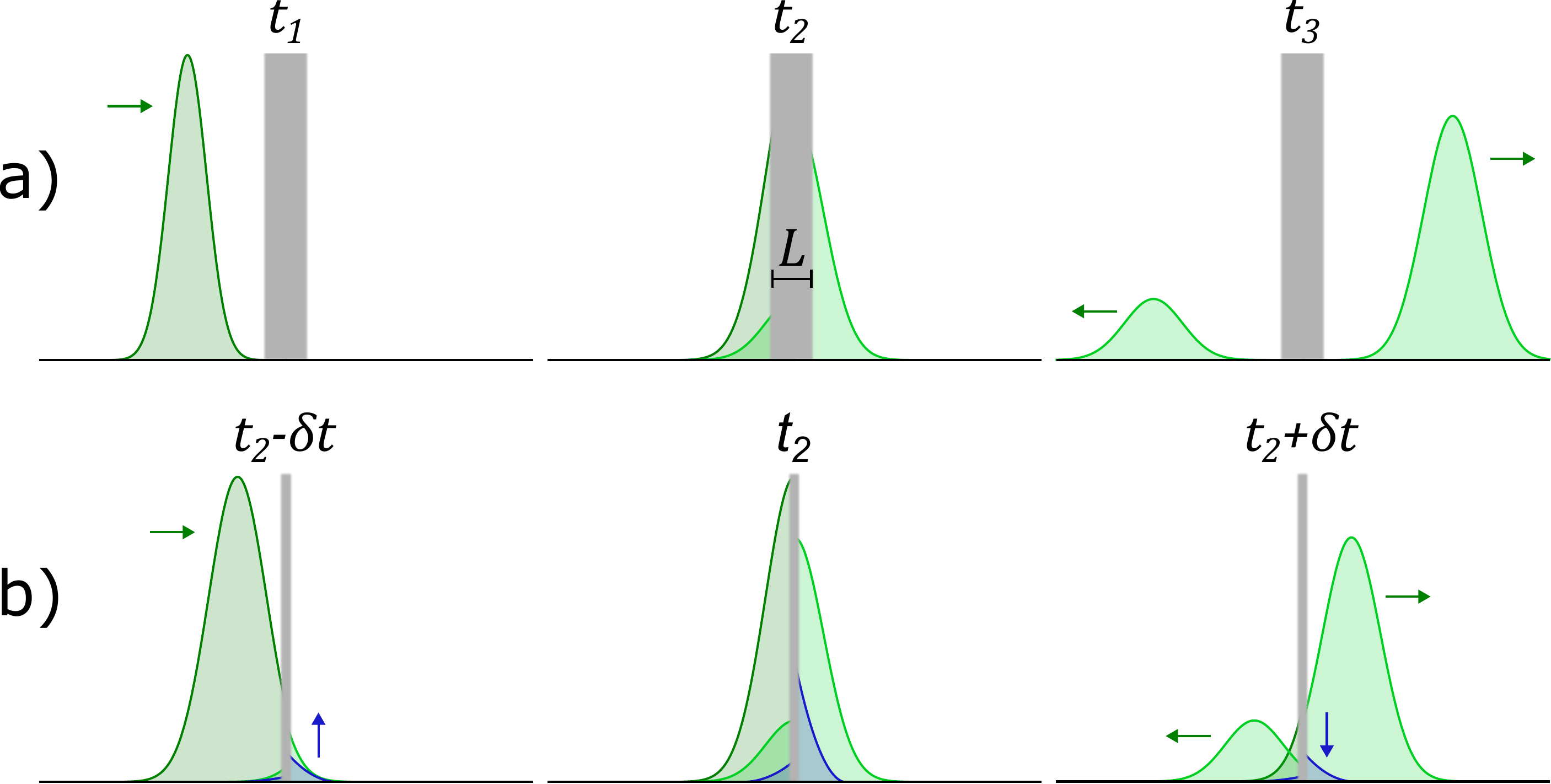}\caption{
Probability conservation at the time of scattering, depending on the size of the scattering region or on the presence of evanescent modes. A Gaussian wave packet is prepared and impinging upon a scattering centre. Unitarity of the $S$-matrix ensures probability conservation before ($t_1$) and after ($t_3$) the scattering in the conductor arms. However, during the scattering process (around~$t_2$), properties beyond unitarity become visible, depending on the situation. If the scattering centre has non-negligible width [panel a)], part of the probability current flows into the scattering region and is temporarily lost for the conductor arms (leading to a finite energy-dependence of $S$). If we consider instead the case of cross-channel scattering with  dispersive evanescent modes [blue real exponentials in panel~b)], probability current temporarily flows into these virtually occupied higher channels, even if the scattering region is short (negligible dwelling time~${\tau_\mathrm{dw}= 1/E_\mathrm{Th}\to 0}$). In this latter case, the finite energy-dependence of the planar part of $S$ is compensated through a constraint related to planar-evanescent (and evanescent-planar) scattering processes, cf. Eq.~\eqref{eq_condition_evanescent} (one of the central results of this work). 
\label{fig:gaussian_scattering} }
\end{figure}
%
%

\noindent
Typically, probability conservation is expressed by the unitarity of the scattering matrix which aggregates and relates the total probability densities well before and after the time of scattering ($t_1$ and $t_3$ in Fig.~\ref{fig:gaussian_scattering}). More generally, however, probability has to be conserved at all times, particularly during the scattering process itself (around $t_2$). By tracking the time-evolution of Gaussian wave packets and integrating over the probability densities, we find that under certain circumstances, there emerge conditions beyond unitarity. The derivation of these conditions is detailed in Appendix~\ref{sec:App_Gaussian_packages}. Below we summarise the results. We consider a single channel scattering problem in two particular limits.

In a first case, we assume that the scattering centre has a finite size~$L$ which, during the actual scattering process, temporarily absorbs a portion of the probability density. This results in a distortion of the Gaussians, the magnitude of which can be related to a term~${\sim \delta S_\mathrm{e} S_\mathrm{e}^* -S_\mathrm{e}\delta S_\mathrm{e}^*}$, where $\delta S_\mathrm{e}$ captures the linear energy dependence of $S_\mathrm{e}$. This case corresponds to the view usually taken in the existing literature, where a finite dwelling time gives rise to a finite energy dependence of the scattering matrix.

We contrast the above with a second case with two important differences. On the one hand, we assume that the scattering region has negligible width ${L=0}$, such that the above argument no longer applies. Conversely, however, we now assume that the momentum $k\left( E\right)$ can no longer be linearized. Throughout this paper, we refer to this as the ``dispersive'' case (or to a regime of ``finite dispersion''). We now still have a finite energy dependence in the scattering problem, this time stemming from a finite dispersion instead of a finite size of the scattering region. This results first of all in an additional non-Gaussian correction term~${\sim S_\mathrm{e}-S_\mathrm{e}^*}$. On the other hand, due to the negligible scattering size, we still have to impose probability conservation everywhere within the conductor arms, leading to an additional condition with respect to the electron scattering matrix $S_\mathrm{e}$ as
\begin{equation}\label{eq_condition_dS}
    \frac{S_\mathrm{e}-S_\mathrm{e}^*}{2\mu}+\delta S_\mathrm{e} S_\mathrm{e}^* -S_\mathrm{e}\delta S_\mathrm{e}^*=0\ .
\end{equation}
We emphasise that while for both cases, the deviation from the Gaussian wave packet is of the exact same shape (cf. Appendix~\ref{sec:App_Gaussian_packages}), the physical origin could not be more different. In the first case, the probability of the electron to reside inside the conductor arms temporarily diminishes, whereas in the latter case it does not -- leading to the extra requirement given in Eq.~\eqref{eq_condition_dS}. Crucially, we show further below, that if the latter is true, such that Eq.~\eqref{eq_condition_dS} applies, the energy-dependence of the scattering matrix cannot open a gap at~${ E=\Delta }$, and the chiral symmetry is preserved even for nonzero $\delta S_e$. 

As an intermediate comment, the reader might wonder, why this discussion is important, as it is common for solid state systems to assume $k_F$ large (thermodynamic limit). However, it is well-known that finite chemical potential effects are important for systems with low charge carrier density. In addition, our argument \textit{also matters} for a system with a large electron density (in the solid state sense) in the multi-channel case as introduced in the dispersion relation given in Eq.~\eqref{eq:dispersion_NC}. Notice that each channel has effectively a different chemical potential, due to the energy contribution of the transversal standing wave $\epsilon_n$. Hence, for a generic multi-channel situation, individual channels may have a noticeable dispersion especially when tuned to a regime where the number of occupied channels changes, as we illustrate in more detail below.

\subsection{Closing the gap at $E=\Delta$}\label{sec_gap_closing}

Up to now, we demonstrated that a finite energy-dependence of the scattering matrix does not necessarily indicate a finite dwelling time of the electrons inside the scattering region, but may also stem from a finite non-linearity in the dispersion relation of the leads. We now show how this fact enters in the behaviour of the ABS spectrum close to the gap.

For this purpose, we return to the chiral symmetry breaking argument presented in Sec.~\ref{sec:detaching_from_continuum}. Since a finite non-linearity of the dispersion relation can create a finite energy-dependence in the scattering matrix, see Eq.~\eqref{eq_condition_dS}, it would now --~at first sight~-- indeed seem that the discussion in Sec.~\ref{sec:detaching_from_continuum} applies, and one has to conclude that the ABS spectrum detaches from the continuum.

However, as we now show, this conclusion is wrong. Namely, in the limit of a finite chemical potential, the form of the Andreev reflection matrix $R$, as given in Eq.~\eqref{eq:planar_Andreev_matrix} is no longer correct, as it requires the assumption that the chemical potential (and the corresponding Fermi wave vector) is infinitely large~\cite{Nazarov_Blanter_2009}. Hence, just like in the derivation of Eq.~\eqref{eq_condition_dS} above, we need to take into account the leading order correction for a finite $\mu$. For a single channel, we get up to first order in $1/\mu$ (for the full expression of all Andreev processes, see Sec.~\ref{sec:recipe} below),
\begin{equation}
    R=\alpha \left(\begin{array}{cc}
-\frac{\Delta}{2\mu} & \left[1+\frac{E}{2\mu}\right]e^{i\widehat{\phi}}\\
\left[1-\frac{E}{2\mu}\right]e^{-i\widehat{\phi}} & \frac{\Delta}{2\mu}
\end{array}\right)\ .
\label{eq:approx_single_channel_R}
\end{equation}
The ordinary Andreev electron-hole receives a small correction ${ \sim \frac{E}{2\mu} }$ , and there is in addition a finite electron-electron (respectively, hole-hole) backscattering, ${ \sim\frac{\Delta}{2\mu} }$. We stress that the finite normal backscattering does not originate from an impurity at the $SN$-interface, but instead from a slight, but finite mismatch of the electron and hole wave vectors -- a consequence of $\mu$ being finite\footnote{In alignment with Sec.~\ref{sec_status_quo}, unitarity only holds for $R$-matrix coefficients when normalising with respect to the (mismatched) electron and hole group velocities.}.

We now use the updated formula for the scattering at the $SN$-interface, and derive a corrected effective Hamiltonian, along the same lines as in Sec.~\ref{sec:detaching_from_continuum}. For $\widehat{\phi}=0$ (where we would normally expect the detaching), we get
\begin{equation}\label{eq_h_eff_planar}
    h_\text{eff}=-i\Delta\left(\begin{array}{cc}
\frac{\Delta}{4\mu}\left(S_{e}-S_{e}^{*}\right) & -E\delta S_{e}^{*}\\
E\delta S_{e} & \frac{\Delta}{4\mu}\left(S_{e}-S_{e}^{*}
\right)\end{array}\right)\ ,
\end{equation}
which differs from Eq.~\eqref{eq_h_eff_nonchiral} by the diagonal terms first order in $1/\mu$. Importantly, it is at this point where the extra condition for the scattering matrix, derived in Eq.~\eqref{eq_condition_dS}, comes into play. If it applies, $h_\text{eff}$ is guaranteed to have eigenvalue $0$ with the corresponding eigenvector
\begin{equation}\label{eq_v0}
    v_0 = \left(\begin{array}{c}
S_{e}^{*}v\\
v
\end{array}\right)\ , 
\end{equation}
where for the here considered 2-by-2 scattering matrix, $v$ is an arbitrary $2$-dimensional vector. Therefore, we can decompose $v$ into two orthogonal vectors, such that actually, $v_0$ represents a set of two eigenvectors with eigenvalue $0$.

We thus conclude that if Eq.~\eqref{eq_condition_dS} applies, the gap closing at $\widehat{\phi}=0$ is conserved. In other words, while a finite energy-dependence of the scattering matrix may have various origins, only an actual finite dwelling time of the electrons inside the scattering region can detach the ABS spectrum from the continuum. If, on the other hand, the energy-dependence stems from finite dispersion (while the actual, physical dwelling time within the scattering region remains negligible) no detaching occurs. This finding is explicitly numerically corroborated in Fig.~\ref{fig:ABS_spectrum_nonlinear}, again with the example of the $L$-junction geometry (see again Appendix~\ref{app_fig2} for details), where the gap closes again (blue dotted curve) when including the dispersive backscattering in $R$.

In what follows, we make two additional general statements. First, we argue that in the presence of multi-channel scattering, a \textit{strong} energy-dependence may occur in the scattering matrix -- yet again, not due to a finite dwelling time, but due to a variant of the same finite dispersive effects involving evanescent modes. Then, we argue why it is nonetheless difficult to generalise the above simple proof of the survival of the gap closing at $E=\Delta$ for $\phi=0$ by means of probability conservation arguments. Nonetheless, after updating the Beenakker formalism to explicitly take evanescent modes into account, we can formulate a new type of constraint that, crucially, does not rely on probability conservation.

\subsection{Cross-channel coupling and evanescent modes}\label{sec_cross_channel_evanescent}

%
\begin{figure}
\centering{}\includegraphics[width=1\columnwidth,height=1\paperheight,keepaspectratio]{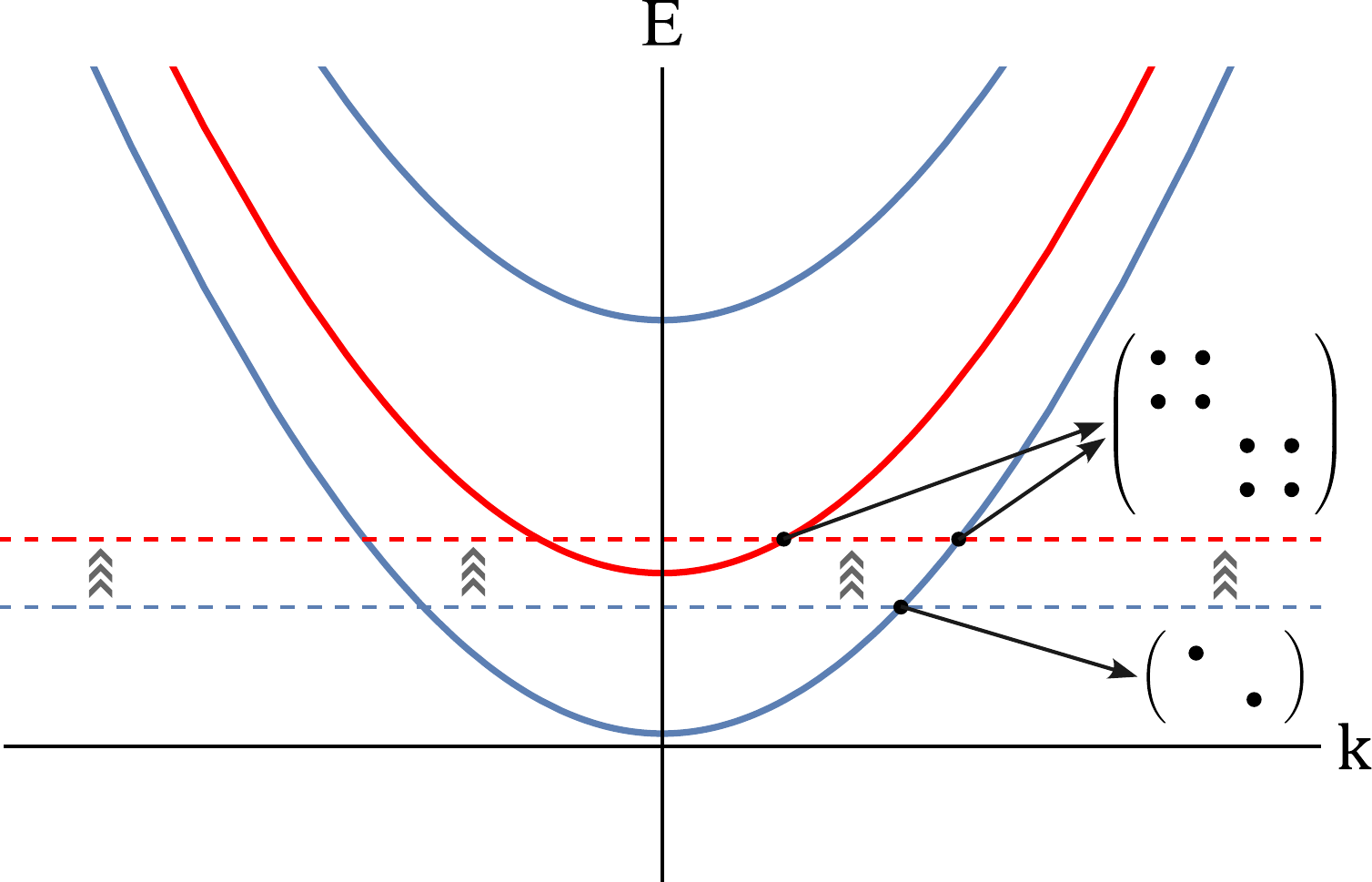}\caption{\label{fig:channel_number}
Single particle dispersion in the normal conductor region. Increasing the chemical potential (blue and red dashed line) beyond the next higher mode (red solid line) changes the channel number and by this the dimension of the (planar) scattering matrix.
}
\end{figure}
%

\noindent 
The previous sections dealt with energy dependence due to a single dispersive channel. We now discuss the energy dependence for multi-channel scattering in more detail, where evanescent modes play a role. For simplicity, but without loss of generality, we consider the case where the chemical potential $\mu$ is chosen such that still only the lowest channel is occupied as a planar mode~${n_\mathrm{P}=1}$, i.e.,~${\xi_{k=0,n=2}=\mu_2>0}$, but now tuned to values where the system is close to occupying the second channel. For $E$ close to zero, we then only have one planar mode (which can propagate forward and backward), and the scattering matrix for electrons (and likewise for holes) has dimensions of two by two (see also blue dashed line in Fig.~\ref{fig:channel_number}). Consequently, this regime would (according to the state of the art in Sec.~\ref{sec_status_quo}) be described by a simple two-by-two scattering matrix.

But now, with the chemical potential tuned close to the second channel, we immediately see that the scattering matrix inevitably must depend strongly on $E$. Very simply, if we increase the energy $E$ beyond the critical value~${\mu_\text{crit}\equiv\xi_{k=0,n=n_\text{P}+1}}$ (which for the current example is the minimum of the second channel, i.e.,~${ \mu_\mathrm{crit}=\mu_2 }$)\footnote{Note that we can easily generalize to similar transitions at higher energy, by $\mu_{n_\text{crit}}\equiv\xi_{k=0,n=n_\text{crit}}$ where $n_\text{crit}$ is the index of the channel undergoing the transition.}, the electron or hole gets two planar channels for its disposal (red dashed line). It stands to reason that the scattering matrix thus changes from two-by-two to four-by-four as a function of $E$, a transition which, in line with the notation adopted in Eq.~\eqref{eq:Se}, can be formally depicted as (see also graphical representation in Fig.~\ref{fig:channel_number})
\begin{equation}
\left(\begin{array}{c} S_e^{1,1}\end{array}\right)\to\left(\begin{array}{cc} S_e^{1,1} & S_e^{1,2}\\ S_e^{2,1} & S_e^{2,2}
\end{array}\right)\text{ .}\label{eq:2x2_to_4x4}
\end{equation}
Importantly, this transition persists even when the scattering region satisfies~${E_\text{Th}\rightarrow \infty}$, that is, when the dwelling time in the scattering region is small. Moreover, it also persists for large $\mu$, in fact, it is very much driven by an increase in $\mu$.

While it is now abundantly obvious that the scattering matrix changes strongly, it is further pivotal to understand how exactly the above transition occurs (e.g., discontinuous versus smooth). There is a straightforward special case where the change is discontinuous. Namely, as stated in the discussion around Eq.~\eqref{eq:product_ansatz}, for a separable scattering potential, channels are independent, such that the scattering matrix is block diagonal with respect to the channel index, that is, $S_e^{1,2}=S_e^{2,1}=0$. Consequently, unitarity is obeyed by each channel separately $\left|T_{n,n}\right|^2 + \left|R_{n,n}\right|^2 =1$. Hence, for energies slightly above $\mu_{n_\text{crit}}$, $S_e^{1,1}$ still has a negligible energy dependence, due to $E\ll \mu$, whereas $S_e^{2,2}$ strongly depends on $E$ due to a small $\mu_\text{crit}$.

In contrast, if translation invariance is broken (which is inevitably the case for strong cross-channel coupling), the situation is significantly more complex. This can already be seen qualitatively by considering the probability conservation condition for the total matrix~$S_e$. Since the off-diagonal cross-channel coupling terms $S_e^{1,2},S_e^{2,1}$ are nonzero, probability conservation of the total  $S_e$ couples all submatrices. Consequently, we can no longer separate the energy dependence of the different channels into separate energy scales $\mu$ (large) and $\mu_{\text{crit}}$ (small). Instead, the small energy scale, $\mu_\text{crit}$, now dominates the energy dependence of the \textit{entire} matrix. 

Importantly, this observation can be made much more concrete and strongly generalised at the same time. Breaking translation invariance means that the wave function inside the scattering region can no longer be separated into longitudinal and transversal components as in Eq.~\eqref{eq:product_ansatz}. Hence, solving the usual continuity and differentiability conditions  of the scattering problem creates interdependencies between coefficients of different channels.
%
\begin{figure}
\centering{}\includegraphics[width=1\columnwidth,height=1\paperheight,keepaspectratio]{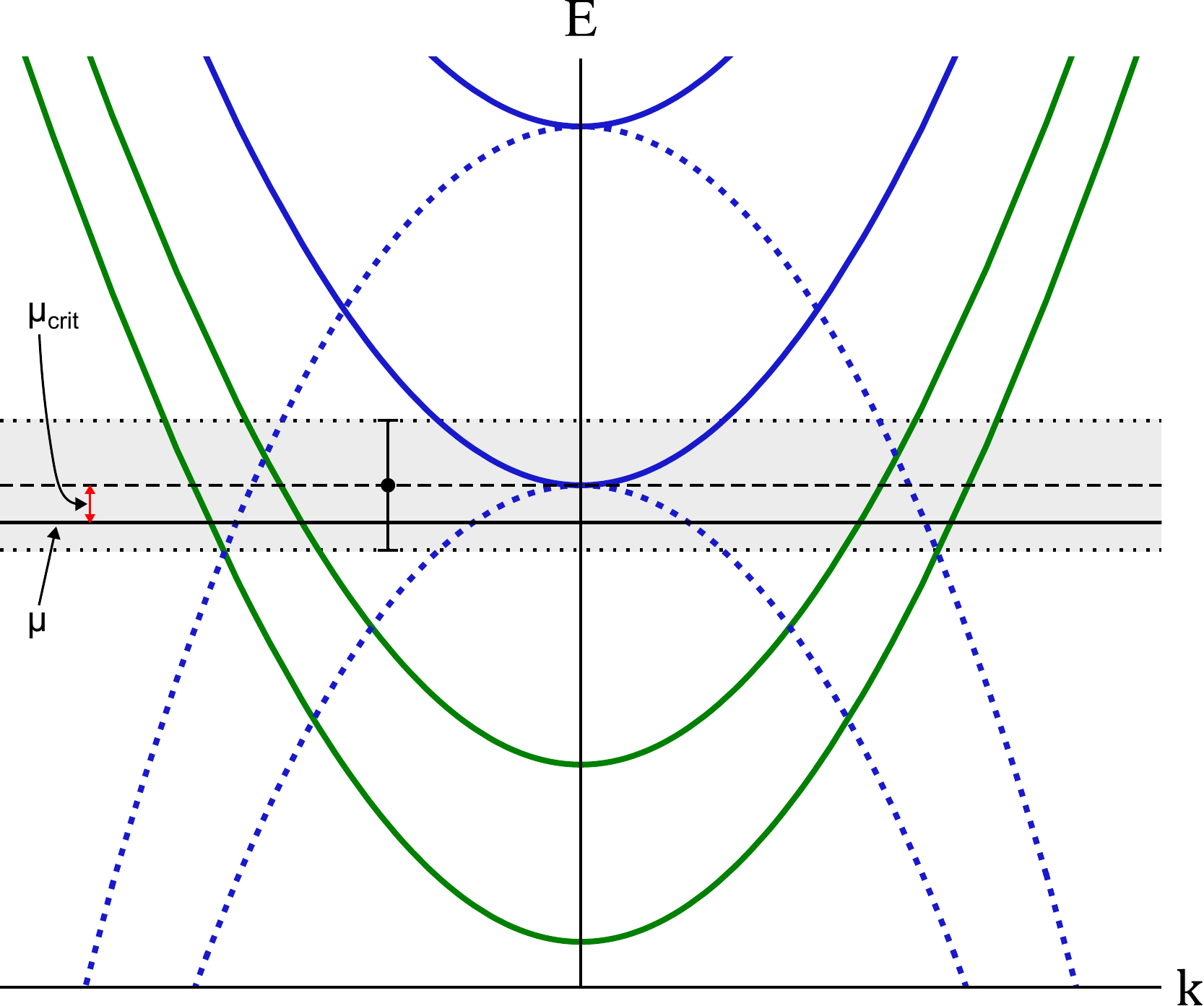}\caption{\label{fig:evanescent_dispersion} 
Dispersion relation of electron bands in the normal conductor. 
Unoccupied higher planar modes (blue curves) turn into virtually occupied evanescent modes with inverted dispersion below~$\mu_\mathrm{crit}$ (dashed blue curves). Scattering is energy sensitive if~$\mu$ is in the vicinity of one of the minima~$\mu_n$ (grey transition region).}
\end{figure}
%
%
Moreover, for this general, non-separable case, it is actually not even enough to only use a pure planar mode ansatz for the scattered wave in the conductor arms, as the continuity and differentiability conditions impose (for a continuous position space) infinitely many conditions, which can in general not be satisfied by the finite number of available planar modes (see also Refs.~\cite{Setiawan_2017, Sticlet_2017}). This problem can only be solved by including \textit{all} channels in the ansatz for the wave functions, including the energetically unavailable modes (with energies $\xi_{k,n}>E$ for all real $k$) as virtually occupied \textit{evanescent modes} with an imaginary wave vector,
\begin{align}
    k_n(E)\rightarrow \pm \mathrm{i}\kappa_n(E) \ .
\end{align}
These real exponentials now satisfy~${ \xi_{\pm i\kappa,n}=E }$ for all ${n>n_P}$, where $n_P$ are the number of planar modes. They can be conveniently represented as inverted dispersion relations (cf. the dashed blue curves in Figure~\ref{fig:evanescent_dispersion}). For a normalisable scattering wave function ansatz, the sign~$\pm$ in the above definition has of course to be chosen such that evanescent modes decay far away from the scatterer. However, as we will explicitly show below, for Andreev bound states, our ansatz must include both decaying and diverging modes due to the additional presence of the SN interface.

Consequently, the full scattering problem in the most general case is not described by a finite dimensional matrix (with dimensionality~$2n_P$) involving only the incoming and outgoing planar modes~${\sim e^{\pm ik_n(E)x}}$~(${n\leq n_P}$), but is, as a matter of fact, an infinite dimensional matrix describing in addition virtual scattering between planar and evanescent modes, ${\sim e^{\pm \kappa_n(E)x}}$ (${n>n_P}$), and even between evanescent and evanescent modes. Again, the details of this will be fleshed out in the subsequent section.

Here, we instead continue with the second important conclusion. As indicated above, the transition points where the number of planar channels change are fully dominated by one small energy scale, $\mu_\text{crit}$, for the channel ${n_\text{P}+1}$ which is evanescent for energies below $\mu_\text{crit}$ and planar beyond $\mu_\text{crit}$. Consequently, momenta can no longer be approximated to be constant~${ k_n\left( E \right) \not\approx \mathrm{const.} }$ if~${ E \approx \mu_n }$. Instead, assuming the dispersion relation given in Eqs.~\eqref{eq:dispersion_NC} and~\eqref{eq:quadratic_dispersion} indicates that momenta~${ k_n\left( E \right)}$ around the local extremum~$\mu_\text{crit}$ are highly energy sensitive, following a square root behaviour. Note that this is true for energies both above and below~$\mu_\text{crit}$. The only difference is that the channel~${n_\text{P}+1}$ either exhibits a strongly varying imaginary wave vector $\mathrm{i}\kappa_{n_\text{P}+1}(E)$ (in the evanescent regime) or a strongly varying real wave vector $k_{n_\text{P}+1}(E)$ (planar). In either case, taking the usual differentiability conditions, momentum drops down~${ \sim k_n\left[ E \right] \cdot \psi }$ and enters in first power. Accordingly, it is the momentum~$k_{n_\mathrm{P}+1}\left( E \right)$ of the channel~${n_\text{P}+1}$ \mbox{--}~and this parameter only~\mbox{--} by which energy dependence enters the scattering problem. If scattering is not translational invariant, as discussed above, the dependence of~$k_{n_\mathrm{P}+1}\left( E \right)$ is projected onto all remaining channels via cross channel interaction, such that the overall energy dependence is dominated by the dispersion of the channel~${n_\text{P}+1}$. Thus, we can already predict on this general level, that the behaviour of the scattering matrix around that transition is given by the square root law~${ \sim \sqrt{\vert 1-E/\mu_\text{crit}\vert } }$.

As a consequence, if the energy is close to the point where additional channels enter the scattering problem, the scattering coefficients smoothly, but strongly, adjust as a function of energy, with the square root power law. That is, there is a region of transition in the vicinity of $\mu_\text{crit}$ where the scattering matrix exhibits a strong algebraic energy dependence even below the point where the next higher channel joins the scattering process (cf. the gray stripe in Figure~\ref{fig:evanescent_dispersion}). 
The full picture, then, is that by approaching~$\mu_\text{crit}$ from below, evanescent modes gradually become increasingly extended within the conductor arms and eventually turn into planar channels
as the next higher mode is populated. In current treatments, it is assumed that evanescent modes exciting the scattering centre quickly fade away and thus can safely be neglected. However, as we show below in more detail, for short junctions this assumption is in general valid only for energies sufficiently far from~$\mu_\text{crit}$. If the chemical potential is tuned to values close to a channel number transition, especially the lowest evanescent mode (whose spatial extension is diverging here) has to be taken into account, as it essentially pierces the SN-interfaces (cf. Figure \ref{fig:junction_limits}). Crucially, this requires a generalisation of the interference condition, Eq.~\eqref{eq:interference_condition}. 

Finally, the above analysis is not only important for the fundamental understanding of weak link physics in the general multi-channel case, but also useful for explicit calculations. As we show below with the concrete example, the square root behaviour can be used as an efficient way to interpolate the scattering matrix at different values of $E$, and thus significantly increase the computational speed.

\subsection{Challenges for the generalisation of the gap closing argument}\label{sec_challenges}

\noindent
As we just demonstrated, when the chemical potential is tuned close to a value where the number of planar channels changes (described by the small energy scale $\mu_\text{crit}$), we have a finite energy dependence of the scattering matrix, in spite of negligible dwelling time. It would therefore seem reasonable to expect that the discussion of Sec.~\ref{sec_gap_closing} applies here as well, in that the finite energy dependence of the scattering matrix would initially open a gap (i.e., detach $E$ from $\Delta$), but that a correction in $R$ would eventually restore the gap closing. And while ultimately, at least parts of that expectation turn out to be true, the pathway towards a full understanding of the near-gap spectral properties of the ABS in this regime is much more complicated. 

First, there is a simple energy scaling argument indicating that we need to explicitly include evanescent modes into the Beenakker framework. Namely, for the single channel case (see Sec.~\ref{sec_gap_closing}) we saw that the only relevant energy scale is~$\mu$, and that the energy-dependant corrections of the scattering matrix~$\sim 1/\mu$ exactly cancelled with the dispersive correction in~$R$ (likewise~$\sim 1/\mu$). If we were to simply apply the same principle to the multi-channel case, but~$\mu$ still tuned to values where there is only one planar channel, we now understand that the planar part of the scattering matrix has corrections that scale with~${1/\mu_\text{crit}\gg 1/\mu}$, and therefore, cannot simply be corrected by a mere dispersive correction to the planar Andreev reflection scaling with~$1/\mu$. Indeed, as we show below, we need to take into account evanescent modes not only within the scattering matrix describing the central scattering region, but also the evanescent equivalent of Andreev reflection (that is, not only~$S$ but also~$R$ needs to be extended). Indeed, for the latter, the energy scale~$\mu_\text{crit}$ emerges again, such that we have at our disposition a process depending on the same, dominant energy scale.

This, however, would not be sufficient, as there is an even more profound challenge. Namely, with Eq.~\eqref{eq_condition_dS}, we derived for the single channel case a constraint on the scattering matrix beyond unitarity (but still rooted in probability conservation), under the condition that the finite energy dependence in $S$ originated from a finite dispersion of the conduction electrons. Indeed, it was precisely this constraint that allowed us to show that $E=\Delta$ remained the correct solution at $\phi=0$ [see discussion around Eq.~\eqref{eq_h_eff_planar}]. It turns out that we can generalise the argument leading to Eq.~\eqref{eq_condition_dS} to the present multichannel case -- but this generalised condition will not be helpful. Namely, one can extend the scattering wave function ansatz of Eqs.~\eqref{eq_wave_packet_ansatz_1} and~\eqref{eq_wave_packet_ansatz_2} to include the dominant evanescent mode in addition to the planar mode, with an evanescent version of a transition and reflection amplitude. However, note that in order for the scattering wave function to be normalisable, we can only include the decaying mode (and not the diverging mode) -- an important caveat, as we will see
in a moment. Now, proceeding in the same way as in Sec.~\ref{sec_dwelling_time_vs_dispersion}, one can derive a very similar type of condition as in Eqs.~\eqref{eq_R_condition} and~\eqref{eq_T_condition}, except that the terms~$\sim (R-R^*)/\mu$ and~${\sim (T-T^*)/\mu}$ would be replaced by terms proportional to the evanescent reflection and transmission amplitudes, divided by~$\mu_\text{crit}$. Interestingly, this new constraint would tell a story in between the previously encountered situations. Here, the dip in the probability of the \textit{planar} mode occupation upon impact of the wave packet at the scattering region is very much real. However, the \textit{total} occupation probability of the electrons inside the conductor arms is still conserved, because at impact, the evanescent mode is temporarily occupied [cf. Figure~\ref{fig:gaussian_scattering}b)].

Yet, crucially, this constraint would not be a suitable condition to understand the properties of the ABS spectrum near the gap, for the simple reason that the ansatz for the full ABS wave function requires (contrary to the scattering wave function) both decaying as well as diverging parts of the evanescent modes (again, for details see below). The updated probability conservation constraint involving decaying modes however would only provide a relationship between the energy-dependence of the scattering amplitudes of the planar modes and the \textit{decaying} evanescent modes, and thus provides only incomplete information. As a last resort, one could try to come up with an even more general ansatz for an incident wave packet including both decaying and diverging evanescent modes. But, it is impossible to construct normalisable wave packets in this way.

By carefully generalising the Beenakker formalism to include evanescent modes below, we will be able to attack this problem from a different angle. In fact, in a sense, we will invert the problem by first providing a perturbative expansion of the generalised Beenakker equation up to first order in $1/\mu_\text{crit}$. Then, instead of providing a condition for the scattering matrix such that a gap closing at $\Delta$ is guaranteed, we \textit{demand} that the gap closing exists, and deduce from that condition a constraint on the generalised scattering matrix (including evanescent modes). Crucially (as indicated above), this constraint cannot follow from probability conservation arguments. Thus, we provide an example where Andreev physics can provide us with fundamental information about the central scattering region, that cannot be obtained otherwise.

Finally, by means of numerical analysis of a specific device geometry, we are able to show this constraint is indeed fulfilled. What is more, it can be shown that the gap closing mechanism even persists beyond the perturbative limit, where higher order terms become relevant, hinting at the exciting outlook that there probably exist higher order constraints, relating a generic scattering problem at two different energies in a nontrivial way.

\section{Beenakker equation including evanescent modes} \label{sec:recipe}

\noindent
In order to include evanescent modes into the ansatz of the ABS wave function, we have to find generalisations for both~$S$ and~$R$ in Eqs.~\eqref{eq:Scattering} and~\eqref{eq:Andreev_ref}. As already indicated, we need to include in general both decaying and diverging evanescent modes within the unproximitised part of the conductor arms. 

\begin{figure}
\centering{}\includegraphics[width=1\columnwidth,height=1\paperheight,keepaspectratio]{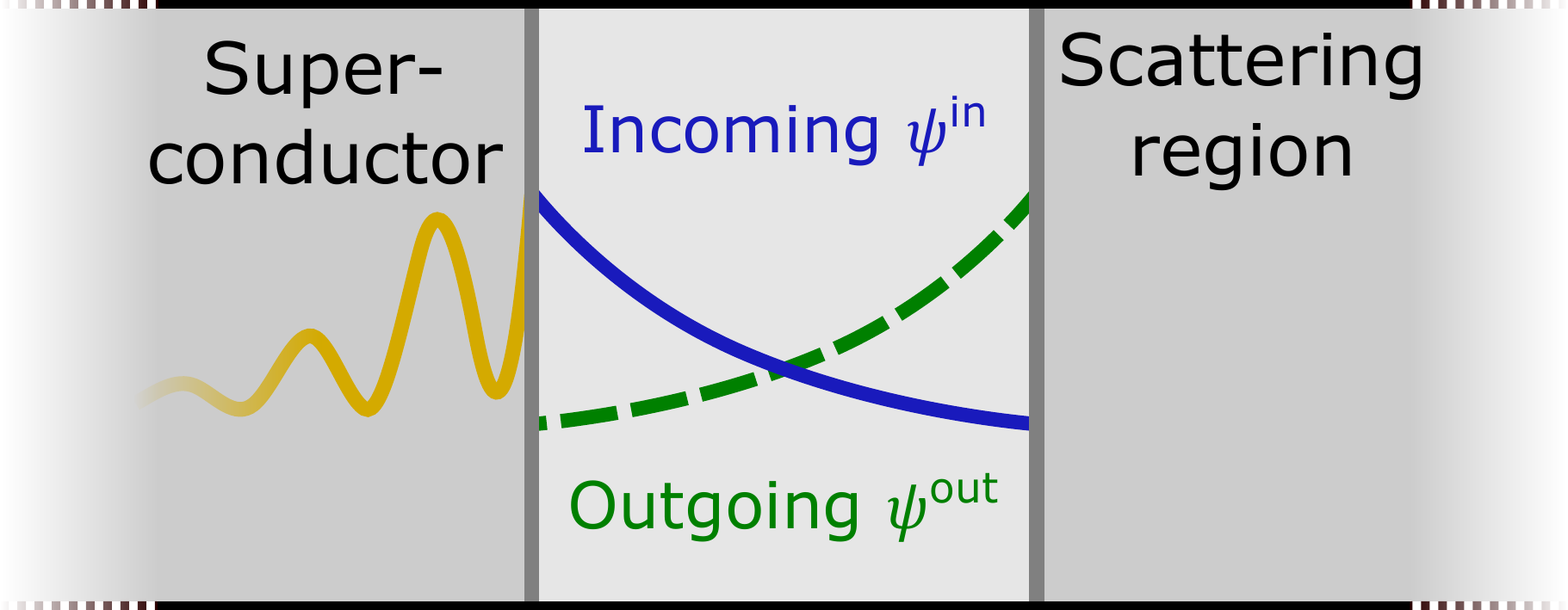}\caption{\label{fig:evanescent_scattering}
Definition of evanescent scattering directions. Transmitted or reflected real exponentials are defined as decaying in direction outgoing (green dotted line) from the scattering region. Diverging modes are defined as incoming (blue solid line). This convention is applied for both electrons and holes. }
\end{figure}

Overall, the wave function now needs to have a generalised notion of incoming and outgoing wave amplitudes, $\psi^\text{in}$ and $\psi^\text{out}$. This distinction is obvious only for planar modes, where one simply distinguishes the propagation direction via the group velocity. For evanescent modes, such a distinction is meaningless, such that we simply \textit{choose} to define the solutions that decay away from the scattering matrix as outgoing, or ``scattered'', evanescent modes (i.e., their amplitudes are added in $\psi^\text{out}$), and those that diverge as incoming, or ``incident'', modes ($\psi^\text{in}$), see Fig.~\ref{fig:evanescent_scattering}. We stress that this choice is arbitrary. The resulting wave function ansatz is nonetheless correct, provided that the scattering at the central scattering region, and at the SN interface, are computed consistently with the ensuing asymptotic boundary conditions away from the scattering region. 

As a first step, the structure of the scattering matrix as shown in Eq.~\eqref{eq:Se} has to be generalised to include evanescent modes. That is, instead of the submatrices~$S^{n,m}$ being limited to indices $n,m$ going from $1$ to $n_\mathrm{P}$, where $n_\mathrm{P}$ is the number of planar modes, $n,m$ now have (at least in general) no upper bound. For illustration purposes, we explicitly separate the total (infinitely dimensional) scattering matrix into subblocks
\begin{align}
S&=\left(\begin{array}{cc}
S_{\text{e}} & 0\\
0 & S_{\text{h}}
\end{array}\right),\quad S_{\text{\ensuremath{\nu}}}=\left(\begin{array}{cc}
S_{\ensuremath{\nu}}^{\text{pp}} & S_{\ensuremath{\nu}}^{\text{pe}}\\
S_{\ensuremath{\nu}}^{\text{ep}} & S_{\ensuremath{\nu}}^{\text{ee}}
\end{array}\right) \ ,
\label{eq:gen_scat_matrix} 
\end{align}
where, for instance, the subblock
\begin{align}
S_{\mathrm{e}}^{\mathrm{ep}}=\left(\begin{array}{cc}
S^{n_\mathrm{P}+1,1} & \dots\\
\vdots & \ddots
\end{array}\right) \ ,
\end{align}
designates in a sense the ``scattering'' from incoming evanescent to outgoing planar modes. Ultimately, all scattering matrix elements are computed such that the resulting wave function is a correct solution of the Schrödinger equation, and all boundary constraints (e.g., continuity and differentiability) are satisfied. We give an operational example of how this is accomplished in the subsequent section.

As just indicated, since there is an infinite number of higher evanescent modes, the S-matrix now becomes infinitely dimensional. Consequently, for explicit calculations, one has to find an appropriate cutoff for the number of modes, to have convergent low-energy solutions. As we see
in a moment, for realistic problems, it actually often suffices to include only the first evanescent mode, whereas all higher modes can be neglected. 

%
%

With evanescent modes included into the model, there now are real exponentials exiting the scattering centre within the conductor arms. Note that conductor arms are often included as fictitious entities, and their size is ultimately set to zero [cf. Fig.~\ref{fig:junction_limits}b)]. However, some experimental device designs have structures that in principle allow for finite conductor arms, see Refs.~\mbox{\cite{Valagiannopoulos_2012, Koelzer_2021, Koelzer_2023}}, such that it is meaningful to also consider long junctions as in Fig.~\ref{fig:junction_limits}a). At any rate, as soon as evanescent modes pierce the proximitised S-region, we in general have to explicitly solve also the evanescent mode version of the Andreev scattering problem at the SN interface\footnote{Since for $S$, we classified modes decaying from the scattering centre as ``outgoing'' (appearing in $\psi^\text{out}$) and diverging modes as ``incoming'' (appearing in $\psi^\text{in}$), we have to apply the same consistent grouping for the boundary conditions at the SN interface as well -- now for both electrons and holes. While the wave function ansatz on the N~side has both decaying and diverging evanescent components, on the S~side, we again have to restrict the ansatz to only decaying modes when receding from the SN interface within the S region (see Fig.~\ref{fig:evanescent_scattering}), to ensure normalisability of the total ABS wave function.}. 
Moreover, in the standard framework, all coefficients were computed in the limit of a strongly linear dispersion relation (Andreev approximation), such that there was no electron-electron (or hole-hole) backscattering, and the Andreev reflection coefficient $\alpha$ did not depend on the channel index. All of these features must now be included. That is, instead of the form given in the standard framework by Eq.~\eqref{eq:planar_Andreev_matrix}, the total reflection matrix at the SN interface now assumes the form
\begin{equation}\label{eq:gen_A_ref_matrix}
    R=\left(\begin{array}{cc}
\widehat{\beta}_{\text{e}} & \widehat{\alpha}_{\text{h}}e^{i\widehat{\phi}}\\
\widehat{\alpha}_{\text{e}}e^{-i\widehat{\phi}} & \widehat{\beta}_{\text{h}}
\end{array}\right) \ .
\end{equation}
The matrices $\widehat{\alpha}_{\nu}$ and $\widehat{\beta}_\nu$ are diagonal matrices whose size is given by the number of channels times the number of terminals, in analogy to the construction of the matrix~${ \text{e}^{-\text{i}\widehat{\phi}} }$ as defined in Eq.~\eqref{eq:phase_operator}. The matrix $\widehat{\alpha}_\nu$ generalises Andreev reflection, whereas $\widehat{\beta}_{\nu}$ captures the aforementioned normal backscattering. The coefficients on the diagonal depend on the channel index (but not the terminal index) due to the non-linearity of each channel's dispersion. As we discuss in detail, there is in particular a marked difference in the behaviour of the coefficients for planar- and evanescent modes. As already discussed in Sec.~\ref{sec_gap_closing}, finite normal backscattering (${\widehat{\beta}\neq 0}$) is important especially if it is the planar channels that exhibit strong dispersion. In what follows, we show that if dispersion instead stems from an evanescent mode, corrections to $\widehat{\alpha}$ will play a pivotal role.

Analogous to the scattering matrix, $\widehat{\beta}_{\nu}$ and~$\widehat{\alpha}_{\nu}$ are decomposed into subblocks distinguishing planar and evanescent modes
\begin{equation}
\widehat{\alpha}_{\nu}=\left(\begin{array}{cc}
\widehat{\alpha}_{\nu}^{\text{pp}} & 0\\
0 & \widehat{\alpha}_{\nu}^{\text{ee}}
\end{array}\right), \quad \widehat{\beta}_{\nu}=\left(\begin{array}{cc}
\widehat{\beta}_{\nu}^{\text{pp}} & 0\\
0 & \widehat{\beta}_{\nu}^{\text{ee}}
\end{array}\right) \ .
\end{equation}
For instance, the subblock
\begin{align}
\widehat{\alpha}_{\mathrm{\nu}}^{\text{ee}}=\left(\begin{array}{cc}
\alpha_{\nu,n_{\mathrm{P}}+1} & 0\\
0 & \ddots
\end{array}\right)\ ,
\end{align}
designates evanescent-evanescent Andreev reflection for electron-hole~(${\nu =\mathrm{e}}$) or hole-electron~(${\nu=\mathrm{h}}$), respectively.

As a matter of fact, it is possible to find explicit analytic expressions for all coefficients appearing in~$R$. To this end, continuity and differentiability at the SN-interface are imposed on the wave function ansatzes for the S- and the N-region. Since we assume that the SN-interface does not couple different channels, we can solve the corresponding boundary conditions in the standard 1D model. The resulting expressions for the Andreev- ($\alpha_{\nu,n}$) and backscattering- ($\beta_{\nu,n}$) coefficients read
\begin{align}
\alpha_{\nu,n}&=2\nu\frac{\Delta}{E}\frac{K_{\nu,n}k_{\mathrm{R},n}}{k_{\mathrm{R},n}\zeta_{n}^{-}+\frac{\Delta}{E}\sqrt{1-\frac{E^{2}}{\Delta^{2}}}\Gamma_{n}^{\nu,+}} \ , 
\label{eq:alpha_factor_evan}  \\
\beta_{\nu,n}&=\nu\frac{k_{\mathrm{R},n}\zeta_{n}^{+}-\nu\frac{\Delta}{E}\sqrt{1-\frac{E^{2}}{\Delta^{2}}}\Gamma_{n}^{\nu,-}}{k_{\mathrm{R},n}\zeta_{n}^{-}+\frac{\Delta}{E}\sqrt{1-\frac{E^{2}}{\Delta^{2}}}\Gamma_{n}^{\nu,+}}  \ , \label{eq:beta_factor_evan}
\end{align}
with
\begin{align}
\Gamma_{n}^{\nu,\pm}=&\left[\left(k_{\mathrm{I},n}\pm K_{\nu,n}\right)\left(k_{\mathrm{I},n}+K_{-\nu,n}\right)+k_{\mathrm{R},n}^{2}\right]\text{ ,}\nonumber \\
\zeta_{n}^{\pm}=&\left(K_{\mathrm{e},n}\pm K_{\mathrm{h},n}\right) \ .
\label{eq:gamma_zeta_expression}
\end{align}
Here, the momenta for planar electron- and hole excitations are denoted as~${ k_{\mathrm{e},n} }$ with~${ \nu=+1 }$ and~${ k_{\mathrm{h},n} }$ with~${\nu=-1}$, respectively. The same notation applies for evanescent momenta~$ {\kappa_{\nu,n} } $. We further use the notation that when $\nu$ is negated, ${\nu \rightarrow -\nu}$, we mean that electron and hole indices are swapped (${-\text{e},-\text{h}\rightarrow \text{h,e}}$). 
The relations~\eqref{eq:alpha_factor_evan} and~\eqref{eq:beta_factor_evan} apply for planar- and evanescent channels respectively, if momentum~$K_{\nu,n}$ is chosen according to
\begin{align}
   K_{\nu,n}\to-\mathrm{i}\nu k_{\mathrm{\nu},n}\,;&	\quad \text{for }n\leq n_{\mathrm{P}}\quad\text{(planar)}\,, \nonumber \\
K_{\nu,n}\to \kappa_{\nu,n}\,;& \quad	\text{for }n>n_{\mathrm{P}}\quad\text{(evanescent)\,.}
\label{eq:momentum_rules}
\end{align}
The planar and evanescent momenta~${ k_{\mathrm{e},n} }$ and~${ k_{\mathrm{h},n} }$ are obtained by considering the dispersion in the N-region  [cf. Eq.~\eqref{eq:dispersion_NC}] and solve it for the momentum~$k_n$, where evanescent channels require the replacement~${ k_n \to -\mathrm{i}\kappa_n }$, such that
\begin{align}
    k_{\nu,n}&=\sqrt{2m\left(\mu-\epsilon_{n}+\nu E\right)} \ , \nonumber \\
    \kappa_{\nu,n}&=\sqrt{2m\left(\epsilon_{n}-\mu- \nu E\right)} \ .
    \label{eq:cont_N_momenta}
\end{align}
Independent of whether we consider planar or evanescent modes in the N-region, the wave vectors in the S-region will be the same standard solutions. Starting from Eq.~\eqref{eq:dispersion_SC} and solving ${ \sqrt{E^{2}-\Delta^{2}}=\xi_{k,n}^{2} }$ for subgap energies~${ \left| E \right| < \Delta }$ we arrive at the complex momentum ${ k_n=\pm k_{\mathrm{R},n}-\mathrm{i}k_{\mathrm{I},n} }$, with~${ k_{\mathrm{R},n}=\widetilde{k}_{+,n} }$ and~${ k_{\mathrm{I},n}=\widetilde{k}_{-,n} }$, where
\begin{align}
    \widetilde{k}_{\pm,n}= \sqrt{m\left[\sqrt{\Delta^{2}-E^{2}+\left(\mu-\epsilon_{n}\right)^{2}}\pm \left(\mu-\epsilon_{n}\right)\right]}\ .
    \label{eq:cont_SC_momenta}
\end{align}
Accordingly, single electron or hole excitations are exponentially suppressed upon entering the S-region within the forbidden subgap regime, as elaborated in Section~\ref{sec_status_quo}.

Now, if dispersion is negligible, i.e.,~$ \left| \mu_n \right| \gg  \Delta, E $, and considering that by definition evanescent (planar) modes have $\mu_n>0$ ($\mu_n<0$), Eqs.~\eqref{eq:cont_N_momenta} and~\eqref{eq:cont_SC_momenta} imply that for evanescent momenta~${ \kappa_{\mathrm{e}}\approx \kappa_{\mathrm{h}}\approx k_{\mathrm{I}} }$ and~${ k_{\mathrm{R}}\approx 0 }$, whereas for planar momenta~${ k_{\mathrm{e}}\approx k_{\mathrm{h}}\approx k_{\mathrm{R}} }$ and~${ k_{\mathrm{I}}\approx 0 }$.
Going to the limit of small $\Delta,E$ with respect to $\mu_n$, the explicit expressions~\eqref{eq:alpha_factor_evan} and~\eqref{eq:beta_factor_evan} can be expanded up to first order as follows. For planar channels $n\leq n_\text{P}$ (i.e., for the respective ``$\text{pp}$'' subblocks) we get the entries 
\begin{align}
    \alpha_{\nu,n}&\approx \alpha\left(1-\nu\frac{E}{2\mu_n}\right) \label{eq:Andreev_planar} \ , \\
    \beta_{\nu,n}&\approx -\nu\alpha\frac{\Delta}{2\mu_n} \label{eq:Andreev_planar_back} \ ,
\end{align}
where $\alpha$ is the reflection coefficient of the conventional framework as defined by Eq.~\eqref{eq:alpha_factor_plane} a result we have already used in Eq.~\eqref{eq:approx_single_channel_R}, see Sec.~\ref{sec_gap_closing}, for a single planar channel. For evanescent channels $n> n_\text{P}$ (the ``$\text{ee}$'' subblocks) we obtain up to the same order
\begin{align}
    \alpha_{\nu,n}&\approx \nu\frac{\Delta}{4\mu_n}\label{eq_Andreev_evanescent} \ , \\ \label{eq_Andreev_evanescent_back}
    \beta_{\nu,n}&\approx 0\ .
\end{align}
Indeed, in the limit of $\Delta/\mu_n\rightarrow 0$, we obtain the standard framework (Sec.~\ref{sec_status_quo}), where both dispersive effects and evanescent modes are irrelevant, as here $\widehat{\alpha}_\nu^\text{pp}=\alpha$, while $\widehat{\alpha}_\nu^\text{ee}=0$ as well as $\widehat{\beta}_\nu =0$. Overall, we clearly see the very distinct behaviour between the planar and evanescent coefficients. For the planar coefficients, both regular Andreev reflection and normal backscattering have a first order correction. For the evanescent coefficient, there only is a weak, nonzero Andreev reflection, whereas normal backscattering is absent. Moreover, both planar coefficients explicitly depend on energy $E$ (note that Eqs.~\eqref{eq:Andreev_planar} and~\eqref{eq:Andreev_planar_back} contain the energy-dependant factor~$\alpha$), whereas evanescent Andreev reflection is constant.

Crucially, the above expansion also establishes a concrete guideline for assessing which modes can be neglected and which ones must be included. For a dispersion relation as shown in Fig.~\ref{fig:evanescent_dispersion} there is only one mode with a small $\mu_n$, namely the one whose energy minimum is closest to the chemical potential $\mu$. When tuning the chemical potential just below the channel number transition, the total number of modes that need to be included is $n_\text{P}+1$, i.e., all planar modes, plus the lowest evanescent mode with strong dispersion. With a chemical potential above the transition, we again only have to take into account the planar modes ($n_\text{P}$), but include finite dispersive effects of the highest planar mode (in analogy to Sec.~\ref{sec_gap_closing}). This very narrow cut-off is insofar surprising as one might have expected that every evanescent mode with a sufficiently long exponential tail (such that it penetrates the SN-interface) would have to be included. For extremely short junctions, this would have concerned a high number of modes. The above expansion result, on the other hand, shows that penetration of the SN-interface only is a necessary condition. In addition, modes also need to have finite dispersion, i.e.~${ \left| \mu_n \right| \approx  \Delta, E }$, in order to be relevant. If not, backscattering is negligible and the planar reflection coefficient~$\alpha_{\nu,n}$ reduces to the conventional form, Eq.~\eqref{eq:alpha_factor_plane} (whereas evanescent Andreev reflection vanishes completely), such that the interplay of energy dependant scattering and evanescent Andreev reflection, as discussed above, does not apply. As argued above, in situations where only one mode is close to a transition point, all modes higher than the first evanescent channel usually can be neglected.
This is a further central result of our work. We illustrate this principle, and its implications, explicitly in the subsequent sections.

To summarise, within the new framework, Andreev reflection and scattering are still governed by the boundary conditions as expressed in Equations~\eqref{eq:Scattering} and~\eqref{eq:Andreev_ref}, such that Eq.~\eqref{eq:interference_condition} remains valid. However, the structure of both~$S$ and~$R$ is changed (both including planar as well as evanescent modes). Consequently, when expressing the interference condition in terms of the submatrix structure, we get an equation of much more general form than, e.g., Eq.~\eqref{eq:Beenakker_eq}. In particular, by including backscattering, the reflection matrix~$R$ loses its block off-diagonal form, as indicated by Eq.~\eqref{eq:gen_A_ref_matrix}. Moreover, in contrast to the state of the art model, the reflection and backscattering coefficients $\widetilde{\alpha}_{\nu,n}$ and $\widetilde{\beta}_{\nu,n}$ are now different for each channel and, thus, can no longer be factored out. Consequently, Eq.~\eqref{eq:Beenakker_eq} takes on the generalised form
\begin{align}
    \det\left[\mathds{1}\!-\!(\mathds{1}\!-\!\widehat{\beta}_{\text{h}}S_{\text{h}})^{-1}\widehat{\alpha}_{\text{e}}e^{-i\widehat{\phi}}S_{\text{e}}(\mathds{1}\!-\!\widehat{\beta}_{\text{e}}S_{\text{e}})^{-1}\widehat{\alpha}_{\text{h}}e^{i\widehat{\phi}}S_{\text{h}}\right]=0 \ .
    \label{eq:gen_Beenakker_equation}
\end{align}
One can immediately see that if backscattering is negligible~${\beta_{\nu,n}=0}$ and if Andreev reflection is channel independent~${\alpha_{\nu,n}=\alpha}$, the Beenakker equation reduces to the original form of the standard framework, Eq.~\eqref{eq:Beenakker_eq}.

\section{Scattering matrix constraint including evanescent modes}\label{sec_constraint_evanescent}

\noindent Both the S- and the R-matrix can be cast into a block form separating explicitly planar and evanescent modes,
\begin{equation}
    S=\left(\begin{array}{cc}
S^{\text{pp}} & S^{\text{pe}}\\
S^{\text{ep}} & S^{\text{ee}}
\end{array}\right) \ ,\qquad R=\left(\begin{array}{cc}
R^{\text{pp}} & 0\\
0 & R^{\text{ee}}
\end{array}\right) \ .
\end{equation}
Since Andreev reflection does not couple different channels, $R$ is block-diagonal in the planar-evanescent subspace.

Starting again from the eigenvalue problem ${ R \cdot S \psi = \psi }$, where $\psi$ likewise has a planar and evanescent subblock, $\psi=(\psi^\text{p},\psi^\text{e})$, we eliminate the evanescent subblock. This yields the equation
\begin{equation}\label{eq_S_eff}
    R^\text{pp}S_\text{eff}^\text{pp}\psi^\text{p}=\psi^\text{p}\ .
\end{equation}
The matrix $R^\mathrm{pp}$ is simply the Andreev reflection matrix of the planar components only. We thus notice that the evanescent modes can be incorporated into an effective pseudo-scattering matrix of the form
\begin{equation}
    S_{\mathrm{eff}}^\text{pp}=S^{\mathrm{pp}}+\underbrace{S^{\mathrm{pe}}\left(\mathds{1}-R^{\mathrm{ee}}S^{\mathrm{ee}}\right)^{-1}R^{\mathrm{ee}}S^{\mathrm{ep}}}_{\text{evanescent  correction}} \ ,
    \label{eq:eff_S_matrix}
\end{equation} 
whose size is given by the number of planar modes. While this matrix still fulfils particle-hole symmetry, $\tau_y\left[S_\text{eff}^{pp}(-E)\right]^*\tau_y=S_\text{eff}^{pp}(E)$ (where $\tau_y$ is the Pauli matrix in Nambu space), probability conservation is not guaranteed, hence, it cannot be considered a true scattering matrix. 

The truncation to planar levels in terms of this pseudo-scattering matrix is nonetheless very instructive to analyse the relationship between evanescent scattering and band touching, in a similar fashion as for purely planar processes in Sec.~\ref{sec_gap_closing}. To that end, note that this effective scattering matrix is no longer block-diagonal in the electron-hole space. This is due to the fact that $R^\text{ee}$ has finite Andreev reflection, due to the finite dispersion of the evanescent mode, Eq.~\eqref{eq_Andreev_evanescent}. It now turns out that the interplay between diagonal and off-diagonal processes (electron-electron versus electron-hole, etc.) allows us to derive a condition relating the energy dependence of the planar part of the scattering matrix ($S^\text{pp}$) to the planar-evanescent ($S^\text{ep}$) and evanescent-planar ($S^\text{pe}$) processes, which is in its nature, crucially, \textit{beyond probability conservation arguments}.

To this end, we proceed similarly as in Sec.~\ref{sec_gap_closing}, but instead of considering the problem perturbatively in the limit where the chemical potential approaches the parabolic trough of the planar mode from above (rendering the dispersion of the planar mode stronger and stronger), we approach the parabola of the leading evanescent mode from below. If we consider for concreteness the case of one planar mode (and all other modes being evanescent) then this means that $|\mu_2|$ tends to zero (while at the same time taking $\Delta$ to be very small, such that the ratio $\Delta/|\mu_2|$ is still $\ll 1$). In this regime, we only take into account the lowest evanescent mode, such that $R^\text{ee}$ has dimensions of 4 by 4 (two entries for the left and right lead, and two for electrons and holes), such that it reduces to [see also Eq.~\eqref{eq_Andreev_evanescent}] 
\begin{equation}
   R^\text{ee}\approx \frac{\Delta}{4\mu_{2}}\left(\begin{array}{cc}
0 & -e^{i\widehat{\phi}}\\
e^{-i\widehat{\phi}} & 0
\end{array}\right)\ .
\end{equation}
Assuming $R^\text{ee}$ small, we expand $S_\text{eff}^\text{pp}$ up to first order in $R^\text{ee}$, and insert this into the modified Beenakker equation, Eq.~\eqref{eq_S_eff}. After following similar steps as in Sec.~\ref{sec_gap_closing} (expanding in leading order in the energy dependence of the scattering matrix around~${E=0}$, and exploiting the fact that $\alpha(E)-\alpha^*(E)\sim i\sqrt{\Delta^2-E^2}$, we arrive again at an effective Schr\"odinger equation, $h_\text{eff}(E)\psi=\sqrt{\Delta^2-E^2}\psi$, this time with the Hamiltonian (evaluated at $\widehat{\phi}=0$)
\begin{equation}
    h_{\text{eff}}\left(E\right)=-i\Delta\left(\begin{array}{cc}
\frac{\Delta}{8\mu_{2}}\Sigma & -E(\delta S_{e}^\text{pp})^{*}\\
E\delta S_{e}^\text{pp} & -\frac{\Delta}{8\mu_{2}}\Sigma^{*}
\end{array}\right)\ ,
\label{eq:h_eff_multi}
\end{equation}
with the matrix
\begin{equation}
    \Sigma=(S_{e}^\text{pp})^{*}S_{e}^\text{pe}(S_{e}^\text{ep})^{*}S_{e}^\text{pp}-(S_{e}^\text{pe})^{*}S_{e}^\text{ep}\ ,
\end{equation}
where (in analogy to Sec.~\ref{sec_gap_closing}) all scattering matrices are evaluated (expanded) around $E=0$. This Hamiltonian shares a striking structural similarity with Eq.~\eqref{eq_h_eff_planar}, except that the subblock $S_e-S_e^*$ (representing the deformation of an incoming planar Gaussian wave packet due to finite dispersion) is here replaced with $\Sigma/2$ (due to the coupling between evanescent and planar modes). Notice furthermore, that $\Sigma$ satisfies the symmetry
\begin{equation}
    \Sigma^{*}=-(S_{e}^\text{pp})^{*}\Sigma S_{e}^\text{pp}\ .
\end{equation}
Consequently, we find that the very same eigenvector as in Eq.~\eqref{eq_v0} is a null eigenvector here, if $E=\Delta$, and the condition
\begin{equation}\label{eq_condition_evanescent}
    \delta S_{e}^\text{pp}(S_{e}^\text{pp})^{*}-S_{e}^\text{pp}(\delta S_{e}^\text{pp})^{*}+\frac{1}{4\mu_{2}}\Sigma^{*}=0 \ ,
\end{equation}
is satisfied. In this condition, the superconducting gap $\Delta$ is likewise eliminated, such that this is a condition \textit{exclusively concerning} the scattering matrix (including planar and evanescent modes), irrespective of the presence or absence of Andreev processes. This is a further central result of our work.
While much of the above derivation works in complete analogy to Sec.~\ref{sec_gap_closing}, the two conditions of Eqs.~\eqref{eq_condition_dS} and~\eqref{eq_condition_evanescent} are fundamentally different in nature for the following reason. In Sec.~\ref{sec_gap_closing}, it was possible to relate the conditions for a gap closing to probability conservation of the scattering process itself. Here, this is explicitly impossible, as the correction in Eq.~\eqref{eq_S_eff} stems from virtual excitations of the electrons into the evanescent states, requiring both a coupling to decaying and diverging modes (i.e., the ``detour'' via evanescent modes involves both $S^\text{pe}$ and $S^\text{ep}$), only the former of which are normalisable (as we already discussed in Sec.~\ref{sec_challenges}). Hence, even though the condition concerns only the properties of the scattering matrix (the parameters of the Andreev reflection have been eliminated), they do not refer to a simple probability conservation principle. Therefore,  as already foreshadowed in Sec.~\ref{sec_challenges}, while it is straightforward to anticipate from qualitative reasoning that there should exist a gap closing due to simple dwelling time arguments, the actual demonstration of this fact is difficult. Here, we have managed to find such a nontrivial relationship for the scattering matrix including evanescent modes by means of Andreev physics.

Finally, we emphasise again that condition~\eqref{eq_condition_evanescent} is guaranteed to be satisfied as $|\mu_2|$ approaches zero. In this regime, the length of the lowest evanescent mode diverges, and thus becomes much larger than the size of the scattering region (which will also contribute to a finite energy dependence of the planar part of the scattering matrix). In the next section, we consider a concrete scattering problem, where we can numerically confirm the validity of Eq.~\eqref{eq_condition_evanescent}. Moreover, we will be able to show that, strikingly, the gap closing also holds beyond the perturbative limit, i.e., when neither $\mu_2$ nor $\Delta$ are small.

\section{Geometric scattering and L-junction model}\label{sec_L_junction}

%
\begin{figure}
\includegraphics[width=1\columnwidth,height=1\paperheight,keepaspectratio]{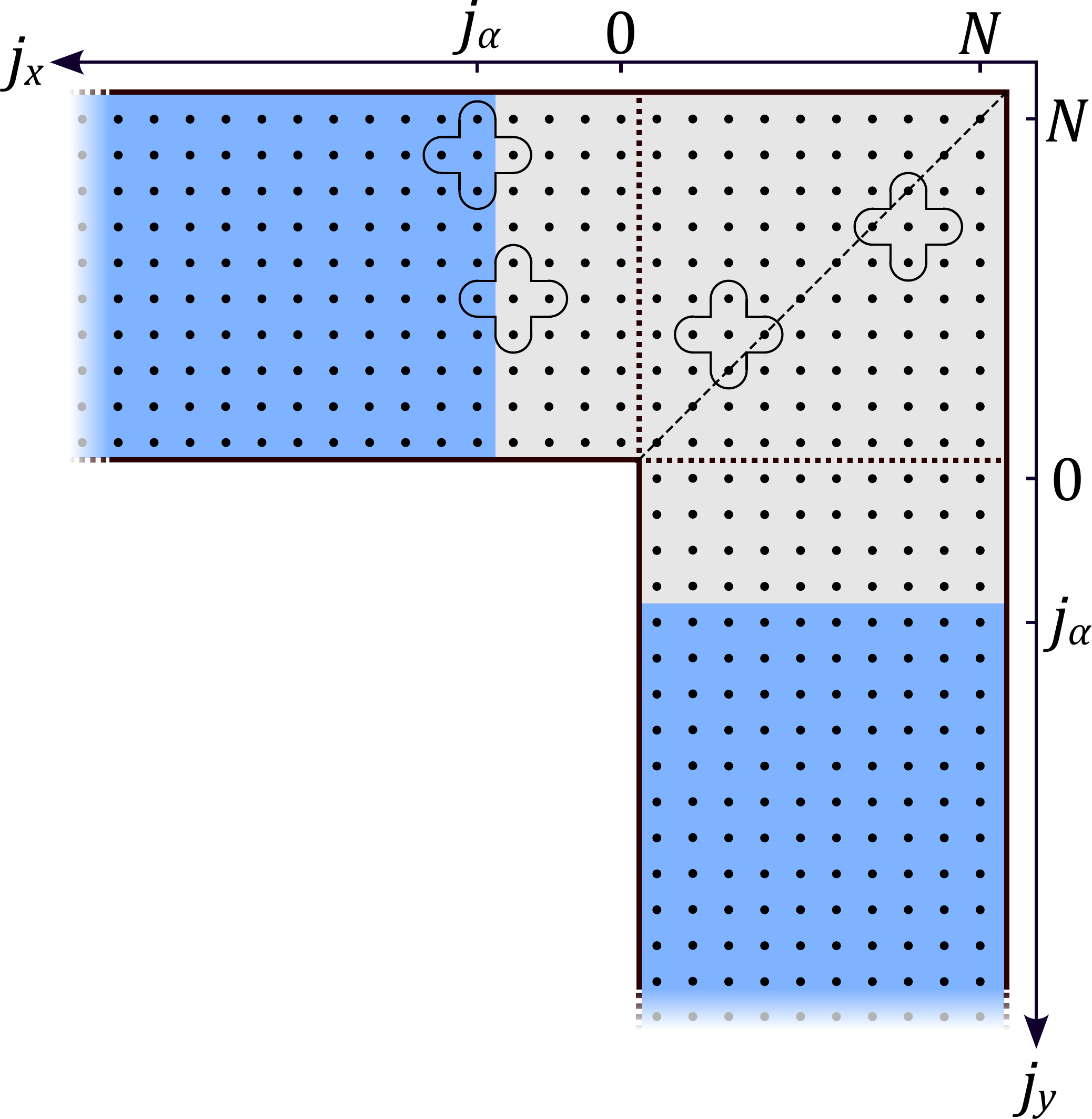}\caption{\label{fig:geometric_junction} Discrete lattice model of a two-terminal junction where scattering occurs across an inverted L-geometry. Owing to the junction geometry, translational invariance is broken, such that there is significant cross-channel coupling. Lattice points that are proximitised by superconducting regions are marked in blue. These regions extend over~${ j_x,j_y \in \left[-\infty, j_\alpha \right] }$, where~$j_\alpha$ denotes the length of the conductor arms in between the scattering centre~(${j=0}$) and the SN-interface~(${j=j_\alpha}$). Transversal wave function components are confined by the width of the conductor arms, where~$N$ is the number of transversal lattice points. In order to compute the scattering matrix, it is useful to consider cross-shaped plaquettes linking neighbouring lattice points due to the tight-binding  tunnel hopping.}
\end{figure}
%

\noindent
As announced, the above framework, as well as the nontrivial aspects regarding the detaching of the ABS spectrum at the superconducting gap $\Delta$ can be illustrated with a simple, yet experimentally highly relevant example device  with non-trivial geometry of the scattering centre. Specifically, we consider a 2D conductor in a two terminal configuration with horizontal and vertical arms, such that scattering occurs at an inverted L-profile (cf. Fig.~\ref{fig:geometric_junction}), which is strongly inspired by recent experimental setups~\mbox{\cite{Koelzer_2021, Koelzer_2023, behner_2023}}. 
In the absence of an additional scalar potential (or impurities), here, the scattering is completely dominated by the geometry of the device (henceforth referred to as \textit{geometric scattering}), as it mixes the longitudinal and transversal components of the wave function at the scattering region. This makes it an ideal and well-controllable example for a scattering scenario where translational invariance is absent. While we here focus on a two-terminal device for simplicity, note that for generic multiterminal junctions, geometric scattering can always be expected to be present.

%
\begin{figure}
\includegraphics[width=1\columnwidth,height=1\paperheight,keepaspectratio]{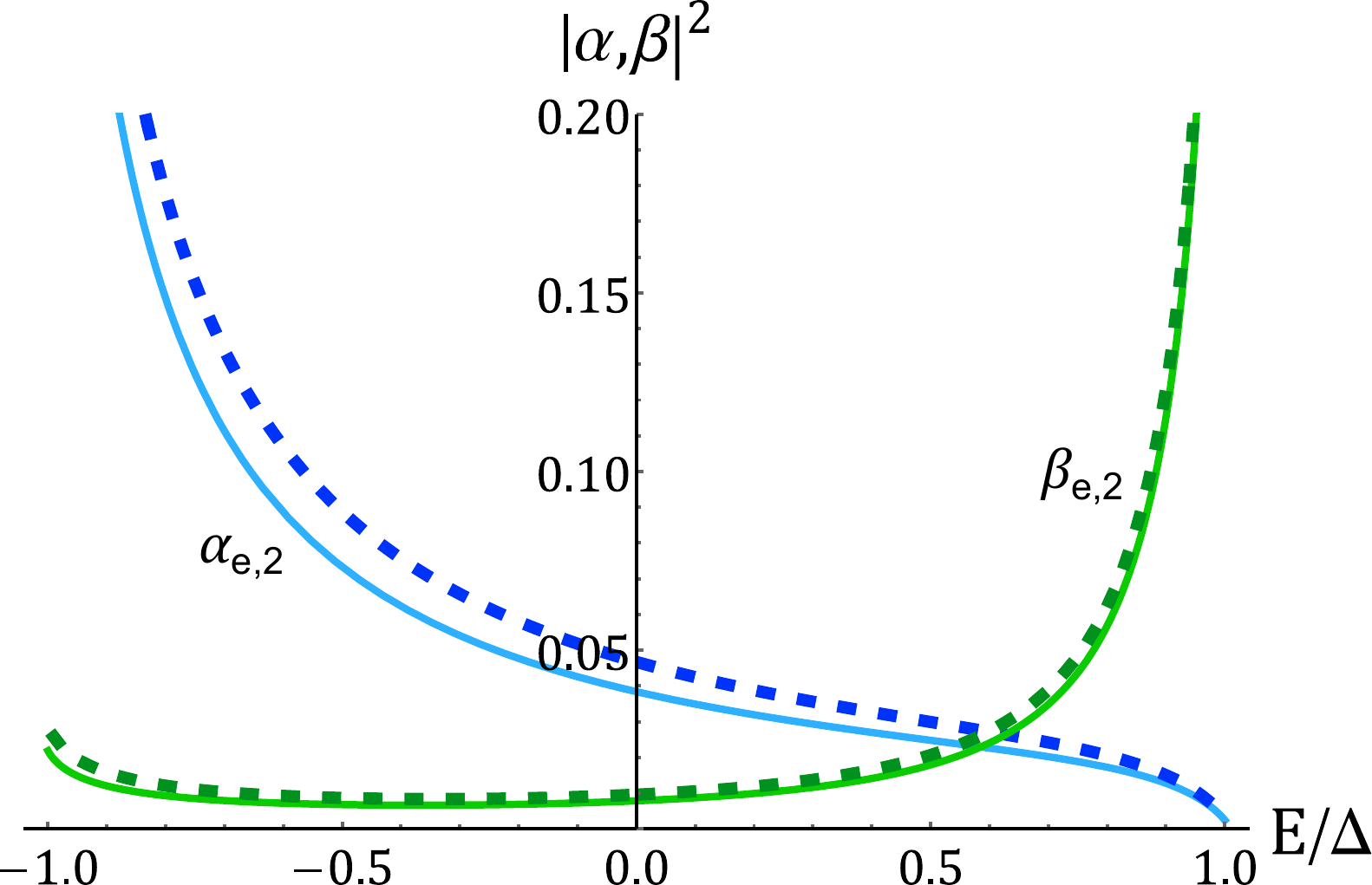}\caption{Comparing the analytic approximation (solid) of the reflection- and backscattering coefficients $\alpha_{\mathrm{e,2}}, \beta_{\mathrm{e,2}}$ to exact numerical computation (dashed) for the first evanescent mode. Here, dispersion is strong in the first evanescent mode~${n=2}$, with~${\mu_2/\Delta=1.01}$,~${\mu_1/\Delta=-72.20}$ and~$N=5$ transversal lattice points. Thus, energy dependence of the reflection coefficients is beyond the perturbative limit, such that the linear expansions, Eqs.~\eqref{eq_Andreev_evanescent} and~\eqref{eq_Andreev_evanescent_back} do not apply. Instead, the coefficients follow the full analytic expressions, Eqs.~\eqref{eq:alpha_factor_evan} and~\eqref{eq:beta_factor_evan}.
}
\label{Fig:analytic_alpha_beta}
\end{figure}
%

%
\begin{figure}
\includegraphics[width=1 \columnwidth,height=1\paperheight,keepaspectratio]{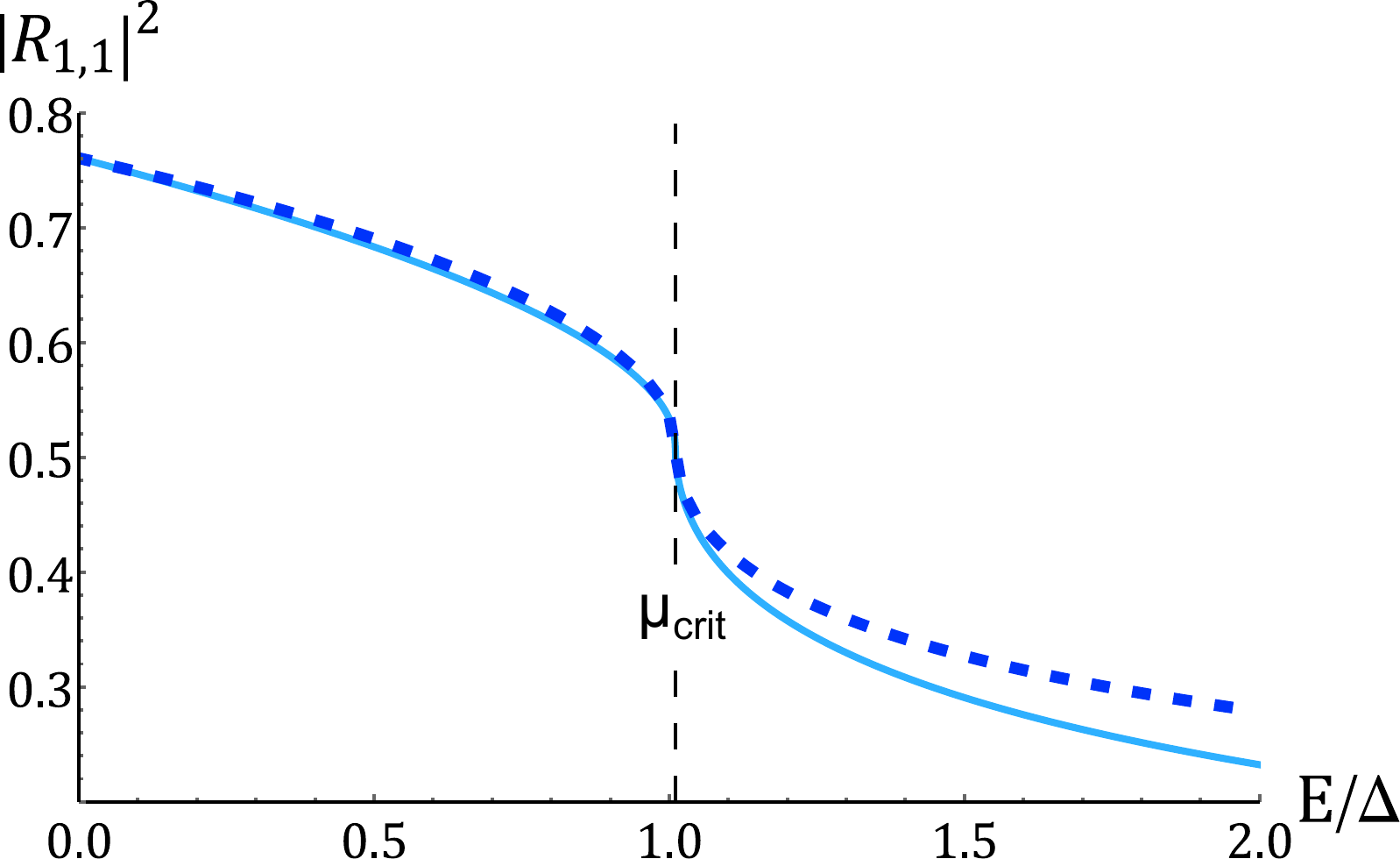}\caption[Caption for LOF]{\label{fig:sqrt-law} Square root behaviour of the planar-planar reflection coefficient as a function of energy. Here, energies are taken with respect to the reference energy scale~$\Delta$\textsuperscript{\ref{foot_fig:sqrt-law}}.
The chemical potential~$\mu/\Delta=98.99$ is tuned such that the second mode is evanescent (${\mu_2=\mu_\mathrm{crit}>0}$) and has strong non-linear dispersion with~$\mu_2/\Delta=1.01$. The planar-planar reflection coefficient is computed numerically (dashed curve) with~$N=5$ transversal lattice points for different energies (the plot shows its absolute squared~$\left| R_{1,1} \right|^2$). Although the planar channel has negligible dispersion ($\mu_1/\Delta=-72.20$), energy dependence is strong due to cross-channel scattering with the second mode and follows a square root law (solid curve), in accordance with Eq.~\eqref{eq:sqrt-relation}. 
}
\end{figure} 
%
%

For convenience, we solve the normal metal scattering problem on a finite discrete lattice. Neglecting orbital motion in the third dimension (valid, e.g., for a \mbox{2-dimensional} electron gas \cite{ Wan_2015, Shabani_2016, Kjaergaard_2016, Hart_2017, Casparis_2018, Moehle_2021, TenHaaf_2024, Wang_2024}), we write the Hamiltonian simply as a tight-binding hopping model in 2D,
\begin{equation}
    H_{\mathrm{N}}=H_{x}\otimes\mathds{1}_{y}+\mathds{1}_{x}\otimes H_{y}-\mu \ .
    \label{eq:n_region_hamiltonian}
\end{equation}
First, the model is solved for the translational invariant case (i.e. infinitely extended conductors) in both the normal- and superconducting regions (grey and blue regions in Fig.~\ref{fig:geometric_junction} accordingly). Normal scattering as well as SN-reflection are then solved numerically by ensuring Schrödinger equation to hold at every lattice point. To this end, the solutions of the translational invariant system are projected onto the diagonal ${j_x=j_y}$ in the scattering region and the SN-interface at ${\left\{ j_x, j_y \right\}=j_\alpha}$. Owing to the hopping Hamiltonian, the resulting matching conditions consist of plaquettes linking adjacent lattice sites at the boundaries, as depicted in Figure~\ref{fig:geometric_junction}. The details of this numerical procedure~\mbox{--}~which can in the following be used as a benchmark for the relations derived above~\mbox{--}~are outlined explicitly in Apps.~\ref{sec_Ljunction_scattering} (scattering) and~\ref{sec:App_R_reflection} (SN-reflection). \footnotetext{\label{foot_fig:sqrt-law}Note that for the quantities considered here,~$\Delta$ does not actually represent a physically relevant energy scale. Nonetheless, we choose it as the reference energy scale, since $\Delta$ is obviously the crucial reference energy for ABS physics (such that all results and data throughout this work can conveniently be compared to each other).}

Alternatively, both the $R$- and $S$-matrices can be computed based on closed analytic expressions. The components of the reflection matrix $R$ can be determined from the explicit expressions for reflection $\alpha_{\nu,n}$ and backscattering $\beta_{\nu,n}$ presented in Sec.~\ref{sec:recipe}, adjusted for the discrete model (cf. Appendix~\ref{sec:App_R_reflection}). The scattering coefficients in the $S$-matrix can be interpolated by applying the square root law introduced in Sec.~\ref{sec_cross_channel_evanescent} (cf. Appendix~\ref{sec_Ljunction_scattering}). As depicted in Figs.~\ref{Fig:analytic_alpha_beta} and~\ref{fig:sqrt-law}, the fitted coefficients agree very well with the full numerical computation. Correspondingly, as we show below, the ABS spectra obtained with these two methods align to very high accuracy.

\begin{table*}

\caption{
Numerical test of the validity of the $S$-matrix  constraint (including evanescent modes) given in Eq.~\eqref{eq_condition_evanescent}. We consider the case where the second channel is evanescent (${\mu_2>0}$) and the number of transversal lattice points is~${N=5}$. As $\mu_2/t$ decreases, dispersive effects become more and more pronounced. The table shows the terms~${\Sigma^{\ast}/\left(4\mu_{2}\right)}$ and~${\Omega=\delta S_{e}^{\text{pp}}(S_{e}^{\text{pp}})^{*}-S_{e}^{\text{pp}}(\delta S_{e}^{\text{pp}})^{*}}$ in Eq.~\eqref{eq_condition_evanescent} and its absolute deviation ${\delta=\Omega/\mathrm{i}-\mathrm{i}\Sigma^{\ast}/\left(4\mu_{2}\right)}$ in relation to the inverse energy scale $1/t$. ${\left\Vert \delta\right\Vert /\left\Vert \Omega\right\Vert }$ is the relative deviation with the matrix norm squared~${ \left\Vert \delta\right\Vert^2= \sum_{n,m} \left| \delta_{n,m} \right|^2  }$. The absolute deviation~$\delta$ due to the finite extension of the scattering region stays constant, whereas the relative deviation~${\left\Vert \delta\right\Vert /\left\Vert \Omega\right\Vert }$ indeed decreases with increasing dispersion. We recall that the full scattering matrix including evanescent modes is evaluated at~${E=0}$, consistent with the discussion after Eq.~\eqref{eq:h_eff_multi}.   
}

\fontsize{8.5}{10}\selectfont
\begin{tabular*}{1\linewidth}{@{\extracolsep{\fill}}@{\extracolsep{\fill}}@{\extracolsep{\fill} }lcccc}
\toprule 
\toprule 
\noalign{\vspace{0.1cm}}
\multicolumn{1}{l}{
$\mu_2/t$} & $t\Omega/\mathrm{i}$  & $\mathrm{i}t\Sigma^{\ast}/\left(4\mu_{2}\right)$  & $t\delta$  & $ {\left\Vert \delta\right\Vert/\left\Vert \Omega\right\Vert} $\tabularnewline
\noalign{\vspace{0.1cm}}
\midrule
\noalign{\vspace{0.05cm}} 
$10^{-3}$  & $\begin{pmatrix}124.8 & 81.1\\
81.1 & 124.8
\end{pmatrix}$  & $\begin{pmatrix}106.5 & 77.0\\
77.0 & 106.5
\end{pmatrix}$  & $\begin{pmatrix}18.2 & 4.2\\
4.2 & 18.2
\end{pmatrix}$  & $0.125$\tabularnewline
\noalign{\vspace{0.05cm}} 
\midrule
\noalign{\vspace{0.05cm}} 
$10^{-4}$  & $\begin{pmatrix}366.0 & 251.8\\
251.8 & 366.0
\end{pmatrix}$  & $\begin{pmatrix}348.0 & 248.1\\
248.1 & 348.0
\end{pmatrix}$  & $\begin{pmatrix}18.0 & 3.8\\
3.8 & 18.0
\end{pmatrix}$  & $0.041$\tabularnewline
\noalign{\vspace{0.05cm}} 
\midrule
\noalign{\vspace{0.05cm}} 
$10^{-5}$  & $\begin{pmatrix}1128.3 & 791.6\\
791.6 & 1128.3
\end{pmatrix}$  & $\begin{pmatrix}1110.5 & 787.9\\
787.9 & 1110.5
\end{pmatrix}$  & $\begin{pmatrix}17.9 & 3.7\\
3.7 & 17.9
\end{pmatrix}$  & $0.013$\tabularnewline
\noalign{\vspace{0.05cm}} 
\midrule 
\noalign{\vspace{0.05cm}} 
$10^{-6}$  & $\begin{pmatrix}3539.2 & 2498.2\\
2498.2 & 3539.2
\end{pmatrix}$  & $\begin{pmatrix}3521.3 & 2494.5\\
2494.5 & 3521.3
\end{pmatrix}$  & $\begin{pmatrix}17.9 & 3.6\\
3.6 & 17.9
\end{pmatrix}$  & $0.004$\tabularnewline
\noalign{\vspace{0.05cm}} 
\bottomrule
\bottomrule
\end{tabular*}

\label{tab:detaching_compensation} 
\end{table*}
%

At this stage, we can explicitly evaluate all scattering amplitudes for the L-junction problem. In particular, this allows us to numerically check relation~\eqref{eq_condition_evanescent}, which followed from the analytical treatment presented in Sec.~\ref{sec_constraint_evanescent}, assuming~${ \Delta/\mu_2\ll 1 }$ while still allowing for~${ \mu_2 \to 0 }$. As established in the same section, if this constraint on the scattering matrix holds, we can deduce that energy solutions at~${ E=\Delta }$ exist in the ABS spectrum at~${ \phi=0 }$~\mbox{--}~\mbox{notably}, without requiring anything beyond solving the scattering matrix itself. To this end, we compute the deformation term~${ \Omega=\delta S_{e}^{\text{pp}}(S_{e}^{\text{pp}})^{*}-S_{e}^{\text{pp}}(\delta S_{e}^{\text{pp}})^{*} }$, the compensation term~${ \Sigma^{\ast}/\left(4\mu_{2}\right) }$ and the difference of the two 
${ \delta=\Omega/\mathrm{i}-\mathrm{i}\Sigma^{\ast}/\left(4\mu_{2}\right) }$. If ${\delta=0}$, then the constraint of Eq.~\eqref{eq_condition_evanescent} would hold exactly. As shown in Table~\ref{tab:detaching_compensation}, the deviation~$\delta$ is small but finite and the relative deviation~${ \left\Vert \delta\right\Vert /\left\Vert \Omega\right\Vert }$ indeed converges with increasing finite dispersion~$\mu_2 \to 0$, readily confirming relation~\eqref{eq_condition_evanescent}. It is interesting to note that the absolute deviation~$\delta$ stays constant. This can be attributed to the finite size of the central scattering region in the L-junction model (cf. Figure~\ref{fig:geometric_junction}), which obviously stays constant when changing the chemical potential. Consequently, a small residual probability density indeed remains within the finite scattering region. Importantly, these two contributions can be distinguished due to their respective behaviour with respect to $\mu_2$ ($\sim 1/\mu_2$ versus constant). Thus, the here presented numerical analysis explicitly confirms one of our main results, Eq.~\eqref{eq_condition_evanescent}, indicating the existence of surprising constraints on the energy-dependant scattering matrix involving evanescent modes, which do not follow from probability conservation. Below, we go even further in the following sense. While Eq.~\eqref{eq_condition_evanescent} follows from perturbative arguments when $\Delta$ is sufficiently small, we show below that the gap closing at ${E=\Delta}$ persists even non-perturbatively. Therefore, in what follows we proceed towards a full calculation of the Andreev bound state spectrum.

%
\begin{figure}
\includegraphics[width=1\columnwidth,height=1\paperheight,keepaspectratio]{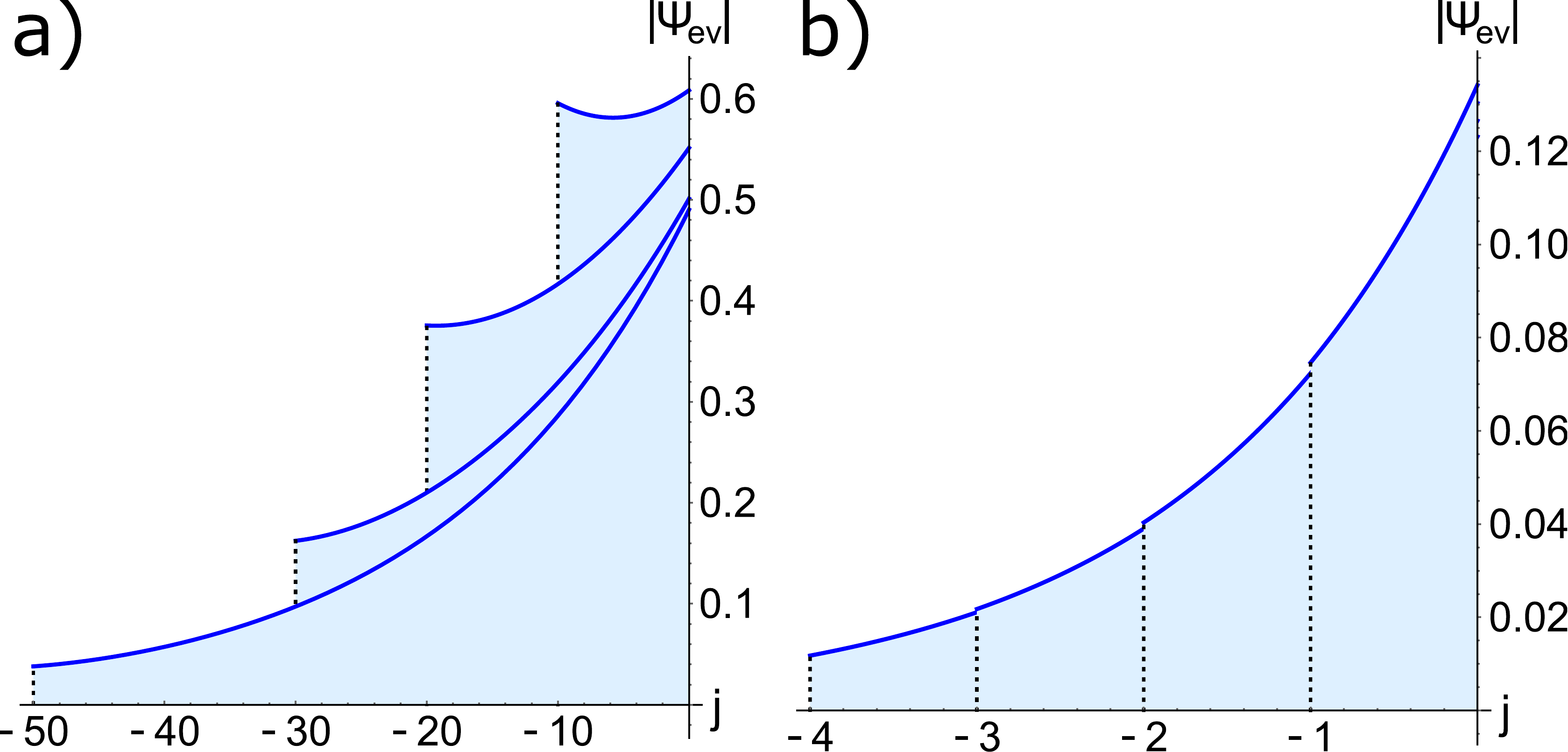}\caption{ Modulus of the first evanescent wave function component in the conductor arms between the scattering region~(${j=0}$) and  the SN-interface~(${j=j_\alpha}$) for different arm lengths~$j_\alpha$. In both panels only the first channel is planar~(${\mu_{n \geq 2}>0}$), the number of transversal lattice points is~${N=5}$ and the hopping potential is chosen as~${t/\Delta=100}$. In panel~a), the chemical potential is chosen, such that there is finite dispersion in the first evanescent channel~(${ \mu_2 / \Delta= 1.01}$ and ${ \mu_1 / \Delta= -72.19}$) and the lengths of the conductor arms are~${j_\alpha \in \left\{ -10, -20, -30, -50 \right\}}$. In Panel~b) dispersion is negligible for all channels, as~(${ \mu_2 / \Delta= 36.6}$ and ${ \mu_1 / \Delta= -36.6}$) with arm lengths~${j_\alpha \in \left\{ -1, -2, -3, -4\right\}}$. The (Andreev reflected) diverging part of the evanescent wave function component becomes stronger for shorter arm lengths~${j_\alpha \to 0}$, if (and only if) dispersion is strong in the corresponding evanescent mode [panel~a)]. In contrast, if dispersion is negligible [panel~b)], evanescent Andreev reflection is absent, such that only the decaying part contributes appreciably to the evanescent component, even though the evanescent mode touches the SN-interface. 
\label{fig:Evanescent Psi}
}
\end{figure}
%

\subsection{Non-perturbative gap closing and long ballistic junctions}

\begin{figure*}
\begin{centering}
\includegraphics[width=1\textwidth,height=0.45\textwidth,keepaspectratio]{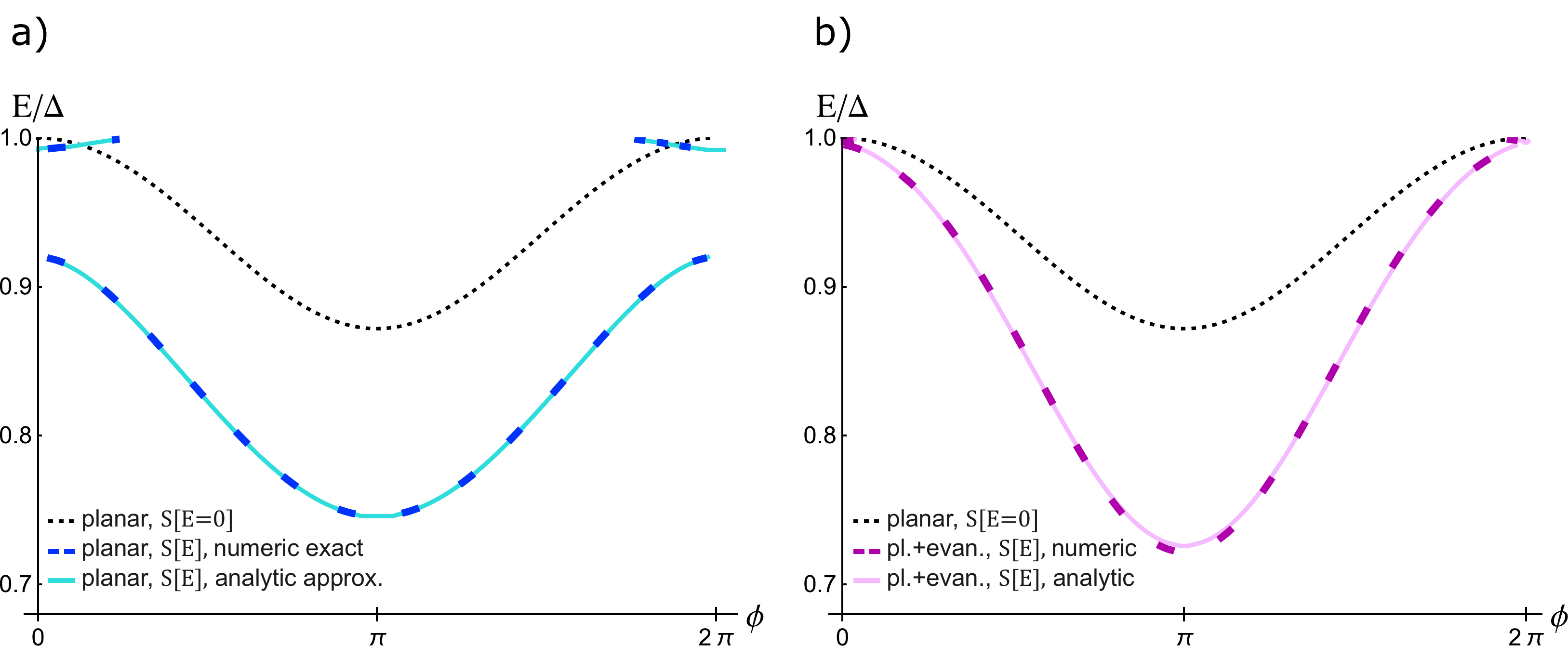}\par\end{centering}
\caption{
\label{fig:ABS_spectrum_short} Impact of strongly dispersive evanescent cross-channel scattering on the bound states spectrum. The ABS spectra are computed in the extreme short junction limit ($j_{\alpha}=0$) with finite dispersion in the first evanescent channel~(${\mu_2/\Delta=1.01}$, ${\mu_1/\Delta=-72.20}$) and~$N=5$ transversal lattice sites. The black dotted curve depicts the standard result with energy dependant scattering and evanescent modes neglected, see Eq.~\eqref{eq:conv_ABS_spectrum}. Panel a): if energy dependence (stemming from evanescent finite dispersion) is fully included in $S$, but evanescent channels are still discarded for $S$ and $R$, the spectra detach from the continuum with very good consistency between the numerical (dashed blue curve) and analytic (solid blue curve) computations. Panel b): if the calculation fully includes evanescent channels explicitly, (both in the scattering matrix~$S$ as well as in the reflection matrix~$R$) the detachment is compensated. Again, the numerical (dashed purple curve) and analytic (solid purple curve) treatment are in very good alignment. 
}
\end{figure*}
%

\noindent
With the formalism for the discrete L-junction model in place, we proceed with calculating the full ABS spectrum. We still focus on the regime where the chemical potential is chosen such that we have only one planar mode, but we are close to the transition (i.e., $\mu_2$ is close to, but above~$0$), where the evanescent mode of the second channel becomes important.

As shown above, see in particular Table~\ref{tab:detaching_compensation}, the scattering matrix satisfies the analytical constraint, Eq.~\eqref{eq_condition_evanescent}, derived in Sec.~\ref{sec_constraint_evanescent}, valid in the perturbative limit of ${\Delta/\mu_2\ll 1}$ (while ${\mu_2\ll \left|\mu_1\right|}$). We could therefore conclude already without explicit calculation of the ABS spectrum, that the gap closes at ${E=\Delta}$ in this perturbative limit. Here, we show by calculating the full ABS spectrum, that, notably, the effect of the gap closing persists  even beyond the above perturbative limit. In fact, the only parameter that can still have an influence on whether a gap closing at $\phi=0$ is present or not is the distance between the scattering region and the SN interface, parameterised by the lattice position $j_\alpha$, see Fig.~\ref{fig:geometric_junction}. Again, we assume for simplicity that there are no impurities within the conductor arms, such that the system for large $j_\alpha$ corresponds to a long ballistic junction.

Computing the normal region wave function, we can track the behaviour of the evanescent component in between the scattering centre and the SN-interface for different arm lengths $j_\alpha$. If~$\psi^\mathrm{in}$ ($\psi^\mathrm{out}$) is the incoming (outgoing) wave vector right at the scattering region~${j=0}$, the first evanescent wave function component~${n=n_\mathrm{P}+1}$ then consists of the superposition ${\psi_{\nu,n}=\psi^\mathrm{in}_{\nu,n}P_{n,n}+\psi^\mathrm{out}_{\nu,n}P^{-1}_{n,n}}$, where~$P$ is the propagator as introduced at the end of App.~\ref{sec:App_R_reflection}. As Fig.~\ref{fig:Evanescent Psi} depicts, for increasing~${ j_\alpha}$, real exponentials exiting the scattering region indeed increasingly drop off before reaching the SN-interface. Moreover, the qualitative behaviour of the wave function strongly depends on the ratio $\Delta/\mu_2$. In the nonperturbative regime $\Delta \approx \mu_2$ [Fig.~\ref{fig:Evanescent Psi}a)], there is a much more pronounced diverging component within the evanescent mode as compared to the perturbative regime $\Delta \ll \mu_2$ [Fig.~\ref{fig:Evanescent Psi}b)], resulting, accordingly, in a much slower decay of the wave function away from the scattering region.

The ABS spectrum in the non-perturbative regime of~${\Delta\approx \mu_2}$ is depicted in Fig.~\ref{fig:ABS_spectrum_short}. First, we consider the extreme short junction limit in the sense that not only the scattering region is small (given by the number of transversal lattice points $N$), but also~$j_\alpha=0$. In fact, it is particularly instructive to compare different frameworks and approximation schemes, to appreciate the impact of the various different mechanisms that have been discussed throughout this work.

If we apply the standard literature framework outlined in Sec.~\ref{sec_status_quo}, i.e., the scattering matrix is approximated for constant energy~$S[E=0]$ and evanescent modes are discarded (such that the matrix $R$ is computed as in Eq.~\eqref{eq:planar_Andreev_matrix}), we arrive at the dotted black curve in Fig.~\ref{fig:ABS_spectrum_short}. Here, the energy spectrum touches the gap $\Delta$ at $\phi=0$ simply due to the unitarity of the scattering matrix. Let us now include the full energy dependence of the scattering matrix~$S\left[ E \right]$, while \textit{still} only retaining planar modes in the computation of the ABS spectrum. In essence, the finite dispersion effects of the evanescent channels are here only indirectly included in the computation of the scattering matrix, resulting in a strongly energy-dependant $S$ when approaching small values for $\mu_2$. Consequently, the bound states spectrum indeed detaches from the continuum at~${E=\Delta}$ (cf. blue dashed curve in Figure~\ref{fig:ABS_spectrum_short}a)), even though the electrons spend very little time within the scattering region. This detaching is in alignment with the chiral symmetry breaking argument brought forth in Sec.~\ref{sec_gap_closing}. However, as amply explained throughout this work, this detaching is here not the correct result, due to the discarding of the evanescent modes in the boundary conditions for the ABS spectrum. Thus, the third and last step consists of fully taking into account the evanescent modes in both~$S$ and~$R$. To repeat, in alignment with the discussion after Eqs.~\eqref{eq_Andreev_evanescent} and~\eqref{eq_Andreev_evanescent_back}, it is perfectly sufficient to only include the first evanescent mode, since all higher modes have vanishing dispersion (In fact, computing the ABS spectrum with two or three evanescent modes included, yields the exact same curves.). This full calculation indeed provides a bound states spectrum where the detaching is fully compensated, as depicted by the purple dashed curve in Fig.~\ref{fig:ABS_spectrum_short}b). Again, we stress that here the revival of the gap closing at energies close to $\Delta$ is shown in the non-perturbative regime of $\Delta\approx \mu_2$ (and thus far beyond the regime considered in Sec.~\ref{sec_constraint_evanescent}). This result therefore very likely indicates the existence of non-perturbative constraints on the scattering matrix, beyond Eq.~\eqref{eq_condition_evanescent}, allowing to relate scattering matrices (including evanescent scattering processes) at different energies, a question which can be pursued in follow-up works.

%
\begin{figure}
\begin{centering}
\includegraphics[width=1\columnwidth,height=1\paperheight,keepaspectratio]{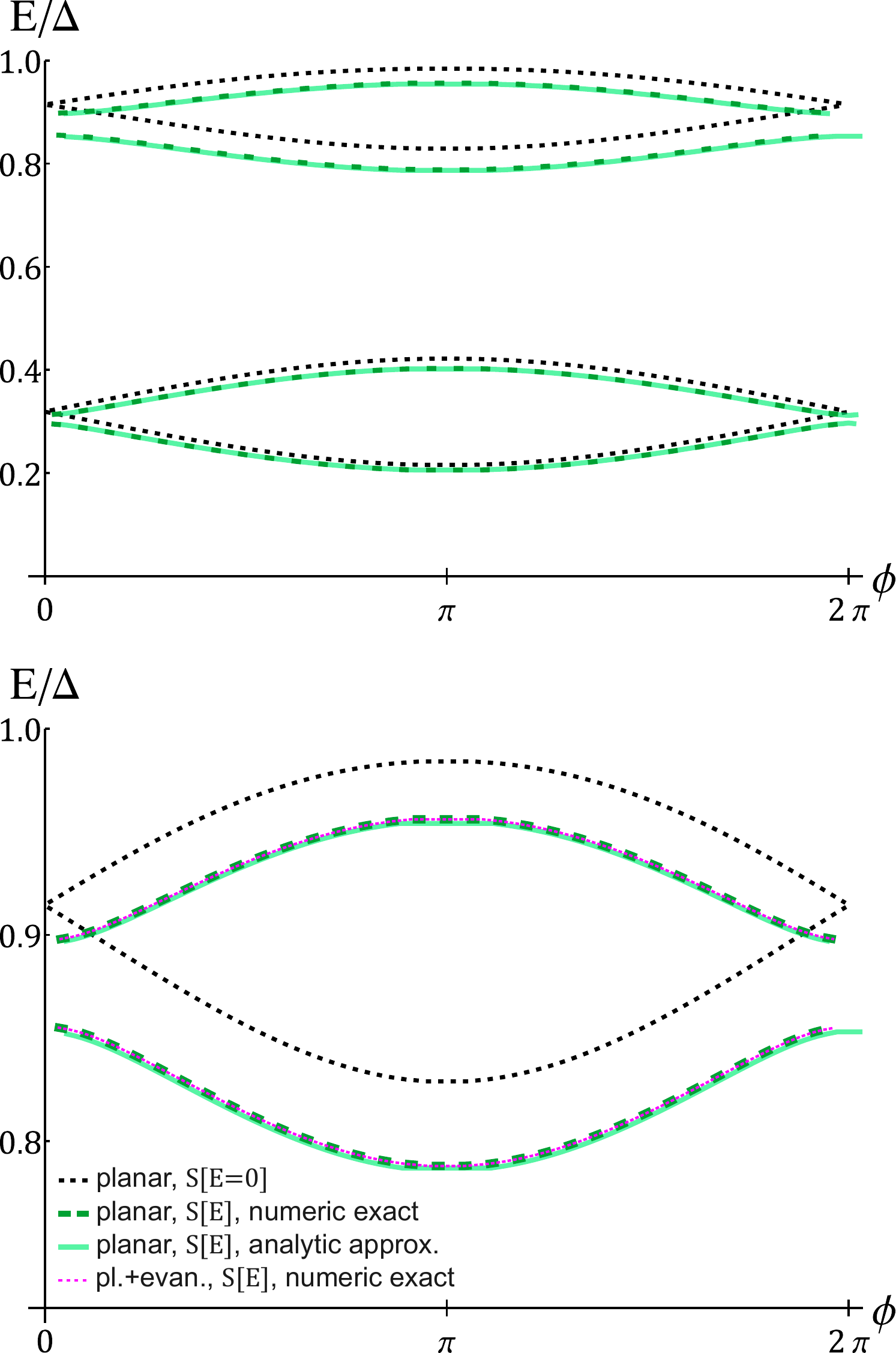}
\par\end{centering}
\caption{\label{fig:ABS_spectrum_long}
ABS spectra in the long junction limit ($j_{\alpha}=-300$) with~${ N=5 }$ transversal lattice points and finite dispersion in the first evanescent channel~(${\mu_2/\Delta=1.01}$, ${\mu_1/\Delta=-72.20}$). The black dotted curve depicts the standard result with energy dependant scattering and evanescent modes neglected, see Eq.~\eqref{eq:conv_ABS_spectrum}. The green dashed (solid) line depicts a numerical (analytic) calculation with only planar modes considered, but energy dependence fully included. The purple dotted curve is a numerical calculation, additionally taking evanescent modes into account. Consequently, in the long junction limit, the only relevant effect to take into account is the energy-dependence of $S$ due to strong dispersion. Contrary to the short junction limit (Fig.~\ref{fig:ABS_spectrum_short}), here, the inclusion of evanescent modes in the $S$- and $R$-matrices does not preserve a gap closing.
}
\end{figure}
%

Having determined the bound states spectra by exact numerical computation, we now compare with the aforementioned analytic approximations. To this end, the $R$-matrix is computed by the analytic expressions for the reflection- and backscattering coefficients $\alpha$,~$\beta$,~Eqs.~\eqref{eq:alpha_factor_evan} and~\eqref{eq:beta_factor_evan}, with applying the discrete dispersion as explained in the previous section. Next, the scattering matrix~${S\left[ E \right]}$ can be interpolated by the square root law,~Eq.~\eqref{eq:sqrt-relation}, along the energy range~${E\in \left[-\Delta, \Delta \right]}$. Solving for the ABS spectrum, the results (solid curves in Figs.~\ref{fig:ABS_spectrum_short} and~\ref{fig:ABS_spectrum_long})  are in extremely good agreement with the numerical calculations (dashed curves) across all regimes. This indicates the validity of a number of helpful approximation schemes that allow for a more efficient computation of the ABS spectra in the here considered more general regimes, including evanescent modes, cross-channel scattering, and finite dispersive effects.

Finally, we discuss the $j_\alpha$-dependence of the ABS spectrum. As pointed out at the beginning of this section, when increasing $j_{\alpha}$, the evanescent modes decay within the conductor arms, see also Fig.~\ref{fig:Evanescent Psi}. As we approach the limit of large $j_\alpha$, two things happen. One the one hand, at each value of $\phi$, we now obtain more than one ABS state. This is a well-known consequence of the finite (ballistic) propagation time inside the conductor arms~\mbox{--}~\mbox{similar} in spirit to the increasing number of eigenenergies within a given energy window of a simple particle in a box. On the other hand, the impact of evanescent modes on the ABS spectrum should decrease since the evanescent modes no longer are able to hit the SN-interface.

And indeed, this expectation can be confirmed as follows. Importantly, note that, even though the number of states increases due to large $j_\alpha$, there still remains a ``shadow'' of the gap closing in the standard literature framework (black dotted line) in Fig.~\ref{fig:ABS_spectrum_long}. However, instead of the ABS energies touching $\Delta$, there instead appear gap closings between the multiple ABS states at energies below $\Delta$, but still at $\phi=0$, see, e.g., Fig.~\ref{fig:ABS_spectrum_long}b) (black dotted lines). If we now, again, include a finite energy dependence, a gap opens at $\phi=0$ (solid and dashed lines). Crucially though, here the compensating effect due to evanescent modes is absent, such that the detaching survives as a real phenomenon in the long junction regime. Consequently, in contrast to the short junction limit, the explicit inclusion (dotted purple line) or omission of evanescent modes in the calculation of the ABS spectrum by means of the determinant equation, Eq.~\eqref{eq:gen_Beenakker_equation}, has no impact whatsoever on the ABS spectrum. The only place where evanescent modes matter here, is indirectly through the proper calculation of the scattering matrix $S$ itself, where they need to be included even if one is only interested in scattering processes between planar waves.

\section{Conclusion and outlook}

\noindent
We provided a framework for the description of superconducting weak links in terms of a generic, energy-dependant scattering matrix, including evanescent modes and cross-channel coupling for general multichannel contacts. We provided an in depth analysis of the different origins of an energy-dependence of the scattering matrix, distinguishing in particular between a finite size of the scattering region itself (where the energy dependence can be linked to the inverse dwelling time of electrons) and a finite dispersion in the conductor arms. In particular, we argued that the latter mechanism can be (and in general \textit{is}) large, when the chemical potential is tuned close to a regime where the number of planar channels changes. We further demonstrated that if energy-dependence in the scattering matrix is the only feature included in Beenakker's determinant equation, then it predicts a detaching of the ABS spectrum between bound and extended quasiparticle states (i.e., at energies equal to $\Delta$). We further show that if the energy-dependence comes dominantly from a finite dispersion (while the actual dwelling time remains negligible), then the detaching is spurious -- and a more complete theoretical treatment still predicts a gap closing. By means of an effective Hamiltonian description of the weak link, we provide a connection between the gapping and the breaking of a chiral symmetry. Depending on whether the dominant dispersive channel is planar or evanescent, there are distinctly different lines of argumentation as to how the chiral symmetry, and thus the gap closing, are restored. For a dispersive planar mode, we argue that there are probability conservation constraints on the scattering matrix beyond unitarity, which can be derived through time-dependant considerations of the scattering process. Crucially, we show that probability conservation arguments are not applicable if the dispersion stems from an evanescent mode, since not all scattering wave functions containing those modes can be normalised. Nonetheless, by careful consideration of the generalised formalism to compute the ABS eigenspectrum, we find new types of constraints on the energy-dependant scattering matrix including coupling to evanescent modes, valid in the perturbative limit of weak dispersion compared with $\Delta$. Finally, this constraint, and the gap closing in a nonperturbative regime are explicitly demonstrated numerically, with the example of a ballistic L-junction. We find that the emergence of gaps in the ABS spectrum at zero phase bias ($\phi-0$) are only physical when increasing the length of the conductor arms, such that evanescent modes decay before they collide with the SN interface.

The effects uncovered here, manifesting as distinct phenomena in the ABS spectra, are ideally suited for experimental verification, which can be carried out using well-established transport measurement techniques or ac spectroscopy tools. For instance, the presence or absence of a detaching in the spectrum shows in the finite-frequency absorption spectra of processes, such as the ejection of a bound quasiparticle state into the continuum~\cite{Kos_2013,Riwar_2015}. Whether or not the ABS spectrum and continuum touch also effects the current as a function of an applied DC voltage~\cite{Michelsen_2010,Eriksson_2017}, as Landau-Zener transitions are known to have a marked dependence on the gap size. In particular, the impact of the lengths of the conductor arms (to control the magnitude of participation of evanescent modes at the SN interface) could be checked by means of geometric junctions with sufficiently high purity, such that transport remains approximately ballistic within the unproximitised part of the device. Note that it is not necessarily required to build very long arms (which could risk bringing the system out of the ballistic regime): the length scale of the evanescent modes is conveniently controlled through in situ tuning of the chemical potential.  

We further note that Majorana-based junctions form an interesting special case within our framework. While the above uncovered mechanisms cannot destroy topological protection of the fractional Josephson effect (for Majorana-based junctions, the chiral symmetry in Sec.~\ref{sec:detaching_from_continuum} refers to the actual energy and not the above introduced pseudo energy $\widetilde{E}$), they are nonetheless of significant relevance for the energy spectrum in the vicinity of the gap -- especially when the Majorana coupling is large (e.g., weak magnetic impurity scattering~\cite{Fu_2008}). We therefore expect our work to be of importance for a precise description of a wide variety of junction types, including Majorana-based junctions.

Overall, we expect this work to be of relevance for multi-pronged future research efforts. On the one hand, energy gaps, especially those close to the continuum, play an important role in the proper description of time-dependently driven junctions, either regarding Landau-Zener transitions coupling the discrete and continuous states (similar in spirit to Ref.~\cite{Eriksson_2017}) or with respect to their dissipative behaviour~\cite{Zazunov_2005,Michelsen_2010}. On the other hand, unitarity and other constraints of the scattering matrix are well-known to play a crucial role in the context of random matrix theory~\cite{Stone_1991,Brouwer_1997,Pichard2001,Nazarov_Blanter_2009,Riwar_2015,Riwar_2016}. This work is to the best of our knowledge the first to consider constraints regarding the \textit{energy-dependence} of the scattering matrix. We therefore consider it likely that our work provides a basis for subsequent studies linking the scattering matrix at different energies in a nontrivial way.

\section*{Acknowledgements}

    \noindent
    We acknowledge interesting and fruitful discussions with Björn Trauzettel, Patrik Recher, Kristof Moors and Abdur Rehman Jalil. This work has been funded by the German Federal Ministry of Education and Research within the funding program Photonic Research Germany (contract No. 13N14891) and by the Bavarian Ministry of Economic Affairs, Regional Development and Energy within Bavaria’s High-Tech Agenda Project “Bausteine für das Quantencomputing auf Basis topologischer Materialien mit experimentellen und theoretischen Ansätzen” (grant No. 07 02/686 58/1/21 1/22 2/23).

\appendix

\section{Computing single channel ABS spectrum with finite dispersion}\label{app_fig2}

%
\begin{figure}
\centering{}\includegraphics[width=1\columnwidth,height=1\paperheight,keepaspectratio]{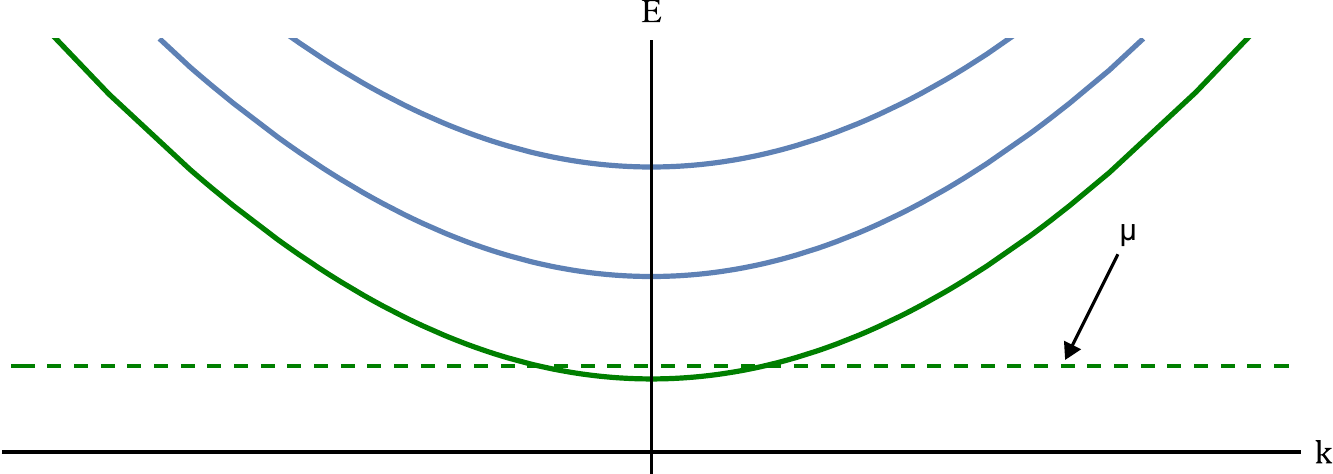}\caption{\label{fig:nonlin_planar_disp}
Finite dispersion in the first planar mode. Only the first channel is occupied as planar mode (solid green curve), where the chemical potential~$\mu$ (green dashed line) is tuned close to the minimum of the channel, such that dispersion is non-linear for this mode. Note that here, for illustrative purposes, the green dotted line is not drawn to scale (the distances from the chemical potential~$\mu$ to the minima of the first two modes actually are ${\mu_1/\Delta=-1.01}$ and ${\mu_2/\Delta=7319.5}$, respectively).
}
\end{figure}
%

\noindent
In Fig.~\ref{fig:ABS_spectrum_nonlinear} the single channel ABS spectrum is calculated with different sets of approximations applied, specifically for the L-junction geometry presented in Sec.~\ref{sec_L_junction}. The regime is such that dwelling is negligible, i.e. the superconductors, as depicted in Fig.~\ref{fig:geometric_junction}, are situated right at the scattering centre,~${j_\alpha=0}$. Additionally, only the first channel is occupied as a planar mode ${n_\mathrm{P}=1}$ (with ${N=5}$ transversal lattice points), exhibiting strong finite dispersion, such that the chemical potential~$\mu$ is tuned very closely above the minimum of that channel ~(${\mu_1/\Delta=-1.01}$, ${\mu_2/\Delta=7319.5}$), as illustrated in Fig.~\ref{fig:nonlin_planar_disp}.

First, the ABS spectrum is calculated using the standard Beenakker framework as outlined in Sec.~\ref{sec_status_quo}, which yields the black dotted curves in Fig.~\ref{fig:ABS_spectrum_nonlinear}. Accordingly, energy dependence is neglected for the scattering matrix and the reflection matrix is determined in the Andreev approximation, i.e. taking the form of Eq.~\eqref{eq:planar_Andreev_matrix}, where the reflection coefficient~$\alpha$ follows Eq.~\eqref{eq:alpha_factor_plane}. With these approximations, the spectrum can be computed by the explicit relation~\eqref{eq:conv_ABS_spectrum}. For the discrete L-junction model, the transmission coefficient~$T$ is determined numerically for zero energy,~${ E=0 }$, by the projection procedure introduced in Sec.~\ref{sec_L_junction} by Eq.~\eqref{eq:projection_procedure}.

Second, in order to compute the green curve in Fig.~\ref{fig:ABS_spectrum_nonlinear}, the energy dependence of the scattering matrix (arising from finite dispersion, not dwelling) is included, but the reflection matrix is still determined within the Andreev approximation. Now, the relation~\eqref{eq:conv_ABS_spectrum} is no longer applicable, such that the Beenakker Equation, Eq.~\eqref{eq:Beenakker_eq}, has to be solved directly, for which both~$R$ and~$S$ need to be computed first. The reflection matrix is determined by Eqs.~\eqref{eq:planar_Andreev_matrix} and~\eqref{eq:alpha_factor_plane}. However, the scattering matrix~${ S\left[ E \right] }$ is now computed for the whole subgap energy range~$E\in \left[-\Delta, \Delta \right]$, again numerically by Eq.~\eqref{eq:projection_procedure}.

Finally, the blue curve is obtained by additionally computing the reflection matrix~$R$ beyond the Andreev approximation. For this, the $R$-matrix takes on a generalised form where normal backscattering submatrices are added on its diagonal, cf. Eq.~\eqref{eq:gen_A_ref_matrix}, and the reflection coefficients~$\alpha$ no longer take on the form~\eqref{eq:alpha_factor_plane}. The reflection matrix can, then, either be solved numerically by a procedure similar to the scattering problem, but by projecting onto the SN-interface instead, Eq.~\eqref{eq:projection_procedure_SN}. Or, by applying the analytic expressions we where able to find for the Andreev- and backscattering coefficients in Sec.~\ref{sec:recipe}, Eqs.~\eqref{eq:alpha_factor_evan} and~\eqref{eq:beta_factor_evan}. For Fig.~\ref{fig:ABS_spectrum_nonlinear} we opted for the numeric version.

\section{Probability conservation for scattering of Gaussian wave packets}\label{sec:App_Gaussian_packages}

\noindent Let us consider a $1\mathrm{D}$ scattering example with two terminals, each with one channel. As for the scattering region, we assume that it extends from~${-L/2<x<L/2}$. Apart from that, we leave the nature of the scattering potential unspecified. Now, instead of connecting each terminal to superconductors, we simply consider conductors in the normal metal state. Hence, in the (now infinitely long) conductor arms, the wave function for a wave incident from the left hand side, is given as
\begin{align}\label{eq_wave_packet_ansatz_1}
  \psi_{x<-L/2}&=e^{ik\left(E\right)x}+R\left(E\right)e^{-ik\left(E\right)x}\\ \label{eq_wave_packet_ansatz_2}
  \psi_{x>L/2}&=T\left(E\right)e^{ik\left(E\right)x}\ .  
\end{align}
Since the central scattering region is arbitrary, we do not have a specific ansatz for the wave function for $-L/2<x<L/2$ (in fact, it is not required for the sake of our argument). The wave function, however, does include the possibility of a generic dispersion relation $k(E)$ and energy-dependant scattering coefficients, $R(E)$ and $T(E)$. Throughout this paper, we adopt the following language with regard to dispersion. If the dispersion relation for all relevant modes can be linearized, then we refer to this as the dispersion-free case, or to dispersion being absent (as wave packets here propagate over time without changing their width). If the linear approximation is invalid, we speak of a dispersive case, or finite dispersion.

To proceed, we expand all of these quantities up to first order, that is, $k(E)\approx k_F+E/v_F$\footnote{It might seem that expansion of $k(E)$ up to linear order is consistent with linear dispersion, but, as we will show shortly, this correction is enough to take into account the lowest order contributions to finite dispersion of wave packets.} (in the single channel limit, we only  have a single Fermi wave vector and Fermi velocity, $k_F$ and $v_F$) and $R(E)\approx R+E\delta R$ (likewise for $T$), where $E$ is still defined with respect to the chemical potential, such that this expansion corresponds to a Taylor series of $S$ in orders of $E$ around ${E=0}$.

We now initialise a Gaussian wave packet, centred around $E=0$,
\begin{equation}
    \psi_{P}\left(x,t\right)=\int_{-\mu}^{\infty}dE\frac{1}{\sqrt{2\pi}\delta E}e^{-\frac{E^{2}}{2\delta E^{2}}}e^{-iEt}\psi\left(x\right)\ .
\end{equation}
Note that the factor $1/\sqrt{2\pi}\delta E$ is added only for convenience, and does not normalise the wave function (as normalisation is not essential here). Up to linear order in $E$ this yields a wave function of the form
\begin{align}
    \psi_{P}\left(x<-L/2,t\right)&\approx e^{ik_{F}x}g\left(x-v_{F}t\right)\nonumber\\&+e^{-ik_{F}x}\left[R+iv_{F}\delta R\partial_{x}\right]g\left(x+v_{F}t\right)\label{eq_gauss_left}  , \\\label{eq_gauss_right}\psi_{P}\left(x>L/2,t\right)&\approx e^{ik_{F}x}\left[T-iv_{F}\delta T\partial_{x}\right]g\left(x-v_{F}t\right)  ,
\end{align}
with the shortcut definition~${g\left(y\right)=\exp[-\delta E^{2}y^{2}/(2v_{F}^{2})]}$ for the Gaussian. 
We see that the linear corrections $\delta R$ and $\delta T$ give rise to a small deviation of the wave packet from a clean Gaussian (depending on the derivative of the Gaussian $\partial_x g$), which emerge upon impact of the wave packet at the scattering region (at around $t=0$). A revealing quantity to look at is the total overlap
\begin{equation}
    \int_{-\infty}^\infty dx \vert\psi_P(x,t)\vert^2=P_S+P_\text{arms} \ ,
\end{equation}
which we separate into the probability of the packet residing within the central scattering region,~$ P_S=\int_{-L/2}^{L/2} dx \vert\psi_P(x,t)\vert^2 $, and the probability of finding the electron in the conductor arms,
\begin{equation}
    P_\text{arms}=\left[\int_{-\infty}^{-L/2} dx+\int_{L/2}^\infty dx\right] \vert\psi_P(x,t)\vert^2\ .
\end{equation}
Obviously, the sum $P_S+P_\text{arms}$ needs to stay constant due to probability conservation. Inserting Eqs.~\eqref{eq_gauss_left} and~\eqref{eq_gauss_right} into $P_\text{arms}$, and expanding in orders of $\delta R,\delta T$, $P_\text{arms}=P_\text{arms}^{(0)}+P_\text{arms}^{(1)}+\mathcal{O}[\delta R^2,\delta T^2,\delta R\delta T]$, we get in zeroth order
\begin{align}
    P_\text{arms}^{(0)}=\int_{-\infty}^{0}dxg^{2}\left(x-v_{F}t\right)&+\left|R\right|^{2}\int_{-\infty}^{0}dxg^{2}\left(x+v_{F}t\right) \nonumber \\&+\left|T\right|^{2}\int_{0}^{\infty}dxg^{2}\left(x-v_{F}t\right) \ ,
\end{align}
whereas first order yields (assuming $k_F$ large -- an assumption which will be important in a moment)
\begin{equation}\label{eq_P_arms_1}
    P_\text{arms}^{(1)}=\frac{iv_{F}}{2}\left[\delta RR^{*}-R\delta R^{*}+\delta TT^{*}-T\delta T^{*}\right]g^{2}\left(v_{F}t\right)\ .
\end{equation}
The zeroth order term is manifestly constant in time if~${\vert R\vert^2+\vert T\vert^2=1}$, which is just a reiteration of the connection between unitarity of the scattering matrix and probability conservation.

Crucially, we note that probability is not only conserved in the asymptotic limit of times long after the scattering event (impact of the Gaussian at the scattering center), but for \textit{all} times. This is where the first order term~${ P_\text{arm}^{(1)} }$ comes into play, as it provides a temporary contribution to the probability (around the time of impact at $t\approx 0$). Note that this term does not vanish due to unitarity of the scattering matrix. Consequently, there is a small dip in the probability of the electron being in the conductor arms, when the wave packet hits the scattering region (at times around $t=0$). Hence, the only solution to conserve the total probability seems to be that there is a finite probability for the electron to temporarily occupy the central region, $P_S\sim g^2(v_F t)$. This is exactly the prevailing picture in the existing literature, and the reason why a finite $E$-dependence of the scattering matrix is commonly associated with a finite dwelling time of the electron inside the scattering region (or, equivalently, a finite Thouless energy).

We here present an important caveat to the above picture. Consider, for instance, a scattering region with infinitesimal size, such as a Dirac delta function potential as a well-documented textbook example, where the scattering coefficients depend explicitly on $k$, and thus, on $E$. Another example will be treated below: nanobridges with nontrivial device geometry. In all of these examples the electron spends manifestly negligible time inside the scattering region, $P_S\rightarrow 0$ (such that it is justified to assume the scattering size $L$ to be zero). Here, the above calculation is at risk of creating a conundrum in the form of a non-conserved probability,  as the temporal ``dip'' in probability $\sim g^2(v_F t)$ in Eq.~\eqref{eq_P_arms_1} would in this case go unaccounted for. But this looming contradiction can be resolved very simply by undoing an approximation we made when deriving Eq.~\eqref{eq_P_arms_1}: assuming $k_F$ infinitely large. Including instead a finite $k_F$, one can see that integrands of the form $\sim e^{\pm i2k_F x}g(x-v_F t)g(x+v_F t)$, which were previously neglected, lead to an important correction of the same order. Namely, we now find (assuming $\delta E\ll v_F k_F$)
\begin{equation}
\begin{split}
    P_\text{arms}^{(1)}=\frac{iv_{F}}{2}\left[\frac{R-R^*}{v_Fk_F}+\delta RR^{*}-R\delta R^{*}\right.\\\left.+\delta TT^{*}-T\delta T^{*}\right]g^{2}\left(v_{F}t\right)\ .
\end{split}
\end{equation}
The new term we neglected before is $\sim R-R^*$. Setting $P_S=0$, probability conservation now imposes a new type of condition,
\begin{equation}\label{eq_R_condition}
    \frac{R-R^*}{v_Fk_F}+\delta RR^{*}-R\delta R^{*}+\delta TT^{*}-T\delta T^{*}=0\ ,
\end{equation}
which thus relates the zeroth order scattering matrix elements, $R,T$, to its first order corrections, $\delta R,\delta T$, in a way which could not possibly be derived from unitarity of the scattering matrix alone (simply because it is less general, requiring $P_\text{S}=0$ as an additional assumption). Note that a similar condition can be derived, when including the state with an incoming wave from the right hand side,
\begin{align}
  \psi^\prime_{x<-L/2}&=T^\prime \left(E\right)e^{-ik\left(E\right)x}\\
  \psi^\prime_{x>L/2}&=e^{-ik\left(E\right)x}+R^\prime\left(E\right)e^{ik\left(E\right)x}\ .  
\end{align}
Assuming a symmetric geometry and time-reversal symmetry, one can set $R^\prime=R$ and $T^\prime=T$. Now, from orthogonality $\int dx \psi_P^*(x)\psi_P^\prime(x)=0$, and yet again neglecting the contribution from the central scattering region, we find in addition 
\begin{equation}\label{eq_T_condition}
    \frac{T-T^*}{v_Fk_F}+\delta TR^{*}-T\delta R^{*}+\delta RT^{*}-R\delta T^{*}=0\ .
\end{equation}
Stitching Eqs.~\eqref{eq_R_condition} and~\eqref{eq_T_condition} together, we can formulate a condition in terms of the full electron scattering matrix as
\begin{equation}
    \frac{S_e-S_e^*}{2\mu}+\delta S_e S_e^* -S_e\delta S_e^*=0\ ,
\end{equation}
where we have used the identity $v_F k_F=2\mu$.

\section{Matching conditions at the scattering centre for the discrete model} \label{sec_Ljunction_scattering}

\noindent
The discrete tight-binding Hamiltonian for $x$- and $y$-direction respectively is
\begin{equation}
    H_{\mu}=-t\sum_{j_{\mu}}\bigg(|j_{\mu}\rangle\langle j_{\mu}-1|+|j_{\mu}-1\rangle\langle j_{\mu}|-2|j_{\mu}\rangle\langle j_{\mu}| \bigg)\ ,
    \label{eq:discrete_Hamiltonian}
\end{equation}
where~${ j_\mu \in \{ j_x,j_y \} }$ are the integer-valued lattice site indices and~${ t=1/(2m) }$ is the hopping amplitude. We added the constant term $-2|j\rangle\langle j|$ for convenience, as it shifts the lower bound of the energy spectrum to $0$, such that the continuum limit can be taken straightforwardly (without redefinition of the chemical potential). The indices $j_\mu$ are limited by the device geometry. We choose the following discrete coordinate system: in the left conductor arm, $j_y=\{1,\ldots,N\}$ ($N$ being the number of transversal lattice sites) and $j_x\leq 0$. In the bottom conductor arm we conversely have $j_x=\{1,\ldots,N\}$ and $j_y\leq 0$. The scattering region is thus a square lattice with $j_{x,y}=\{1,\ldots,N\}$. 
The eigenenergies of $H_N$ thus are
\begin{equation}
    \xi_{k,n} =	\underbrace{-2t\left[\cos\left(k\right)-1\right]}_{\epsilon_{k}}\underbrace{-2t\left[\cos\left(\frac{\pi n}{N+1}\right)-1\right]}_{\epsilon_{n}}-\mu \ ,
\label{eq:disc_dispersion_NC}
\end{equation}
in analogy to Eq.~\eqref{eq:dispersion_NC}, where the quantised transversal contribution~$\epsilon_{n}$ (due to hard wall boundary conditions) gives rise to channels. Note that just like in the continuum case we also here define~${\mu_n=\epsilon_n-\mu}$. The $n$th component of the eigenstate for the horizontal arm is
\begin{align}
|k_n\rangle\otimes|n\rangle = \sum_{j_{x}}\sum_{j_{y}}\underbrace{e^{\mathrm{i}k_{n}j_{x}}}_{\mathrm{long}} \underbrace{\sin\left(\frac{\pi n}{N+1}j_{y}\right)}_{\mathrm{trans}}|j_{x}\rangle\otimes|j_{y}\rangle   \ , \label{eq:disc_eigenstates_NC}
\end{align}
where $k_{n}$ is energy-dependant, and can be obtained by inverting the dispersion, Eq. \eqref{eq:disc_dispersion_NC}, such that 
\begin{align}    \cos\left(k_{n}\right)=\frac{\mp E-\mu_{n}+2t}{2t} \ .
\label{eq:disc_N_momenta}
\end{align}
Here, choosing the negative sign constitutes the electron momentum $k_{\mathrm{e},n}$ whereas the positive sign corresponds to the hole momentum $k_{\mathrm{h},n}$.
For the vertical arm the eigenstate is the same but with interchanging $x\leftrightarrow y$.

The scattering problem is now solved as follows. We take a generic ansatz as a superposition of the eigensolutions in Eq.~\eqref{eq:disc_eigenstates_NC} for the two conductor arms (left and vertical), which corresponds to a given scattering process (e.g., incoming planar from the left, or incoming evanescent from the vertical arm, and so forth). The ansatzes in either arm, then, consist of a superposition of different modes and will have the following forms, accordingly
\begin{align}
    |\psi_{n}^{\mathrm{\left(0\right)}}\rangle=&\underbrace{|k_{n}\rangle\otimes|n\rangle}_{\text{incoming}}+\underbrace{\sum_{m}R_{n,m}|-k_{m}\rangle\otimes|m\rangle}_{\text{reflected}} \ , \nonumber \\|\psi_{n}^{\mathrm{\left(1\right)}}\rangle=&\underbrace{\sum_{m}T_{n,m}|-k_{m}\rangle\otimes|m\rangle}_{\text{transmitted}} \ .
    \label{eq:disc_scattering_ansatz}
\end{align}
Here, if an electron is incident from the left arm, the ansatz in the left (vertical) arm is denoted as $|\psi_{n}^{\mathrm{\left(0\right)}}\rangle$ ($|\psi_{n}^{\mathrm{\left(1\right)}}\rangle$), whereas for the opposite process the two need to be interchanged. The coefficient $R_{n,m}$ ($T_{n,m}$) denotes reflection (transmission) from the~$n$th to the~$m$th channel.  For evanescent channels, momentum needs to be replaced as~$ {k_n \to \mathrm{i}\kappa_n }$\footnote{The same convention will also be applied below for both Eq.~\eqref{eq:discrete_continuity_cond} and Eq.~\eqref{eq:discrete_diff_cond}.}. 
Note that in the coordinate system used here, modes incoming from the left arm are transmitted towards negative $y$-direction in the vertical arm and therefore are associated with negative momentum, and vice versa (as depicted in Figure~\ref{fig:geometric_junction}).

Also, here and in the following, wave functions will be presented in terms of electron excitations only, for the sake of simplicity. Whenever considering planar hole modes, momentum has to switch sign~${ k_{\mathrm{e},n} \to -k_{\mathrm{h},n} }$, since within the framework (positively charged) holes are treated as negative charges with propagation in opposite direction. Very importantly, however, evanescent modes do not flip sign~${ \kappa_{\mathrm{e},n} \to \kappa_{\mathrm{h},n} }$, since sign convention is such that real exponentials always decay in direction associated with reflection or transmission, as discussed in Sec.~\ref{sec:recipe} and depicted in Figure~\ref{fig:evanescent_scattering}. Ultimately, the hole scattering matrix $S$ (including evanescent modes) must still fulfil the particle-hole symmetry $S^*(-E)=S(E)$, as it did for the case of exclusively planar modes. 

In order to solve for the scattering, the wave functions in the normal conductor arms can be propagated right up to the diagonal~${ j_x=j_y }$ (cf. Fig.~\ref{fig:geometric_junction}), that is, the ansatz of the left arm can be simply continued up to the diagonal for all lattice points $j_x\leq j_y$, and the ansatz for the vertical arm is continued for all lattice points $j_y\leq j_x$. To understand this, note that the tight-binding Hamiltonian, Eq.~\eqref{eq:discrete_Hamiltonian}, essentially links lattice sites in a cross-shaped arrangement (which we refer to as plaquettes, see Fig.~\ref{fig:geometric_junction}), representing tunnel hopping in $x$ and $y$ direction. If the wave function values at four out of the five lattice points of a given plaquette are known, then the fifth can be inferred from the Schrödinger equation. At the diagonal itself, the two arms must be matched with two independent matching conditions. The first condition demands~${ \left\langle j_x,j_y\right| \psi_{n}^{\mathrm{\left(0\right)}} \rangle = \left\langle j_x,j_y\right| \psi_{n}^{\mathrm{\left(1\right)}} \rangle  }$ at the diagonal which yields the following set of conditions
\begin{align}
    &e^{\mathrm{i}k_{n}j}\sin\left(\frac{\pi j}{N+1}\right)  \nonumber \\
    &=\sum_{m=1}^Ne^{-\mathrm{i}k_{m}j}\sin\left(\frac{\pi mj}{N+1}\right)\left(T_{n,m}-R_{n,m}\right) \ ,
    \label{eq:discrete_continuity_cond}
\end{align}
where~${ n \in \left\{ 1,\dots,N \right\} }$. Thus, for each of the~$N$ lattice points on the diagonal, the above condition is demanded to hold for all of the~$N$ channels, amounting to a total of~$N^2$ conditions. A second set of an additional~$N^2$ conditions can be obtained by demanding Schrödinger equation to hold locally and projecting onto the individual lattice sites of the diagonal~${ j_x=j_y }$ such that
\begin{equation}
       \langle j,j | H_\mathrm{N}	|\psi_{n}\rangle=\langle j,j|E|\psi_{n}\rangle \ . \label{eq:projection_procedure} 
\end{equation}
Again, the Hamiltonian as defined by Eqs.~\eqref{eq:n_region_hamiltonian} and~\eqref{eq:discrete_Hamiltonian} connects adjacent lattice points, such that for different parts of the resulting plaquettes (depicted in Fig.~\ref{fig:geometric_junction}) ansatz $|\psi_{n}^{\mathrm{\left(0\right)}}\rangle$ or $|\psi_{n}^{\mathrm{\left(1\right)}}\rangle$ is valid respectively, whereas right on the diagonal both ansatzes apply (as per Eq.~\eqref{eq:discrete_continuity_cond}). The resulting set of conditions reads

\begin{align}
0&=  e^{\mathrm{i}k_{n}j}\left[-\mathrm{i}\,\sigma\!\left(k_{n}\right)\sin\left(k_{n}\right)\sin\left(\frac{\pi nj}{N+1}\right)\right.\nonumber\\
 &+ \left.\sin\left(\frac{\pi n}{N+1}\right)\cos\left(\frac{\pi nj}{N+1}\right)\right]\nonumber\\
 &+ \sum_{m}\left(T_{n,m}+R_{n,m}\right)e^{-\mathrm{i}k_{m}j}\left[\mathrm{i}\,\sigma\!\left(k_{n}\right)\sin\left(k_{n}\right)\sin\left(\frac{\pi mj}{J+1}\right)\right.\nonumber\\
 &+ \left.\sin\left(\frac{\pi m}{N+1}\right)\cos\left(\frac{\pi mj}{N+1}\right)\right],\label{eq:discrete_diff_cond}
\end{align}
%
%
where $\sigma( k_n )=k_{n}^{2}/|k_{n}|^{2}$
assigns a positive or negative sign depending on whether momentum is real or imaginary.
By the conditions~\eqref{eq:discrete_continuity_cond} and~\eqref{eq:discrete_diff_cond} all~$2N^2$ parameters~$T_{n,m}$ and~$R_{n,m}$ can be determined. This allows for numerically computing the total scattering matrix for different energies within the subgap region~${ \left| E \right| < \Delta }$. We note that in the continuum limit, the two conditions given in Eqs.~\eqref{eq:discrete_continuity_cond} and~\eqref{eq:discrete_diff_cond} correspond to continuity of the wave function and its derivative, respectively.

 By the above numerical procedure, the scattering matrix needs to be evaluated for each energy $E$ separately. However, we can significantly reduce computation time by exploiting the square root behaviour deduced in Sec.~\ref{sec_cross_channel_evanescent}. Thus, the scattering problem only needs to be solved for the energies~${ E = 0 }$ and~${ E=\mu_\text{crit} }$, and the region in between, then, can be interpolated by the square root law. Consequently, the coefficients of the electron scattering matrix can be inferred as a function of energy by the following analytic relation
\begin{align}
     \left| S^{i,j}_\mathrm{e} \left[ E \right] \right| = a^{i,j}+b^{i,j} \sqrt{\left| 1- \frac{E}{\mu_\text{crit}} \right|} \text{ ,} \label{eq:sqrt-relation}
\end{align}
where the matrices
\begin{align}
    a^{i,j}= & \left| S^{i,j}_\mathrm{e} \left[ E=\mu_\text{crit} \right] \right| \, , \nonumber \\ 
    b^{i,j}= & \left| S^{i,j}_\mathrm{e} \left[ E=0 \right] \right| - \left| S^{i,j}_\mathrm{e} \left[ E=\mu_\text{crit} \right] \right| \, ,
\end{align}
explicitly capture the deviation across the whole energy range from ${E=0}$ to $\mu_\text{crit}$. This holds in complete analogy for ${ \mathrm{Arg}\left[ S^{i,j}_\mathrm{e} \left[ E \right] \right] }$, such that the full electron scattering matrix can be assembled as ${ S^{i,j}_\mathrm{e} \left[ E \right] = \left| S^{i,j}_\mathrm{e} \left[ E \right] \right| \cdot \mathrm{exp} \left[ \mathrm{Arg}\left[ S^{i,j}_\mathrm{e} \left[ E \right] \right] \right] }$.
The hole components of the scattering matrix can be calculated by using particle-hole symmetry. In Fig.~\ref{fig:sqrt-law}, we explicitly illustrate the very good agreement between the square root behaviour and the numerical calculation, with the example of the planar-planar reflection coefficient $R_{1,1}$. Here, the regime is such that the first channel is planar~$\mu_1/\Delta \ll 0$, the second channel is evanescent with finite dispersion~$\mu_2/\Delta \gtrapprox 0$, and the width of the arms is~${N=5}$ lattice sites.

\section{Matching conditions at the SN-interface for the discrete model} \label{sec:App_R_reflection}

\noindent
The reflection processes at the SN-interfaces can, in the discrete model, be solved by a projection procedure similar to that of the normal scattering above. However, we now need to additionally consider the superconductor regions which are subject to the Bogoliubov-de Gennes Hamiltonian
\begin{equation}
H_\mathrm{BdG}=	\left(\begin{array}{cc}
H_{N} & \text{e}^{\text{i}\phi}\Delta\\
\text{e}^{-\text{i}\phi}\Delta & -H_{N}
\end{array}\right) \ ,
\label{eq:BdG_Hamiltonian}
\end{equation}
where~$H_{N}$ is the single particle normal conductor Hamiltonian such that~$\Delta=0$ for~$j>j_\alpha$. In the discrete limit,~$H_{N}$ is defined by Eqs.~\eqref{eq:n_region_hamiltonian} and~\eqref{eq:discrete_Hamiltonian}, and the eigenenergies~$\mathcal{E}_{k,n}$ are still in accordance with Eq.~\eqref{eq:dispersion_SC} but with~${ \xi_{k,n}}$ defined by Eq.~\eqref{eq:disc_dispersion_NC}. Consequently, for subgap energies~${ |E|<\Delta }$ we have
\begin{align}
  \xi_{k_{\mathrm{R}},k_{\mathrm{I}}}=&-2\text{i}t\sin\left(k_{\mathrm{R}}\right)\sinh\left(k_{\mathrm{I}}\right) \nonumber \\
  =&-\mathrm{Sign}\left(k_{\mathrm{R}}\right)*\mathrm{Sign}\left(k_{\mathrm{I}}\right)*\text{i}\sqrt{\Delta^{2}-E^{2}} \ ,
\end{align}
where momentum turns fully complex $k=k_{\mathrm{R}}-\text{i}k_{\mathrm{I}}$ as discussed above.
In analogy to Eq.~\eqref{eq:cont_SC_momenta}, solving the discrete version of the dispersion for the real and imaginary momentum components yields~${ k_{\mathrm{R},n}=\widetilde{k}_{+,n} }$ and~${ k_{\mathrm{I},n}=-\mathrm{i}\widetilde{k}_{-,n} }$, where
\begin{align}
    \mathrm{sin}\left(\widetilde{k}_{\pm,n}\right)&=\mathrm{sign}\left(\widetilde{k}_{\pm,n}\right)*\sqrt{\frac{1}{2}P\pm\sqrt{\left(\frac{P}{2}\right)^{2}-Q}}\text{ ,}
    \label{eq:disc_SC_momenta}
\end{align}
with
\begin{align}
    P & =-\frac{\mu_{n}\left(4t+\mu_{n}\right)}{4t^{2}}+Q\ , \nonumber \\
Q & =-\frac{\Delta^{2}-E^{2}}{4t^{2}} \ .
\end{align}
Since Andreev reflection is assumed to be translational invariant, as discussed in Section~\ref{sec_status_quo}, the wave function can be separated and the transversal component can be discarded, such that reflection at the SN-interface effectively reduces to a $1\mathrm{D}$~problem.  Accordingly, the longitudinal component of the eigenstates in the S-regions is 
\begin{align}
 \left(\begin{array}{c}
u_{k,n}\\
v_{k,n}
\end{array}\right)|k_n\rangle 
\ ,
\end{align}
with the Bogoliubov coefficients
\begin{align}
    \left(\begin{array}{c}
u_{k,n}\\
v_{k,n}
\end{array}\right)=&\left(\begin{array}{c}
\mathrm{Sign}\left[E\right]\sqrt{1+\mathrm{Sign}\left[E\right]\frac{\xi_{k,n}}{\sqrt{\Delta^{2}+\xi_{k,n}^{2}}}}\\
\text{e}^{-\text{i}\phi}\sqrt{1-\mathrm{Sign}\left[E\right]\frac{\xi_{k,n}}{\sqrt{\Delta^{2}+\xi_{k,n}^{2}}}}
\end{array}\right)\text{ .}
\end{align}
In order to solve for the Andreev reflection taking place at the SN-interfaces ${ j=j_\alpha }$, let us now consider the superconductor to extend over~$\left[-\infty,j_{\alpha}\right]$ as depicted in Figure~\ref{fig:geometric_junction}. The ansatz for the wave function corresponding to the superconductor side of the interface (for subgap energies), then, is
\begin{align}
|\psi^\mathrm{SC}_n\rangle=&	\left[\frac{A_{n}}{\sqrt{2}}\left(\begin{array}{c}
u_{-k_{\mathrm{R}},n}\\
v_{-k_{\mathrm{R}},n}
\end{array}\right)|-k_{\mathrm{R},n}-\text{i}k_{\mathrm{I}
,n}\rangle 
\right. \nonumber \\
+&	\left.\frac{B_{n}}{\sqrt{2}}\left(\begin{array}{c}
u_{k_{\mathrm{R},n}}\\
v_{k_{\mathrm{R},n}}
\end{array}\right)|k_{\mathrm{R},n}-\text{i}k_{\mathrm{I},n}\rangle 
\right] \ . \label{eq:disc_ansatz_SC}
\end{align} 
Accordingly, the above ansatz represents a superposition of plane waves propagating to the right and left, enveloped by a real exponential decaying towards~$j\to -\infty$. For the normal conductor side, let us consider the example of an incident electron
\begin{align}
|\psi_{n}^{\mathrm{NC}}\rangle=\left(\begin{array}{c}
|-k_{\mathrm{e},n}\rangle+r_{n}|k_{\mathrm{e},n}\rangle\\
r_{\alpha,n}|-k_{\mathrm{h},n}\rangle
\end{array}\right) 
\ ,
\label{eq:disc_ansatz_SN_N_region}
\end{align}
where $r_{n}$ is the coefficient for normal electron-electron backscattering and $r_{\alpha,n}$ is the electron-hole Andreev reflection coefficient. As discussed in the remarks following Eq.~\eqref{eq:disc_scattering_ansatz}, the hole reflected in positive $j$-direction still has negative momentum $-k_{\mathrm{h},n}$, since it is treated as negative charge with propagation in opposite direction. The case of an incident hole works completely analogously. Likewise, for evanescent Andreev reflection, the momenta have to be replaced as ${k\to\pm \kappa}$ where the sign has to be chosen such that reflected components always drop off by which incoming modes always pierce the interface with the tail of the exponential, as discussed above after Eq.~\eqref{eq:disc_scattering_ansatz} and depicted in Figure~\ref{fig:evanescent_scattering}. 
Projecting onto the SN-interfaces of each contact at~${ j_\mu=j_\alpha }$, see also Fig.~\ref{fig:geometric_junction}, yields the following set of $2N$~conditions
\begin{align}
    \langle j_\alpha|H_\mathrm{BdG}	|\psi_{n}\rangle=\langle j_\alpha|E|\psi_{n}\rangle \ ,
    \label{eq:projection_procedure_SN}
\end{align}
which correspond to the differentiability constraints in the continuum limit. Depending on which side of the interface is involved, either the ansatz for the superconductor~$|\psi^\mathrm{SC}_n\rangle$ or the normal conductor region~$|\psi_{n}^{\mathrm{NC}}\rangle$ is applied, accordingly. With the SN-interface being situated a finite distance of lattice points~$j_\alpha$ from the scattering region, the wave function must be propagated in between by an operator~$P$. In the absence of impurities (our default assumption for simplicity), this propagator~$P$ consists of a diagonal matrix with elements~${P_{n,n}\left[ j \right]=\mathrm{e}^{-K_{\nu,n} j}}$, where the momentum~${K_{\nu,n}}$ is determined as defined by Eq.~\eqref{eq:momentum_rules}. For instance, if $\psi^\mathrm{in}$ is the incident wave vector right at~${j=0}$, it would be subject to the eigenvector equation $PRP^{-1}S\psi^\mathrm{in}=\psi^\mathrm{in}$. In the standard Beenakker formalism as well as in our generalised framework above, the propagator is usually absorbed into the $S$-matrix (or, alternatively, in~$R$).

While it is possible to derive a closed expression for the coefficients of the discrete SN interface, the respective formulas are impractically bulky (which is why they are not explicitly derived here, and instead the lattice model coefficients are determined numerically). For comparison and in order to simplify the calculation, we can, however, use the analytic Andreev reflection- and backscattering coefficients~$ \alpha_{\nu,n}$, $\beta_{\nu,n} $  from the continuous model presented in Sec.~\ref{sec:recipe}, cf. Eqs.~\eqref{eq:alpha_factor_evan} and~\eqref{eq:beta_factor_evan}, and simply insert the respective dispersion relations for electrons and holes from the lattice. It turns out that even if taking a conductor arm width of just~${N=5}$ lattice points the analytic approximation (solid curves) and the numerical computation (dashed curves) are in very good agreement (see Fig.~\ref{Fig:analytic_alpha_beta}). This is due to the fact that the chemical potential is chosen close to a change of the channel number, where the dispersion relation of the relevant channel (here the second channel, $n=2$) is to a very good accuracy parabolic.


\bibliographystyle{longapsrev4-2}
\bibliography{sn-bibliography}

\end{document}